\documentclass[twocolumn,amsmath,amssymb,aps,prd,10pt,superscriptaddress,nofootinbib,noeprint,preprintnumbers,floatfix,notitlepage]{revtex4-2}

\pdfoutput=1

\usepackage{graphicx}
\usepackage{amssymb}
\usepackage{amsmath}
\usepackage{mathtools}  
\usepackage{multirow}
\usepackage{soul} 
\usepackage{amsfonts}
\usepackage{graphicx}
\usepackage{color}
\usepackage{bm}
\usepackage[utf8]{inputenc}
\usepackage{hyperref}
\usepackage{subdepth} 
\usepackage{bbold}

\usepackage[letterspace=-10]{microtype} 

\graphicspath{{./figs/}}

\definecolor{jlab_red}{RGB}{192,39,45}
\definecolor{jlab_orange}{RGB}{249,102,0}
\definecolor{jlab_blue}{RGB}{47,122,121}
\definecolor{jlab_green}{RGB}{65,125,10}

\newcommand{\cm}{\ensuremath{\mathsf{cm}}}

\renewcommand{\star}{\ensuremath{\ast}}

\newcommand{\ccz}{\ensuremath{\chi_{c0}}}
\newcommand{\cco}{\ensuremath{\chi_{c1}}}
\newcommand{\cct}{\ensuremath{\chi_{c2}}}
\newcommand{\cce}{\ensuremath{\chi_{c3}}}

\newcommand{\etap}{\ensuremath{\eta^\prime}}

\newcommand{\DD}{\ensuremath{{D\bar{D}}}}
\newcommand{\DsDs}{\ensuremath{{D_s\bar{D}_s}}}
\newcommand{\DDst}{\ensuremath{{D\bar{D}^\star}}}
\newcommand{\DstDst}{\ensuremath{{D^\star\bar{D}^\star}}}
\newcommand{\DsDsst}{\ensuremath{{D_s\bar{D}_s^\star}}}
\newcommand{\DsstDsst}{\ensuremath{{D^\star_s\bar{D}_s^\star}}}
\newcommand{\DDz}{\ensuremath{D\bar{D}^\star_0}}
\newcommand{\psiom}{\ensuremath{{\psi\omega}}}
\newcommand{\psiphi}{\ensuremath{{\psi\phi}}}
\newcommand{\etce}{\ensuremath{{\eta_c\eta}}}
\newcommand{\etcep}{\ensuremath{{\eta_c\eta^\prime}}}

\newcommand{\SLJ}[3]{\ensuremath{{\:\!}^{{#1}\!}{#2}_{#3}}}
\newcommand{\SLJc}[3]{\ensuremath{ { \{  {\:\!}^{{#1}\!}{#2}_{#3}  \}  } } }
\newcommand{\SLJC}[3]{\ensuremath{ { \{  {\:\!}^{{#1}\!}{#2}_{#3}  \}  } } }

\hypersetup{%
pdftitle = {charmonium scattering 840},
pdfauthor = {Hadron Spectrum Collaboration},
colorlinks = {true},
filecolor = {black},
linkcolor = {jlab_blue},
menucolor = {black},
citecolor = {jlab_green},
urlcolor = {jlab_green},
}{}

\begin{document}

\title{Charmonium $\chi_{c0}$ and $\chi_{c2}$ resonances in coupled-channel scattering from lattice QCD}

\author{David~J.~Wilson}
\email{d.j.wilson@damtp.cam.ac.uk}
\affiliation{DAMTP, University of Cambridge, Centre for Mathematical Sciences, Wilberforce Road, Cambridge, CB3 0WA, UK}
\author{Christopher~E.~Thomas}
\email{c.e.thomas@damtp.cam.ac.uk}
\affiliation{DAMTP, University of Cambridge, Centre for Mathematical Sciences, Wilberforce Road, Cambridge, CB3 0WA, UK}
\author{Jozef~J.~Dudek}
\email{dudek@jlab.org}
\affiliation{\lsstyle Thomas Jefferson National Accelerator Facility, 12000 Jefferson Avenue, Newport News, VA 23606, USA}
\affiliation{Department of Physics, College of William and Mary, Williamsburg, VA 23187, USA}
\author{Robert~G.~Edwards}
\email{edwards@jlab.org}
\affiliation{\lsstyle Thomas Jefferson National Accelerator Facility, 12000 Jefferson Avenue, Newport News, VA 23606, USA}

\collaboration{for the Hadron Spectrum Collaboration}

\date{\today}


\begin{abstract}
\noindent
In order to explore the spectrum of hidden-charm scalar and tensor resonances, we study meson-meson scattering with $J^{PC}=0^{++}, 2^{++}$ in the charmonium energy region using lattice QCD. Employing a light-quark mass corresponding to $m_\pi \approx 391$ MeV, we determine coupled-channel scattering amplitudes up to around 4100 MeV considering all kinematically relevant channels consisting of a pair of open-charm mesons or a charmonium meson with a light meson. A single isolated scalar resonance near 4000 MeV is found with large couplings to $D\bar{D}$, $D_s \bar{D}_s$ and the kinematically closed $D^* \bar{D}^*$ channel. A single tensor resonance at a similar mass couples strongly to $D\bar{D}$, $D\bar{D}^*$ and $D^* \bar{D}^*$. We compare the extracted resonances to contemporary experimental candidate states, previous lattice results and theoretical modeling. In contrast to several other studies, we do not find any significant feature in the scalar amplitudes between the ground state $\chi_{c0}(1P)$ and the resonance found around 4000~MeV.
\end{abstract}
                
\maketitle

\section{Introduction}                                     

\noindent
In 2003, the discovery of the $X/\cco(3872)$ thrust hadron spectroscopy into a new era. Beginning with the \mbox{$B$-factories} and later at BES-III and LHCb, many hadrons with masses consistent with containing a charm-anticharm quark pair have been found in places that were not expected within existing models of charmonium. To date, no theoretical picture has explained the complete pattern of observed states, which have been dubbed the ``XYZ''~\cite{Brambilla:2019esw}. 

One key question concerns the relationship between newly observed hadrons and nearby hadron-hadron thresholds. Many states are found in close proximity to thresholds, but is this merely a coincidence or is it the presence of the threshold that drives the existence of the state? One might anticipate that the simplest place to begin to answer this would be close to the lowest open-charm threshold, i.e. where $D\bar{D}$ is produced. However presently neither the experimental landscape, nor our theoretical understanding of this energy region is clear. In the $\ccz$ and $\cct$ channels, where isoscalar $\DD$ interact in $S$ and $D$-wave respectively, several candidate states have been reported experimentally.

In the $\ccz$ channel, above the unambiguous ground state $\ccz(1P)$ at 3415~MeV, while simple $c\bar{c}$ quark models would expect an isolated $2P$ state near 3920 MeV~\cite{Godfrey:1985xj}, recent experimental analyses suggest multiple candidate states. The lightest, $\ccz(3860)$, is so far claimed by only one experiment~\cite{Belle:2017egg}, in $e^+e^-\to J/\psi\,\DD$, appearing as a rather broad resonance. Suggestions of a similar feature have also been made for the $\gamma\gamma\to D\bar{D}$ process measured at both Belle~\cite{Belle:2005rte} and BaBar~\cite{BaBar:2010jfn}, although the resonant interpretation is ambiguous, with much of the structure apparently driven by the Born-term. 
The $\ccz(3860)$ is \emph{not} seen in a recent evaluation of $B$ decays to $D^+D^-K^+$ by the LHCb experiment~\cite{LHCb:2020pxc}, where such a state might be expected to play significant role. Inclusive production of $\DD$ close to threshold shows an enhancement close to $D^0\bar{D}^0$ threshold, but this is explained in terms of ``feed-down'' from decays of the $X/\cco(3872)$~\cite{LHCb:2019lnr}, and no additional scalar resonance is needed to describe the data.
Theoretical re-analyses of the experimental data, in particular $\gamma\gamma\to\DD$, suggests the energy-dependence used to motivate a broad resonance $\ccz(3860)$, may actually belong to a \emph{sub-threshold} pole that behaves like a bound-state in the $D\bar{D}$ channel~\cite{Wang:2020elp,Deineka:2021aeu}. 

Above 3900~MeV there are strong experimental signals for new resonances, although it is not clear how features observed in different hadron-hadron channels relate to each other. The LHCb study of the $D^+D^-K^+$ final state identifies overlapping narrow $J^{PC}=0^{++}$ and $2^{++}$ resonances decaying to $D^+D^-$ with masses 3924 MeV and 3927 MeV~\cite{LHCb:2020pxc} respectively. Analysis of three-body final states requires the cross-channel amplitudes, in this case $DK$, to be modeled, and it is not obvious how sensitive the need for both scalar and tensor $\chi_c$ resonances is to the details of this modeling. 
A recent LHCb analysis proposes a state decaying to $\DsDs$ around 3960~MeV~\cite{LHCb:2022aki}, which is suggested to be a separate state to the one reported in $D^+ D^-$, although other studies suggest the $\DD$ and $\DsDs$ enhancements could be related to a single resonance pole~\cite{Guo:2022zbc,Ji:2022vdj}.
Earlier results from Belle~\cite{Belle:2009and} indicate a low-statistics enhancement in $\gamma \gamma \to J/\psi\:\omega$ around 3915~MeV. This final state can be populated in $S$-wave by either $J^{PC}=0^{++}$ or $2^{++}$ owing to the vector nature of the $J/\psi$ and the $\omega$. 

While the experimental situation for excited $\chi_{c0}$ states, as described above, is currently rather unclear, with even the number of states not settled, the situation for $2^{++}$ is a little better. The single $\chi_{c2}(3930)$ claimed in $\gamma\gamma\to \DD$ is the leading candidate~\cite{BaBar:2010jfn}, although whether some of the enhancements currently assigned to $0^{++}$ could actually be due to $2^{++}$ remains to be seen. 


Experimental analyses are typically performed final-state by final-state, with descriptions of the resonance content of a particular $J^{PC}$ being inferred from that single data set, sometimes with inspiration from observations in other final-states, but typically not with analysis of multiple final-states simultaneously. 
Theoretically the relevant fundamental object is a partial-wave \emph{scattering amplitude}, which is a matrix in the space of coupled multi-hadron channels. The enhancements seen in experiment for real values of the scattering energy correspond to poles of this scattering amplitude present at complex energies, and it is the pole locations and residues that provide a model-independent description of the resonance content and channel couplings. The scattering matrix is subject to important fundamental constraints, such as unitarity and analyticity, which are not always respected in practical data analysis, and which can give rise to important effects, particularly at kinematic thresholds.

Quantum Chromodynamics (QCD) is the fundamental theory of hadrons, but connecting the strong interactions of quarks and gluons to the presence of resonances in meson-meson scattering is not simple, and for want of a rigorous approach, models have been developed that incorporate some of the features of QCD. While the simplest approaches have only $c\bar{c}$ bound-states, other approaches include compact tetraquark constructions, or molecular meson-meson bound-states, and in these pictures, many more states are expected.
Early suggestions of the importance of meson-meson contributions were put forward long before any XYZ states were discovered~\cite{DeRujula:1976zlg,Tornqvist:1993ng}, and the various possibilities have been discussed in several recent reviews~\cite{Esposito:2016noz,Guo:2017jvc,Brambilla:2019esw,JPAC:2021rxu,Bicudo:2022cqi}. While models are useful to inform our understanding of the possible mechanisms at work, ultimately they are not QCD, and we must turn to a first-principles approach like \emph{lattice QCD}.

In recent years, approaches for computing scattering processes using lattice QCD have undergone rapid development, reviewed in Ref.~\cite{Briceno:2017max}, such that we are now in a position to consider the challenging $\ccz$ and $\cct$ systems. We benefit from several recent breakthroughs, including the ability to compute \emph{coupled-channel} scattering amplitudes~\cite{Dudek:2014qha,Wilson:2014cna}, and to consider final states with \emph{mesons of nonzero spin}~\cite{Briceno:2014oea,Woss:2018irj,Woss:2019hse,Woss:2020ayi}.


\vspace{3mm}

This paper reports on a computation in QCD of the coupled-channel scattering matrix in the energy region where the above-mentioned resonance candidates lie. We use an approximation in which $c\bar{c}$ annihilation is forbidden, and a larger-than-physical light-quark mass corresponding to $m_\pi\approx 391$~MeV. With these choices, no hadron in the energy region we will consider can decay to more than two lighter hadrons.
The scattering amplitudes resulting from this approach can be analytically continued to complex energies to determine resonance poles and their channel couplings.

The mass scale of charmonium systems brings in several difficulties that increase the complexity of this calculation with respect to calculations considering lighter hadrons composed primarily of light and strange quarks. The relevant discrete spectra are compressed relative to light quark spectra, and the small energy gap between $D$ and $D^*$ mesons means that several significant thresholds open almost simultaneously. Considering also closed-charm channels,  involving a charmonium meson and a light meson, we are forced to account for physics in a large number of coupled-channels. Much of this article is dedicated to disentangling those channels in which strong scattering effects occur from those which are decoupled and weak.

We will report coupled-channel amplitudes with ${J^{PC}=0^{++}}$ and $2^{++}$, constrained using large numbers of discrete energy levels taken from three lattice volumes, in both the rest frame and moving frames. One key new feature of this work is computation of a ``complete'' \mbox{$S$-matrix} in these quantum numbers, one which includes all kinematically accessible channels, with no a priori assumptions of which might be ``weak enough to ignore''.

Comparing with previous attempts to study this scattering system in lattice QCD~\cite{Lang:2015sba,Prelovsek:2020eiw}, we find that previously claimed near-threshold features at $\DD$ and $\DsDs$ do not appear in the present study. We find no broad resonance low in the $S$-wave $\DD$ amplitude, nor any bound state between the ground state $\ccz(1P)$ and the first hadron-hadron threshold.

A quite simple picture arises from our study, with only a single relatively narrow resonance in each of $J^{PC}=0^{++}$ and $2^{++}$. Each is found to have large couplings to several open-charm $D$-meson decay modes, channels consisting of pairs of open-charm mesons, with only relatively small couplings to closed-charm channels such as $J/\psi\, \omega$. 

This article is organized as follows. In Section~\ref{sec:lattice}, we describe the lattices and methods used to compute the finite-volume spectra, present the masses of the stable scattering hadrons, give the partial-waves that feature in the irreducible representations of the lattice symmetry, and outline the operators that are required to access the spectrum. The determined finite volume energy levels are presented in Section~\ref{sec:spectra}. Section~\ref{sec:luescher} explains how these energies are translated into scattering amplitudes through extensions of the L\"uscher formalism. In Section~\ref{sec:amps} we describe how the amplitudes are determined, in increasing complexity, beginning with just a few energies at rest, before ultimately making use of more than 200 energy levels including systems with nonzero total momentum. We discuss the pole singularities present in the determined scattering amplitudes in Section~\ref{sec:poles}. In Section~\ref{sec:interp}, we offer some interpretations and comparisons of our results to experiment, prior lattice calculations, and other theoretical approaches. We conclude with a brief summary in Section \ref{sec:summary}. A concise description of this work may be found in Ref.~\cite{Wilson:2023_short}.

\pagebreak
\section{Lattice QCD setup}                                
\label{sec:lattice}

Within lattice QCD, the discrete spectrum of eigenstates of QCD in a finite-volume can be obtained from the time-dependence of two-point correlation functions. For the current calculation, we use $N_f=2+1$ flavours of dynamical quarks with exact isospin symmetry, and opt to work with a larger-than-physical value of the degenerate $u,d$ quark mass, while the strange quark mass is approximately at the physical value. The quenched charm quark mass is tuned to approximately reproduce the physical $\eta_c$ mass~\cite{Liu:2012ze}. We disallow $c\bar{c}$ annihilation so that low-lying charmonia like the $\eta_c$ and $J/\psi$ are absolutely stable.\footnote{In effect we implement two degenerate non-dynamical charm quarks, and study only the charm-isospin=1 sector. In this way the approximation is self-consistent.}

The quark dynamics is implemented by an anisotropic Wilson-clover action as described in Ref.~\cite{Edwards:2008ja,Lin:2008pr}, with a temporal lattice spacing $a_t$, finer than that in space\, $a_s$, by a factor $\xi = a_s/a_t \approx 3.5$. Distillation is used to smear the quark fields and enable efficient computation of all contributions, including those where light quarks or strange quarks annihilate~\cite{Peardon:2009gh}. The three lattice volumes used are summarized in Table~\ref{tab:lattices}, where $L$ and $T$ are the spatial and temporal extent of the lattice respectively. Correlation functions are averaged over several ($N_\mathrm{tsrcs}$) source timeslices. The lattice scale is set using the physical $\Omega$ baryon mass, leading to $a_t^{-1}=5667$~MeV~\cite{Edwards:2012fx}. The pion mass is determined to be $a_t m_\pi=0.06906(13)$, corresponding to $m_\pi\approx 391$ MeV~\cite{Dudek:2012gj}.

In the finite cubic volume defined by the lattice, continuous rotation symmetry is broken, and states are characterized as lying in irreducible representations (irreps) of the cubic group at rest, and of relevant little groups 
at non-zero momentum, rather than by spin, $J$. Charge-conjugation, $C$, remains a good symmetry, but states are only of definite parity, $P$, at rest. The finite periodic boundary implies that momentum is quantized,
$\vec{p}=\frac{2\pi}{L}\vec{n}$, where $\vec{n}=(i,j,k)$ is a triplet of integers (we will often use a shorthand notation $\vec{p}=[ijk]$).
 
The finite-volume quantization condition which relates the discrete spectrum to continuous scattering amplitudes is sensitive to the total momentum of the scattering system, and as such we compute spectra in moving frames (nonzero overall momentum $\vec{P}$) as well as in the rest frame to obtain more constraint on the scattering amplitudes.
We refer to irreps at rest with the labels $[000]\Lambda^{P}$, and to moving-frame irreps as $[ijk]\Lambda$.

For each irrep, we compute a matrix of correlation functions constructed using a wide range of operators resembling single--, two-- and three--hadron constructions. The resulting correlation matrix as a function of Euclidean time, $C_{ij}(t)=\big<0 \big|\mathcal{O}_i(t)\mathcal{O}_j^\dagger(0) \big|0\big>$, is then analyzed variationally to obtain the discrete spectrum contributing to these correlation functions~\cite{Michael:1985ne,Luscher:1990ck,Blossier:2009kd}, with our implementation described in Ref.~\cite{Dudek:2010wm}. The analysis takes the form of solving a generalized eigenvalue problem on each timeslice,
\begin{align}
\bm{C}(t) \,v^\mathfrak{n}=\lambda_\mathfrak{n}(t)\, \bm{C}(t_0)\, v^\mathfrak{n}\,,
\label{eq:gevp}
\end{align}
where the discrete spectrum $\{E_\mathfrak{n}\}$ is obtained from the time dependence of the eigenvalues.
The eigenvectors can be used to construct optimized operators as described below. They also provide helpful qualitative information by indicating which states are produced dominantly by particular operator constructions via overlap factors,
\begin{align*}
Z_i^\mathfrak{n} & =\bigl<\mathfrak{n}|\mathcal{O}^\dagger_i|0\bigr> = \sqrt{2E_\mathfrak{n}}\: e^{E_\mathfrak{n} t_0/2}v_j^{\mathfrak{n}\ast}C_{ji}(t_0) \, .
\end{align*}

The construction of single-hadron-like operators, as fermion bilinears using gamma matrices $\bm{\Gamma}$ and up to three gauge-covariant derivatives, is as described in Refs.~\cite{Dudek:2009qf,Dudek:2010wm,Thomas:2011rh}. These are of the form $\mathcal{O}^\dagger\sim \bar{q}\:\bm{\Gamma} \overleftrightarrow{D}...\overleftrightarrow{D}\:q$ with a definite continuum $J$ that is projected into cubic group irreps $\Lambda$.

The construction of two-hadron-like and three-hadron-like operators is as described in Ref.~\cite{Dudek:2012gj} and Ref.~\cite{Woss:2019hse} respectively. 
Both leverage \emph{optimized} single-hadron operators that have reduced excited state contamination (when compared to using only a single operator).
A variationally-optimal single-hadron operator
for meson $M_1$ is formed from a linear combination of a basis of single-hadron operators with $M_1$ quantum numbers where the coefficients are given by the eigenvectors from the variational analysis, $\Omega_{M_1}^\dagger \sim \sum_i v^\mathfrak{n}_i \mathcal{O}^\dagger_i$. These are then used in product pairs to form two-hadron operators, 
\begin{align*}
\mathcal{O}^\dagger_{M_1 M_2}(\vec{p})\sim \sum_{\vec{p}_1,\vec{p}_2} \:\mathrm{CGs}\:\Omega^\dagger_{M_1}(\vec{p}_1)\:\Omega^\dagger_{M_2}(\vec{p}_2) \, ,
\end{align*}
for the $M_1 M_2$ channel with overall momentum $\vec{p}$ where the sum is over all momenta related by an allowed lattice rotation such that $\vec{p}=\vec{p}_1+\vec{p}_2$ and ``CGs'' represents the necessary lattice Clebsch-Gordan coefficients to project to the appropriate quantum numbers.
A recursive approach can be adopted to form three-hadron-like operators from optimized single-hadron and two-hadron operators.

\begin{table}
\begin{tabular}{c|c|c|c|c|c|c}
$L/a_s$ & $T/a_t$ & $N_\mathrm{cfg}$ & $N_\mathrm{vec}$ & $N_\mathrm{tsrcs}$ & $L/\mathrm{fm}$ & $m_\pi L$ \\
\hline
16      & 128     & 478 &     64    & 8--16    & 1.9  & 3.8 \\
20      & 256     & 288 &    128    & 4--8     & 2.4  & 4.8 \\
24      & 128     & 553 &    160    & 2--4     & 2.9  & 5.7
\end{tabular}
\caption{Lattices used: $N_\mathrm{cfg}$ is the number of gauge configurations, $N_\mathrm{vec}$ is the number of distillation vectors, and $N_\mathrm{tsrcs}$ is the number of time sources.}
\label{tab:lattices}
\end{table}

\subsection{Stable hadrons}     \label{sec:stable_hadrons}

\begin{figure}[b]
\includegraphics[width=0.95\columnwidth]{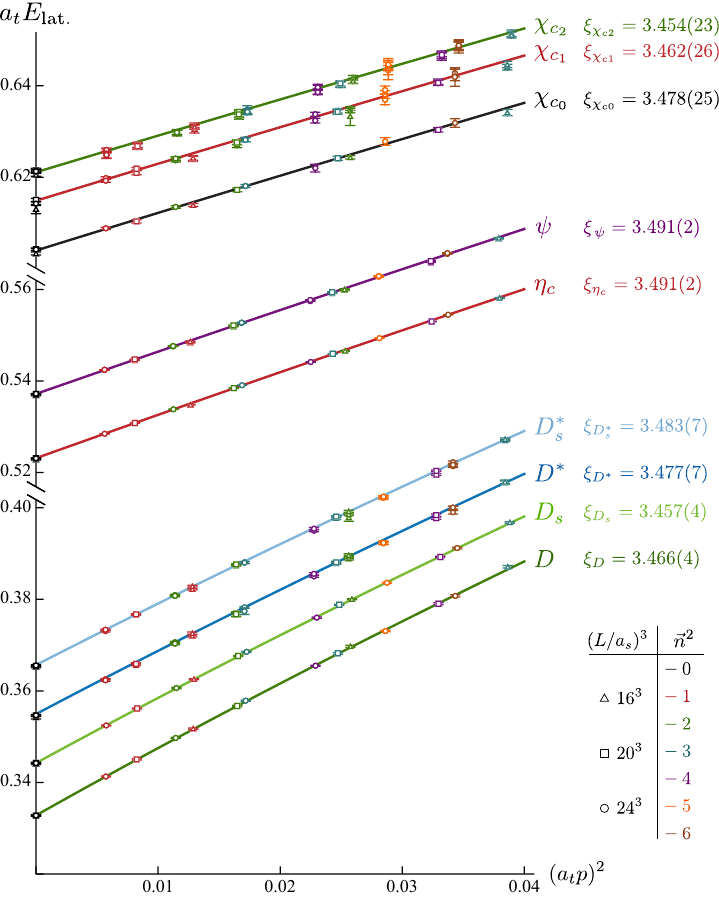}
\caption{Lattice-determined rest mass and moving frame energies of stable hadrons fitted with a relativistic dispersion relation, Eq.~\ref{eq:disp}. Energy levels taken from $\{16^3,20^3,24^3\}$ lattices using momenta corresponding to $|\vec{n} |^2\le \{3,4,6\}$  respectively.}
\label{fig_disp}
\end{figure}

The systems of coupled-channel scattering we will consider feature a number of hadrons which are stable against strong decay on the lattices used in this study.
Their energies as a function of momentum are determined using spectra extracted from matrices of correlation functions.\footnote{The current calculation makes use of a $20^3\times 256$ lattice~\cite{Briceno:2017qmb,Hansen:2020otl}, while earlier calculations with charm quarks~\cite{Moir:2016srx,Cheung:2020mql,Lang:2022elg} used a shorter time-extent, $20^3\times 128$, lattice. }
Figure~\ref{fig_disp} shows dispersion relations for a selection of the stable hadrons used in this study, along with fits using the relativistic expression,
\begin{align}
\big(a_t E \big)^2 = \big( a_t m \big)^2 + \big|\vec{n} \big|^2 \left(\frac{2\pi}{\xi \, L/a_s}\right)^2
\label{eq:disp}
\end{align}
from which the rest mass $a_t m$, and the anisotropy $\xi$, are determined separately for each hadron. 
The masses resulting from such fits are presented in Table~\ref{tab:masses} along with relevant kinematic thresholds for isospin--0, $C=+$ scattering channels.

\begin{table}[b]
\begin{center}
\begin{tabular}{l|l|l}
hadron         & $J^{P(C)}$ & $a_t m$\\
\hline
\hline
$\pi$          & $0^{-+}$ & 0.06906(13)\\
$K$            & $0^{-}$  & 0.09698(9)\\
$\eta$         & $0^{-+}$ & 0.10364(19)\\
$\eta^\prime$  & $0^{-+}$ & 0.16410(100)\\
$\sigma$       & $0^{++}$ & 0.1316(9)$^\ddagger$\\
$\omega$       & $1^{--}$ & 0.15541(29)\\
$\phi$         & $1^{--}$ & 0.17949(21)\\
[0.2ex]
\hline
$D$            & $0^-$    & 0.33281(9)$^\dagger$\\
$D^\star$      & $1^-$    & 0.35464(14)$^\dagger$ \\
$D_s$          & $0^-$    & 0.34424(11)$^\dagger$ \\
$D_s^\star$    & $1^-$    & 0.36566(14)$^\dagger$  \\
$D_0^\star$    & $0^+$    & 0.40170(18)$^\ddagger$\\
$D_{s0}^\star$& $0^+$    & 0.4200(5)$^\ddagger$ \\[0.2ex]
\hline
$\eta_c$       & $0^{-+}$ & 0.52312(4)$^\dagger$\\
$\psi$         & $1^{--}$ & 0.53715(5)$^\dagger$\\
$h_c$          & $1^{+-}$ & 0.61662(26)$^\dagger$\\
$\ccz$         & $0^{++}$ & 0.60422(25)$^\dagger$\\
$\cco$         & $1^{++}$ & 0.61488(46)$^\dagger$\\
$\cct$         & $2^{++}$ & 0.62110(28)$^\dagger$\\
$\eta_c^\prime$& $0^{-+}$ & 0.64160(55)$^\dagger$\\
$\psi^\prime$  & $1^{--}$ & 0.64566(111)$^\dagger$\\[0.2ex]
\hline
$\Omega$       & $\tfrac{3}{2}^+$ & 0.2951(22)
\end{tabular}
\hspace{0.5cm}
\begin{tabular}{l|l}
channel & $a_t E_{\mathrm{thr}.}$ \\
\hline\hline
$\eta_c\eta$    & 0.6268(2)   \\
$\eta_c\pi\pi$  & 0.6612(2)   \\
$\eta_c\etap$   & 0.6872(10)  \\
$\psi\omega$    & 0.6926(3)   \\
$\ccz\eta$      & 0.7079(3)   \\
$\psi\phi$      & 0.7166(2)   \\
$\eta_c K\bar{K}$ & 0.7171(1) \\
$\cco\eta$      & 0.7185(5)  \\
$\cct\eta$      & 0.7247(3)  \\
$\eta_c\pi\pi\pi$& 0.7303(3)  \\
$\eta_c \eta\eta$ & 0.7304(3) \\
$\psi K\bar{K}$ & 0.7311(1) \\
$\ccz\pi\pi$    & 0.7424(3)   \\[0.2ex]
\hline
$\DD$           & 0.6656(1)   \\
$\DDst$         & 0.6875(2)   \\
$\DsDs$         & 0.6885(2)   \\
$\DstDst$       & 0.7093(2)   \\
$\DsDsst$       & 0.7099(2)   \\
$\DsstDsst$     & 0.7313(2)   \\
\hline
$\eta_c\sigma$     & 0.6547(9)$^\ddagger$  \\
$\DDz$             & 0.7345(2)$^\ddagger$  \\
$\ccz\sigma$       & 0.7358(9)$^\ddagger$  \\
\end{tabular}

\end{center}
\caption{Stable hadron masses (left panel) and $I=0$, $C=+$ multi-hadron decay channels (right panel). 
 For hadrons labelled with $^\dagger$, the mass is obtained from a dispersion relation fit using energies computed on the $24^3$ lattice, while $^\ddagger$ indicates a bound-state pole mass from a meson-meson scattering amplitude fit~\cite{Briceno:2016mjc,Briceno:2017qmb,Moir:2016srx,Cheung:2020mql}. The remaining masses are obtained from dispersion relations using energies computed on all three volumes. In the right panel $^\ddagger$ indicates a channel formed from a bound-state with a significant two-hadron contribution where it follows that three-hadron effects may be present.}
\label{tab:masses}
\end{table}

In the dispersion relation fits, points are observed to be scattered around the mean with a deviation beyond what would be expected from statistical fluctuations alone -- comparisons can be seen in Fig.~\ref{fig:disp_L20} in Appendix~\ref{app:disp_extras}.
The most significant deviations are seen for the $\eta_c$, $\ccz$, and the $\cct$, with largest values of ${a_t \delta E \approx 0.00030, 0.00050, 0.00100}$ respectively. These are tiny in absolute terms, but relatively large on the scale of the very small statistical uncertainties. We choose to treat these deviations as an additional systematic uncertainty on the energy levels to be added to the statistical uncertainties when used in energy level fits to determine scattering amplitudes.
In practice, we include an additional systematic uncertainty on the $E_\cm$ energies of $a_t \delta E_{\mathrm{syst.}} = 0.00050$ when the amplitude we wish to determine is $J=0,1$. If the amplitude we wish to determine has $J \ge 2$, we use $a_t \delta E_{\mathrm{syst.}} = 0.00100$. 

Similar effects have been observed in other lattice studies with charm quarks by other groups. There is a consensus that this needs to be accounted for in some way. An alternative used in Refs.~\cite{Piemonte:2019cbi,Prelovsek:2020eiw,Xing:2022ijm} is to apply an energy shift configuration by configuration to force these numbers into agreement. We consider the approach adopted in the current paper to be more conservative, treating the difference as a systematic uncertainty that will be propagated through into scattering amplitude determinations.  Further details are given in Appendix~\ref{app:disp_extras}.

The anisotropies, $\xi=a_s/a_t$, obtained for different stable hadrons differ somewhat, and we choose to increase the uncertainty on the value obtained for the pion, ${\xi_\pi=3.444(6)}$~\cite{Dudek:2012gj}, to account for such deviations, using in practice $\xi=3.444(50)$, which spans the extracted values for $\eta_c$ and $J/\psi$.\footnote{This range is almost exactly the same as used in Ref.~\cite{Woss:2019hse} that was chosen to account for the observed differences in the helicity components of the vector $\omega$.} This uncertainty is propagated through when center-of-momentum frame energies are obtained from computed lattice frame energies.

\subsection{Resonance expectations and partial-waves}     \label{sec:partial-waves}

The goal of this calculation is to obtain coupled-channel partial wave amplitudes with $J^{PC}=0^{++}$ and $2^{++}$ up to around 4100 MeV, $a_t E_\cm = 0.72$ in lattice units. This runs to slightly above the $\DstDst$ threshold, and corresponds to an energy region where resonant features have been observed experimentally. 

\begin{figure*}
\includegraphics[width=1.0\textwidth]{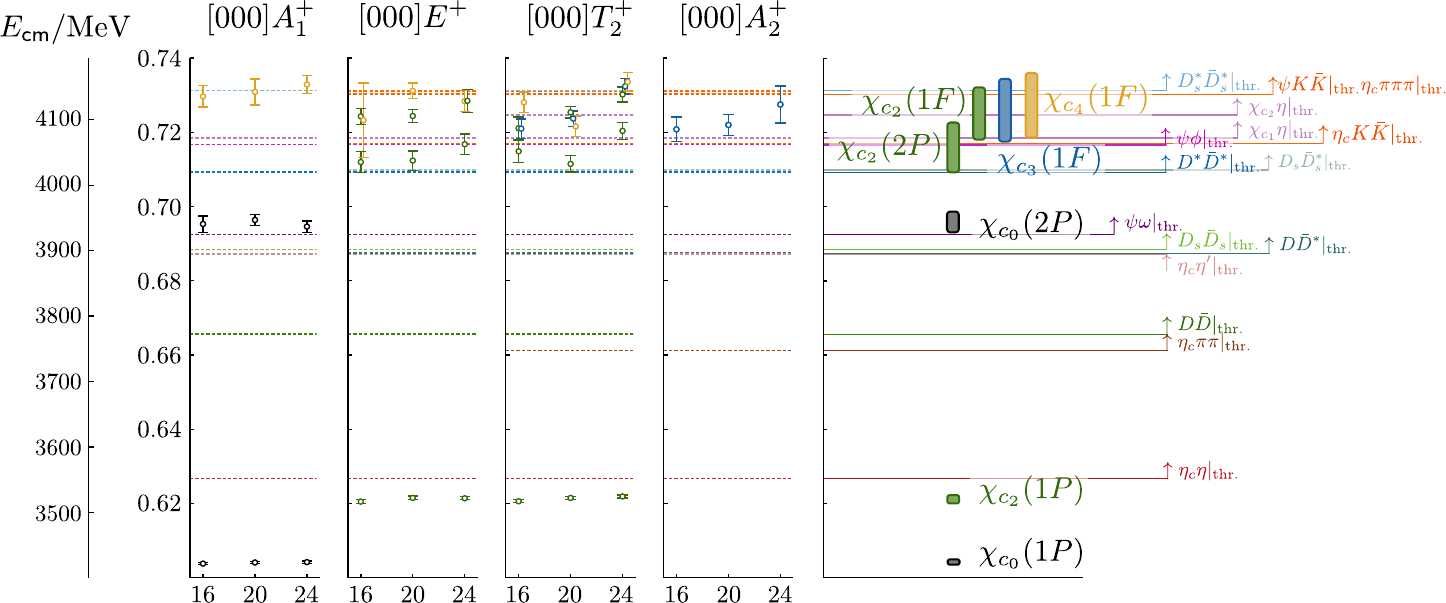}
\caption{
Energy levels obtained from diagonalising correlation matrices constructed using \emph{only} $c\bar{c}$-like operators in irreps relevant for $J^{PC}=0^{++},2^{++},\ldots$ with zero overall momentum. The locations of hadron-hadron thresholds are marked. The right panel shows a summary of the observed levels based on the pattern of levels across irreps, and the quark-model--like labeling follows from Ref.~\cite{Liu:2012ze}.}
\label{fig:spec_qqbar}
\end{figure*}

An earlier lattice QCD calculation of the spectrum on the current lattices using \emph{only single-hadron-like operators}~\cite{Liu:2012ze} found results that suggest narrow resonant states may appear. Updating these calculations using more time sources and a longer time-extent for the $20^3$ lattice, leads to the spectra presented in Fig.~\ref{fig:spec_qqbar} in irreps relevant for $J^{PC} = 0^{++}, 2^{++}, \ldots$. 
The pattern is reminiscent of the $J=0$, 2, 3 and 4 members of $q\bar{q}$ quark-model multiplets, $nL =$ $1P$, $2P$ and $1F$, and the overlap of states onto operators subduced from particular $J^{PC}$~\cite{Liu:2012ze} is in agreement with this. Working up to $a_t E_\cm \approx 0.72$ appears to be sufficient to capture the $2P$-like $\chi_{c0}, \chi_{c2}$ states in this energy region.\footnote{We will reserve comment on the $J^{PC}=1^{++}$ member for a future study. It will not contribute in any of the irreps used in this work.}

As shown in Fig.~\ref{fig:spec_qqbar}, the expected $\chi_{cJ}$ states lie above a number of kinematical thresholds, and hence should appear as \emph{resonances} in meson-meson scattering amplitudes. We label each meson-meson channel according to its total spin $S$ (combining the spin quantum numbers of the two scattering hadrons), orbital angular momentum $\ell$, and total angular momentum $J$, using the standard notation, $\SLJ{2S+1}{\ell}{J}$. Meson-meson partial wave amplitudes grow at threshold according to their orbital angular momenta $\ell$, via $k_i^{2\ell}$ where $k_i$ is the $\cm$ momentum for meson-meson pair $i$. This is a relevant property since it establishes a hierarchy among partial-waves whereby the lowest $\ell$ dominate, unless disturbed by some nearby singularity.

Due to the presence of scattering mesons with nonzero spin, for a given $J^{PC}$, amplitudes with more than one $\ell$ can contribute. The $J^{PC}=0^{++}$ amplitudes are relatively simple, consisting of pairs of pseudoscalars and vectors in a relative $S$-wave ($\SLJ{1}{S}{0}$). The lowest threshold is $\eta_c\eta$, followed by $\DD$, $\eta_c\eta^\prime$, $\DsDs$, $\psi\omega$, and $\DstDst$. The first contributions from  partial waves with $\ell > 0$ arise from $\psi\omega$ in $\SLJ{5}{D}{0}$, and $\cco\eta$ in $\SLJ{3}{P}{0}$. There can be no contributions from vector-pseudoscalar channels such as $\DDst$. The $J^{PC}=2^{++}$ amplitudes contain pairs of pseudoscalars in $\SLJ{1}{D}{2}$ combinations, while vector-vector channels can arise in $S$-wave through $\SLJ{5}{S}{2}$. Furthermore, vector-pseudoscalar channels such as $\DDst$ now contribute, the lowest combination being $\SLJ{3}{D}{2}$.
We summarise the meson-meson partial wave contributions relevant at low energy for each $J^{PC}$ considered in this study in Table~\ref{tab:pwa_chans}.

\begin{table*}
\begin{center}
\renewcommand{\arraystretch}{1.5}

\begin{tabular}{c|l}
$J^{PC}$ & hadron-hadron channels below $a_tE_\cm\approx0.72$\\
\hline
$0^{-+}$   &  $\eta_c f_0$, $\ccz\eta$ $\SLJC{1}{S}{0}$; $\psi\omega$, $D\bar{D}^\ast$, $D^\ast\bar{D}^\ast$ $\SLJC{3}{P}{0}$; $\cct\eta\SLJC{5}{D}{0}$; \\
$0^{++}$   &  $\eta_c\eta$, $D\bar{D}$, $\eta_c\eta^\prime$, $D_s\bar{D}_s$, $\psi\omega$, $D^\ast\bar{D}^\ast$, $\psi\phi$ $\SLJC{1}{S}{0}$; $\cco\eta$ $\SLJC{3}{P}{0}$;  $\psi\omega$, $D^\ast\bar{D}^\ast$, $\psi\phi$ $\SLJC{5}{D}{0}$;  \\ 
$1^{-+}$   &  $\eta_c\eta$, $\eta_c\eta^\prime$, $\psi\omega$, $\psi\phi$ $\SLJC{1}{P}{1}$; $\psi\omega$, $\DDst$, $\DstDst$, $\psi\phi$ $\SLJC{3}{P}{1}$; $\psi\omega$, $\psi\phi$ $\SLJC{5}{P}{1}$; $\cco\eta$ $\SLJC{3}{S}{1}$; $\cco\eta$ $\SLJC{3}{D}{1}$; $\cct\eta$ $\SLJC{5}{D}{1}$;  \\
$1^{++}$   &  $\psi\omega$, $D\bar{D}^\ast$ $\SLJC{3}{S}{1}$; $\psi\omega$, $\DDst$ $\SLJC{3}{D}{1}$; $\psi\omega$, $\DstDst$ $\SLJC{5}{D}{1}$; $\ccz\eta$, $\eta_c f_0$ $\SLJC{1}{P}{1}$; $\cco\eta$ $\SLJC{3}{P}{1}$; $\cct\eta$ $\SLJC{5}{P}{1}$; \\
$2^{-+}$   &  $\psiom$, $\DDst$, $\DsDsst$, $\psi\phi$ $\SLJC{3}{P}{2}$; $\cct\eta$ \SLJC{5}{S}{2}; $\cco\eta$ $\SLJC{3}{D}{2}$; $\cct\eta$ $\SLJC{5}{D}{2}$;  $\eta_c f_0$ $\SLJC{1}{D}{2}$ \\
$2^{++}$   &  $\psi\omega$, $\DstDst$, $\psi\phi$, $\DsstDsst$ $\SLJC{5}{S}{2}$;
$\eta_c\eta$, $\DD$, $\eta_c\eta^\prime$, $\DsDs$, $\psi\omega$, $\DstDst$, $\psi\phi$ $\SLJC{1}{D}{2}$; \\
&
$\psi\omega$, $\DDst$, $\DsDsst$, $\DstDst$, $\psi\phi$ $\SLJC{3}{D}{2}$;
$\cco\eta$ $\SLJC{3}{P}{2}$; $\cct\eta$ $\SLJC{5}{P}{2}$;\\
$3^{-+}$   &  $\psi\omega$, $\psi\phi$ $\SLJC{5}{P}{3}$; $\cco\eta$ $\SLJC{3}{D}{3}$; $\cct\eta$ $\SLJC{5}{D}{3}$; $\eta_c\eta$, $\eta_c\eta^\prime$, $\psi\omega$, $\psi\phi$ $\SLJC{1}{F}{3}$; $\psi\omega$, $\DDst$, $\DstDst$, $\psi\phi$ $\SLJC{3}{F}{3}$; $\psi\omega$, $\psi\phi$ $\SLJC{5}{F}{3}$; \\
$3^{++}$   &  $\DDst$, $\DsDsst$, $\psi\omega$, $\psi\phi$ $\SLJC{3}{D}{3}$; $\DstDst$, $\DsstDsst$, $\psi\omega$, $\psi\phi$ $\SLJC{5}{D}{3}$; \\
           &  $\eta_c f_0$ $\SLJC{1}{F}{3}$; $\cct\eta$ $\SLJC{5}{P}{3}$; $\ccz\eta$ $\SLJC{1}{F}{3}$; $\cco\eta$ $\SLJC{3}{F}{3}$; $\cct\eta$ $\SLJC{5}{F}{3}$; \\
$4^{-+}$   &  $\psiom$, $\DDst$, $\DstDst$, $\DsDsst$ $\SLJC{3}{F}{4}$;\\
$4^{++}$   &
$\psi\omega$, $\DstDst$, $\psi\phi$, $\DsstDsst$ $\SLJC{5}{D}{4}$; $\cco\eta\SLJC{3}{F}{4}$; $\cct\eta\SLJC{5}{F}{4}$; 
\end{tabular}

\end{center}
\caption{The lowest few $I=0, C=+$ hadron-hadron channels for each $J^{PC}$ considered, labeled by spin $S$, orbital angular momentum $\ell$ and total angular momentum $J$, in the notation $\SLJ{2S+1}{\ell}{J}$. Because of $k_i^\ell$ threshold suppression, the lowest $\ell$ are typically the most relevant at low energies. 
}
\label{tab:pwa_chans}
\end{table*}

\subsection{Irreps and operators}     \label{sec:irrep_op_selection}

In order to constrain the coupled-channel scattering amplitudes for $J^{PC} = 0^{++}$ and $2^{++}$ we will compute finite-volume spectra in several irreps.

Working at zero overall momentum, the $[000]A_1^{+}$ irrep constrains $0^{++}$ with contaminations from $4^{++}$ and higher. $J^{PC}=2^{++}$ information can be obtained from $[000]E^{+}$ and $[000]T_2^{+}$, where the second of these also receives contributions from $3^{++}$. In order to constrain this $3^{++}$ component, it is advantageous to consider $[000]A_2^{+}$ where it is the leading contribution.

When working at nonzero overall momentum, partial waves of both parities appear. For example, in the moving frame $[ijk]A_1$ irreps we have contributions from $0^{++},1^{-+}$ and higher. In the moving frame $[00i]B_{1,2}$ irreps where $2^{++}$ is present, $2^{-+}, 3^{\pm+}$ and higher also appear. In order to determine the $1^{-+}$, $2^{-+}$ and $3^{-+}$ amplitudes, we also consider the rest frame irreps $[000]T_1^{-}$, $[000]E^{-}$, $[000]T_2^{-}$ and $[000]A_2^{-}$.

In summary, our selection of irreps enables us to determine scattering amplitudes for $J^{PC}=0^{++}, 1^{-+}, 2^{\pm+}$ and  $3^{\pm+}$.

\vspace{2mm}

Within each irrep we establish a basis of operators, including all single-hadron-like operators with up to 3 derivatives at-rest and up to 2 derivatives in moving frames. These are supplemented with all two and three-hadron operators expected to be relevant in the energy region of interest based on their corresponding non-interacting energy. For $N_{\mathrm{had.}} = 2$ or 3 hadrons, this is determined from
\begin{align}
a_t E_\mathrm{n.i.}=\sum_{a=1}^{N_{\mathrm{had.}}}\left( (a_t m_a)^2 + |\vec{n}_a|^2 \left(\frac{2\pi}{\xi L/a_s}\right)^2\right)^{1/2}
\label{eq:nonint}
\end{align} 
where $n_a=(i,j,k)$ is a vector of integers and $m_a$ is the scattering hadron mass. If this `lattice-frame' energy, when boosted into the $\cm$ frame, lies below ${a_t E_\cm =0.743}$, the corresponding operator is included in the basis.\footnote{This upper limit corresponds to the $\ccz\pi\pi$ threshold.}$^,$\footnote{Two exceptions are an $\eta_c[012]\eta[001]$ in $[002]B_1$ on the $L/a_s=16$ volume, and very high-lying $\cct\pi\pi$ operators that would be expected to produce levels above the energy limits used in the scattering analyses below.}
Full lists of operators used for the presented results can be found in the supplemental material.\footnote{No \emph{local} four-quark operator constructions of the form found to be irrelevant in Ref.~\cite{Cheung:2017tnt} are included.}
When analysing the correlation matrices, the operator basis is varied to explore the sensitivity to the precise selection, and in this process some of the highest lying operators are discarded as their presence does not affect low-lying levels.

Three-meson operator constructions are only expected to be relevant at relatively high energies, owing to the large light quark mass and prohibition of $c\bar{c}$ annihilation in this calculation. There are very few relevant three-hadron operators in this energy region, and no four-hadron operators. The lowest three-hadron operators arise from $\eta_c\pi\pi$ combinations where $\pi\pi \sim \sigma$ and there is relatively high orbital angular momentum between the `$\sigma$' and the $\eta_c$ (typically an $F$-wave). These operators are constructed as described for the $\mathbb{RM}$ operators in Sec.~II.C of Ref.~\cite{Woss:2019hse}. The projection coefficients for the near-threshold $\sigma$ are obtained from the analysis performed in Ref.~\cite{Briceno:2017qmb} and these are combined with the single-hadron $\eta_c$ optimized operator. If there were strong interactions in the $\eta_c\pi$ subsystems, it is possible that these operators alone may not be sufficient.

Further three-hadron contributions arise from $\chi_{cJ}\pi\pi \sim \chi_{cJ}\sigma$--like combinations. In $[000]A_1^{+}$, one might naively expect a level at the threshold energy $m_{\chi_{c0}}+ 2m_\pi$. However, we know from Refs.~\cite{Briceno:2016mjc,Briceno:2017qmb} that on these lattices the $\sigma$ channel has a level below threshold with a large volume dependence owing to a bound-state $\sigma$ strongly coupled to $\pi\pi$. We will see that this feature survives the addition of a $\chi_{c0}$ operator, producing a level below $m_{\chi_{c0}}+2m_\pi$ with an energy that slowly rises with $L/a_s$ (similar to the $\sigma$ in $\pi\pi$ scattering in $[000]A_1^+$). Further details are given in the next section.
A few other three-hadron channels are listed in Table~\ref{tab:masses}. These are not expected to produce levels in the energy region of interest. When determining scattering amplitudes we will not utilize any energy levels found to have large overlaps with three-meson operators. 

\section{Finite-volume spectra}                            

\label{sec:spectra}

\begin{figure*}
\includegraphics[width=0.75\textwidth]{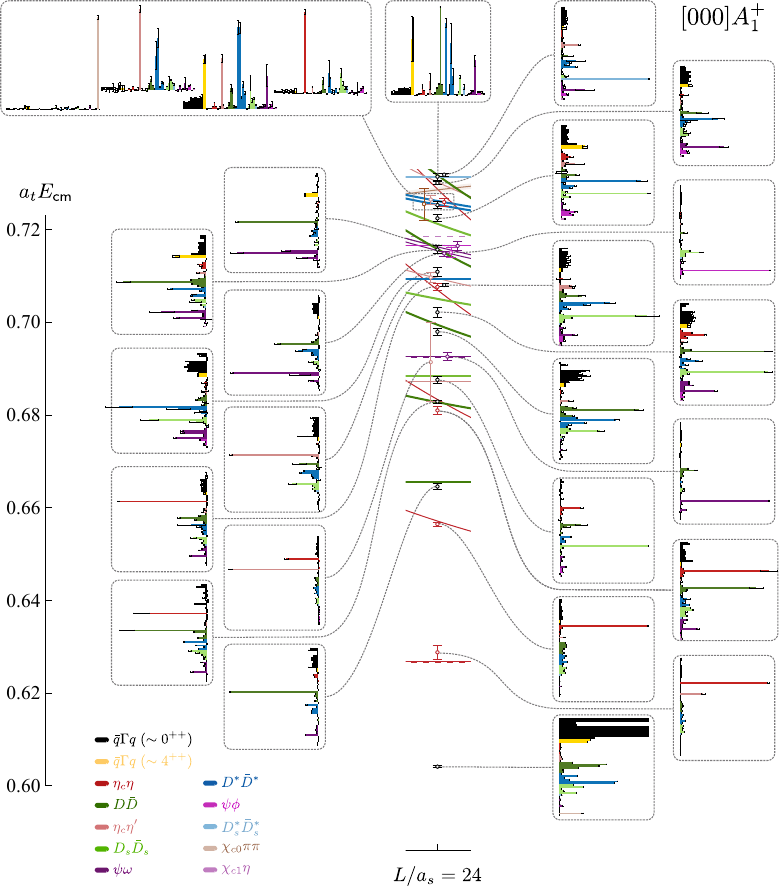}
\caption{The spectrum and normalised operator overlaps $Z_i^\mathfrak{n}$ for the $[000]A_1^{+}$ irrep on the $L/a_s=24$ volume.
The spectrum obtained from the lattice QCD calculation is shown in the center, colored according to the largest operator overlap as described in the text. Solid curves in the center show the non-interacting energies. The magnitudes of operator overlaps $Z_i^\mathfrak{n}$ for each state, normalised as described in the text, are shown in the dotted boxes.
}
\label{fig:histos_A1pP_L24}
\end{figure*}

The operator bases described in the previous section are used to compute a matrix of correlation functions for each irrep and these are then analyzed variationally as discussed above. The resulting spectrum in the $[000]A_1^+$ irrep on the $24^3$ volume is presented in Fig.~\ref{fig:histos_A1pP_L24}. The plot also shows histograms of the overlap factors, $Z_i^\mathfrak{n}$, for each state where these have been normalized such that the largest value for a particular operator, considered across all states, takes value 1. Clear patterns emerge, and in several cases levels are dominated by a single operator construction. These are often close to a non-interacting energy level, as determined by Eq.~\ref{eq:nonint}, with a potential explanation of there being a decoupled channel with only weak interactions.

\begin{figure*}[!h]
\includegraphics[width=0.9\textwidth]{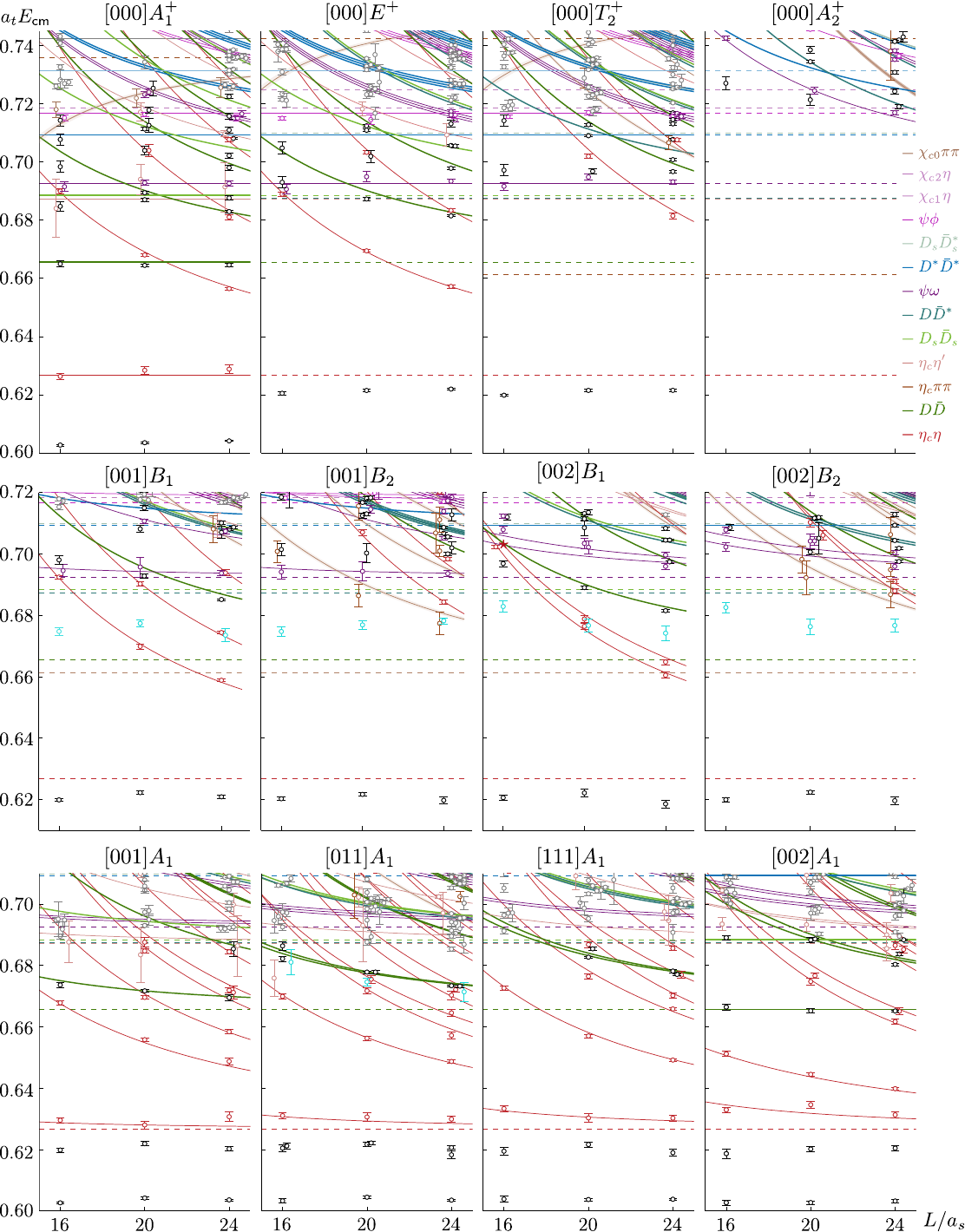}
\caption{Finite-volume spectra in irreps with zero overall momentum and positive parity, and irreps with non-zero overall momentum. Points with uncertainties are the energy levels determined from lattice QCD correlation matrices, colored according to the dominant operator overlap as described in the text. Solid thin curves are non-interacting energies, thick light-brown curves are ``2+1'' non-interacting levels as described in the text, and dashed horizontal lines are hadron-hadron thresholds. A single red star in $[002]B_1$ for $L/a_s=16$ indicates an $\eta_c[012]\eta[001]$ operator not included in the basis.}
\label{fig:specs}
\end{figure*}

\begin{figure*}[htp]
\includegraphics[width=0.90\textwidth]{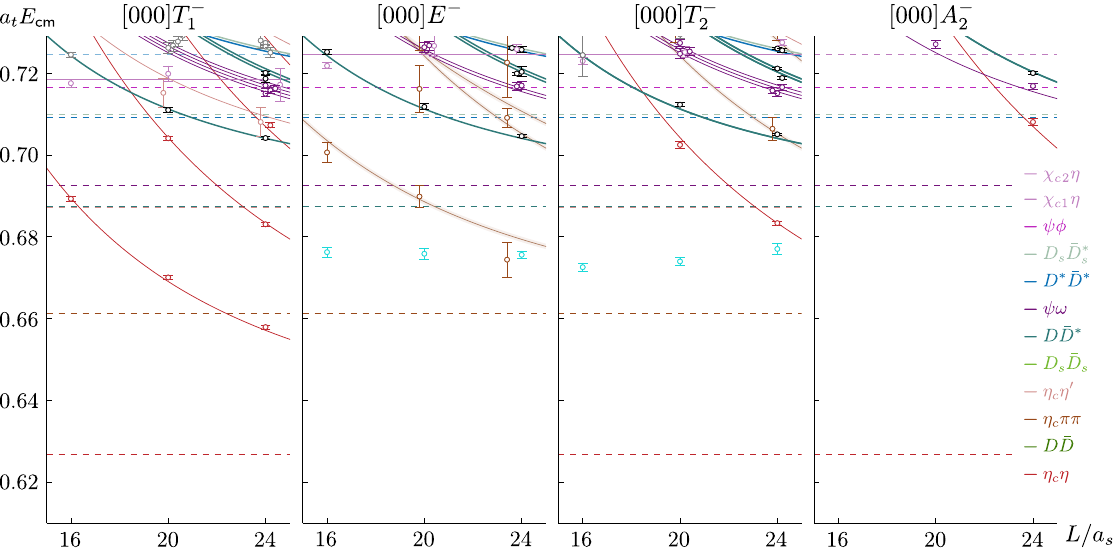}
\caption{As Figure~\ref{fig:specs} but for irreps with zero overall momentum and negative parity.}
\label{fig:specs_mP}
\end{figure*}

In Figs.~\ref{fig:specs} and~\ref{fig:specs_mP} we plot all of the finite-volume energy levels extracted from the variational analysis to be used to constrain scattering amplitudes. 
The uncertainties shown are estimated using jackknife. In several cases, in particular where there is a significant variation in the extracted energy for different $t_0$ values (in Eq.~\ref{eq:gevp}) or timeslice fit ranges and forms, the uncertainties are enlarged to provide a conservative estimate of the energy value. We also vary the operator bases by adding and removing operators when possible to ensure the spectrum is stable with respect to small and reasonable changes.\footnote{The plots do not show the additional systematic uncertainty $a_t\delta E_{\mathrm{syst}}$ discussed in Section \ref{sec:stable_hadrons} that is added to every level. This will be shown in later plots as an additional error bar on each point.}

In Figs.~\ref{fig:histos_A1pP_L24}, \ref{fig:specs} and~\ref{fig:specs_mP} we choose a presentation scheme where those states having a single dominant operator overlap are colored to indicate which operator is dominant. Black points show levels that are dominated by single-hadron-like operators with $C=+$, and/or operators constructed from a pair of $D$-mesons -- as seen in Fig.~\ref{fig:histos_A1pP_L24}, the mixing between these sectors appears to be large. Levels shown in cyan have dominant overlap with single-hadron-like operators subduced from $J^{PC} = 2^{-+}$ (we will later associate them with a bound $2^{-+}$ state). 

Non-interacting energies are shown by the continuous curves, colored according to the meson-meson combination, and when a non-interacting level is degenerate it is shown by a repeated curve, slightly displaced vertically above.
Wide brown bands indicate three-body combinations of $\eta_c\pi\pi$ and $\chi_{cJ}\pi\pi$ where the $\pi\pi$ part is taken from the $\sigma$ channel, similar to the ``2+1'' non-interacting energies described in Sec.~II.C of Ref.~\cite{Woss:2019hse}. 
At this light quark mass, the $\sigma$ appears as a near-threshold bound state that exerts an influence over a relatively wide region of energy, both above and below threshold, owing to its strong coupling to the $\pi\pi$ channel. The curves are determined from
\begin{align}
E_{\textrm{n.i.}}^{(2+1)} = E_\mathfrak{n,\,\cm}^{\sigma}(\Lambda^P,L) + \Bigl(m_{3}^2 + |\vec{p}_3|^2\Bigr)^{\frac{1}{2}} \, ,
\end{align}
where $m_3$ and $\vec{p}_3$ are the mass and momentum of the $\eta_c$ or $\chi_{cJ}$, and this $E_{\textrm{n.i.}}^{(2+1)}$ in the lattice frame is then boosted back to the $\mathsf{cm}$-frame.

The (almost) volume independent levels lying below the lowest two-meson threshold, $\eta_c\eta$, near $a_tE_\cm\approx 0.63$ in all irreps where $0^{++}$ and/or $2^{++}$ contributes correspond to the stable $\ccz(1P)$ and $\cct(1P)$ states. The impact of such bound-states on the scattering amplitudes above threshold is modest and can be modeled as smooth ``background''. We will not explicitly include description of these subthreshold energy levels as part of our amplitude analysis.\footnote{We will perform a limited analysis to check that they can indeed be neglected in Appendix~\ref{app:bs_pole}.}

At the highest energies, the extracted spectra can become rather dense, and levels can overlap at the level of the statistical uncertainty (although they are distinguished by their orthogonal eigenvectors in the variational approach). This high density is to be expected, given the relatively small mass splittings in the charmed meson sector. The fact that vector mesons appear in scattering channels leads to a large number of possible spin-combinations and these can often subduce into irreps in several different ways for a given momentum combination. In practice, we will make only limited use of the densest parts of the spectrum.

Some indications of the likely resonant content can be read off from the gross structure of the extracted spectra, particularly on the smallest volume. In the $[000]\Lambda^{+}$ irreps which have contributions from $J=0,2$, there are clear additional levels (beyond the number of expected non-interacting levels) around $a_tE_\cm\approx 0.7$, and these have large overlaps onto operators with single-hadron-like and $\DD$-like constructions (see also Fig.~\ref{fig:histos_A1pP_L24}). This could be interpreted as indicating that something remains of the picture inferred from the spectrum found using only single-hadron-like operators, with these states possibly being $0^{++}$ and $2^{++}$ resonances with $\DD$ decays. In addition, the $[000]A_2^{+}$ irrep shows a pattern of levels that could indicate a single isolated $3^{++}$ resonance around $a_t E_\cm\approx0.73$.

On the other hand, aside for a low-lying volume-independent level likely interpretable as a $2^{-+}$ bound state, the $PC=-+$ irreps feature only levels lying on the non-interacting curves, suggesting the absence of any significant scattering strength at these energies.

In this first calculation, we do not consider the energy region above $a_t E_\cm\approx 0.72$, and will not address the possible presence of a $J^{PC} = 4^{++}$ state, nor a second $J^{PC}=2^{++}$ state that might be a member of the $q\bar{q}$ quark-model $1F$ multiplet.

To make more quantitative and robust conclusions about the resonant content, in the next section we use the L\"{uscher} approach to relate the finite-volume spectra to infinite-volume scattering amplitudes.

\section{Scattering amplitudes from finite volume spectra} 

\label{sec:luescher}

In order to translate the finite-volume spectra into infinite-volume scattering amplitudes we make use of the extensions of L\"uscher's finite volume formalism to coupled-channel hadron-hadron scattering~\cite{Briceno:2014oea}. In terms of the infinite-volume scattering $t$--matrix, the phase-space matrix, $\bm{\rho}$, and a matrix of known kinematic functions $\bm{\mathcal{M}}$, this can be expressed as
\begin{align}
\det[\bm{D}]&=0 \, ,\nonumber\\
\bm{D}&= \bm{1} + i\,\bm{\rho}\cdot\bm{t}\cdot\left(\bm{1}+i\,\bm{\mathcal{M}}\right)\,.
\label{eq:det}
\end{align}
The matrices exist in a space of coupled meson-meson channels and partial-waves~\cite{Briceno:2017max}. Given some parameterized $t$-matrix, the roots of Eq.~\ref{eq:det} yield the finite-volume spectrum in a given volume and irrep, $\{E_\mathfrak{n}^{\mathrm{par.}}\}$. The so-obtained spectra can matched level-by-level with the lattice QCD obtained spectrum, $\{E_\mathfrak{n}\}$, and a correlated $\chi^2$ formed, as defined in Eq.~8 of Ref.~\cite{Wilson:2014cna}. Minimization of this $\chi^2$ under variation of the free parameters in $\bm{t}(E_\mathsf{cm})$ then gives a lattice QCD constrained scattering amplitude. Finding the roots of Eq.~\ref{eq:det} in the case of many coupled channels and/or partial-waves can be efficiently achieved by making use of an eigenvalue decomposition of $\bm{D}$, where the eigenvalues can be seperately searched for zeros~\cite{Woss:2020cmp}. The corresponding eigenvectors are also useful, as described below. 

In our previous applications of this approach to lattice QCD data, matching finite-volume energy levels between the lattice calculation and the parameterization has not been a significant issue. The simplest algorithm is to pair levels by their energy order starting from the lowest first. Another straightforward algorithm is to pair levels working from the smallest energy \emph{differences} first. A third option, which suffers from combinatoric growth with the number of levels, is to compute all possible pairings and choose the combination that produces the smallest $\chi^2$. The somewhat novel case encountered in this study features a relatively high density of states high in the spectrum, and here the algorithms above prove to be somewhat imperfect, yielding ambiguous level matching that does not yield a smooth variation of the $\chi^2$ over parameter space. 

An improved approach makes use of the eigenvector information obtained in the decomposition of $\bm{D}$. For small changes in the scattering amplitude, the eigenvectors at each eigenvalue zero vary relatively slowly and can thus be used to help match the spectra obtained from two evaluations with similar parameter values. One method is to insist that the dot product of the eigenvectors for a given energy level for slightly different parameter values is significantly far from zero. 
We have found that under a $\chi^2$ minimization procedure, this helps to ensure a smooth evolution of $\chi^2$ value with changing amplitude parameters, and provides well-defined minima even for very dense spectra.\footnote{An extension of this use of the eigenvectors of $\bm{D}$ helps to identify levels that can be associated with decoupled channels. These typically have overlap only onto meson-meson operators of a particular structure (see Fig.~\ref{fig:histos_A1pP_L24}). In some very limited cases, the dominance of one channel component in the eigenvector of $\bm{D}$ is matched with dominance of one overlap, in order to associate a lattice level with a zero of $\det[\bm{D}]$.}

\vspace{3mm}

Parameterizations of coupled-channel partial-wave \mbox{$t$--matrices} are required that exactly respect unitarity,
which was assumed in the derivation of Eq.~\ref{eq:det}. In this study, we make use of forms that include the flexibility for there to be bound-states and resonances in the $s$--channel scattering process, in particular the $K$-matrix expressions,
\begin{align}
K_{ij} &= \sum_p \frac{g^{(p)}_i g^{(p)}_j}{m_p^2-s} + \sum_a \gamma^{(a)}_{ij} s^a \nonumber \\
[{\bm t}^{-1}]_{ij} &={(2k_i)^{-\ell_i}}[{\bm K}^{-1}]_{ij}{(2k_j)^{-\ell_j}} +{\bm I}_{ij} \, ,
\label{eq:kmat_poles_poly}
\end{align}
where $\bm K$ is a real symmetric matrix for real $s= E_\mathsf{cm}^2$, and $g^{(p)}_i$, $m_p$ and $\gamma^{(a)}_{ij}$ are real parameters. The factors $(2k_i)^{\ell_i}$ implement the behavior at threshold required by angular momentum conservation.

The matrix ${\bm I}$ is diagonal and has imaginary part $\mathrm{Im}\, I_{ij}=-\rho_i=-2k_i/\sqrt{s}$, which precisely accounts for $s$--channel unitarity in the scattering process. The real part of $I_{ij}$ can be fixed to zero, which we will sometimes refer to as a ``simple'' phase space. Alternatively, a dispersion relation can be used to generate a real part from the known imaginary part, which we will refer to as a ``Chew-Mandelstam'' phase space.\footnote{See Appendix B of Ref.~\cite{Wilson:2014cna} for implementation details. The resulting function has the same logarithms as found from scalar loop integrals often implemented in amplitude modeling and effective field theory approaches, see Appendix B of Ref.~\cite{Gayer:2021xzv}.}
The resulting integral features a subtraction, and the location of the subtraction point is a free choice, with convenient choices being the kinematic threshold, or one of the pole locations, $s=m_p^2$.

These amplitude parameterizations do not directly parameterize physics in the $t$-- or $u$--channels that can generate ``left-hand cuts'' which might appear in the energy region considered. We will return to a discussion of this point in Section~\ref{sec:poles}.

\section{Scattering amplitude determinations}              
\label{sec:amps}

\begin{figure}
\includegraphics[width=\columnwidth]{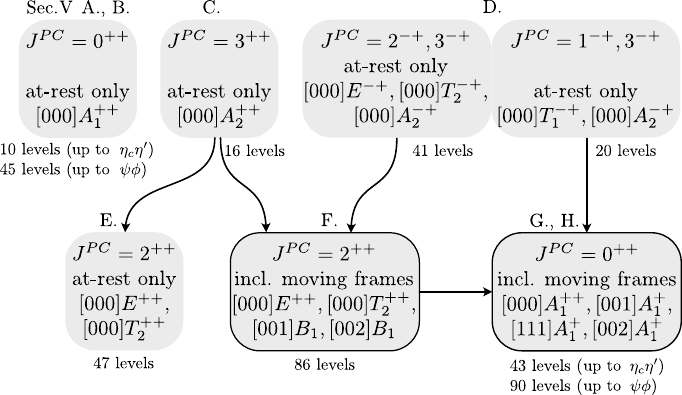}
\caption{A summary of the scattering amplitude extractions carried out in this work, indicating the dependencies of later calculations on fixed ``background'' waves (arrows). The number of energy levels used in each amplitude determination is indicated. We make use of more than 200 energies in total to ultimately determine the $J^{PC}=0^{++}$ and $2^{++}$ amplitudes.}
\label{fig:workflow}
\end{figure}

Our approach to determining scattering amplitudes constrained by the spectra presented in Section~\ref{sec:spectra} will be to proceed systematically, beginning with description of rest-frame irreps receiving contributions from a minimal set of partial waves. Setting the contributions of higher partial waves using rest-frame irreps in which they are leading, we will then be able to analyse moving frame energies with these waves fixed. The workflow is presented in Fig.~\ref{fig:workflow}, where each grey box represents a subsection below.

\subsection{$J^{PC}=0^{++}$ below $\etcep$ and $\DsDs$ threshold from the rest-frame $[000]A_1^+$ irrep}
\label{sec:Swave_rest_DDthreshold}

\begin{figure*}
\includegraphics[width=0.75\textwidth]{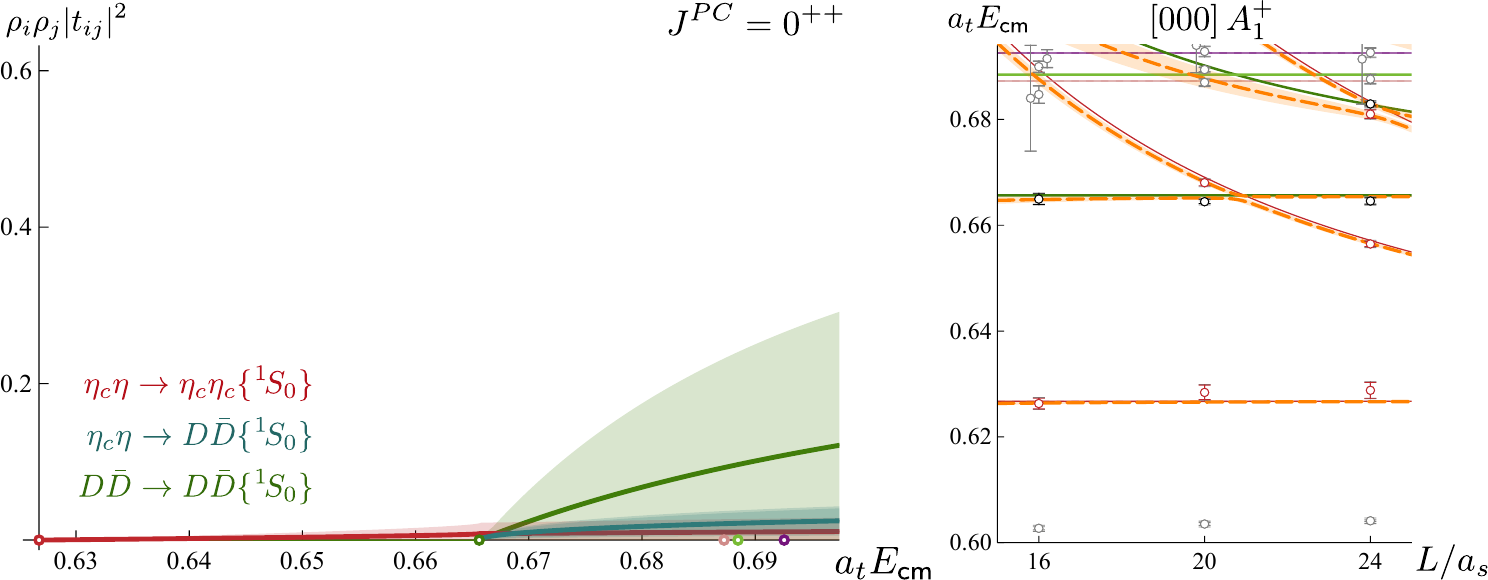}
\caption{Left panel: $J^{PC}=0^{++}$ scattering amplitudes corresponding to Eq.~\ref{fit_rest_ee-DD}. 
Amplitudes are only determined up to the $\etcep$ threshold indicated as the pink circle on the horizontal axis. 
Right panel: The finite volume spectrum in $[000]A_1^+$ from Fig.~\ref{fig:specs} (points) plotted with the solutions of Eq.~\ref{eq:det} with the scattering amplitude as defined by Eq.~\ref{fit_rest_ee-DD} (orange dashed lines with bands). 
The effect of the ``additional'' systematic uncertainty applied before determining the amplitudes as described in Sec.~\ref{sec:lattice} is shown by the outer grey error bars on each energy level (in this case, for most points it is barely visible).}
\label{fig:amp_spec_Swave_two_channel}
\end{figure*}

At the lowest energies, $0^{++}$ is a coupled-channel system of $S$-wave closed-charm $\eta_c \eta$ and open-charm $D\bar{D}$ scattering. In the $[000]A_1^+$ irrep these are the only kinematically open channels below $a_t E_\cm=0.684$, a cutoff selected to lie some way below the $\etcep$ and $D_s\bar{D}_s$ thresholds.

Figure~\ref{fig:specs} indicates only small departures from the relevant non-interacting energies on all three volumes, with possibly very mild attraction at the $D\bar{D}$ threshold. Nothing in the spectrum suggests a near-threshold resonance or bound-state.

The presence in the spectrum of an energy level near $a_t E_\cm=0.604$ on each volume is explained by the stable ground state $\ccz$. Such a deeply bound state will have no direct effect on the scattering amplitudes above threshold, so its presence is not included in amplitude parameterizations.\footnote{In Appendix~\ref{app:bs_pole} we briefly explore amplitudes in which such a bound-state is explicitly included.}

The 10 energy levels in this energy region can be described by a constant $K$-matrix implemented with a threshold-subtracted Chew-Mandelstam phase space, with best-fit parameters, 

\begin{center}
{\small
\begin{tabular}{rll}
$\gamma_{\eta_c\eta\to\eta_c\eta}$ &=  $(0.34 \pm 0.23 \pm 0.09)$ & \multirow{3}{*}{ 
$\begin{bmatrix*}[r]   1.00 &   0.77 &  -0.24\\
&  1.00 &  -0.22\\
&&  1.00\end{bmatrix*}$ } \\ 
$\gamma_{\eta_c\eta\to\DD} $       &= $(0.58 \pm 0.29 \pm 0.05)$ & \\
$\gamma_{\DD \to\DD}$              &= $(1.39 \pm 1.19 \pm 0.24)$ & \\[1.3ex]
&\multicolumn{2}{l}{ $\chi^2/ N_\mathrm{dof} = \frac{5.65}{10-3} = 0.81$\,,}
\end{tabular}
\vspace{-0.5cm}
\begin{equation}\label{fit_rest_ee-DD}\end{equation}
}
\end{center}
\noindent where the first error is statistical, and the second represents the result of varying the stable hadron masses and anisotropy within their errors.\footnote{We perform four additional determinations of the amplitudes, two using the hadron masses at their mean $\pm 1\sigma$ values from Table~\ref{tab:masses}, and two from varying the anisotropy to $\xi_-=3.438$ and $\xi_+=3.450$, the $\pm 1\sigma$ values determined from the pion.} The matrix on the right gives the correlations between parameters. The resulting amplitude is presented in Fig.~\ref{fig:amp_spec_Swave_two_channel}, where it is clear that the system in this energy region is compatible with there being no significant scattering, and there certainly being no near-threshold $\DD$ bound-state. The simplicity of the spectrum indicates no need for more elaborate amplitude parameterizations.

\subsection{$J^{PC}=0^{++}$ up to and including $\psi\phi$ threshold from the rest-frame $A_1^+$ irrep}
\label{sec:Swave_scattering_coupled_rest}

Extending description of the $[000]A_1^+$ irrep to higher energies requires inclusion of the $\eta_c\eta^\prime$, $D_s\bar{D}_s$ and $\psi\omega$ channels which appear in $S$-wave almost simultaneously.\footnote{$\psi\omega$ also produces $\SLJ{5}{D}{0}$ and $\SLJ{5}{D}{4}$ waves that can contribute in $[000]A_1^+$ but these are expected to be suppressed at energies close to threshold.}

In Fig.~\ref{fig:specs} we see that levels with large overlap onto $\eta_c\eta^\prime$ operators tend to be compatible with the corresponding non-interacting energies, but only within rather large uncertainties across all three volumes. 

An ``extra'' level, beyond the counting expected from non-interacting energies, is observed on each volume slightly above $\psi\omega$ threshold, at an energy close to that seen in the spectrum obtained using only single-hadron-like operators presented in Fig.~\ref{fig:spec_qqbar}. As shown in Fig.~\ref{fig:histos_A1pP_L24}, this level has large overlaps onto both the single-hadron-like operators in the basis and the operator resembling $D^\ast_{[000]} \bar{D}^\ast_{[000]}$, motivating the inclusion of the kinematically-closed $D^\ast \bar{D}^\ast\SLJc{1}{S}{0}$ channel into our analysis.\footnote{When including moving frames later, we will consider a more limited energy region below $\DstDst$ threshold in Appendix~\ref{app:amps_Swave_below_DstDst} where the $\DstDst$ channel can be neglected.}

We proceed by considering a system of coupled $\eta_c\eta,\, D\bar{D}, \, D_s\bar{D}_s, \, \eta_c\etap, \, \psi\omega, \, D^\ast\bar{D}^\ast$ and $\psi\phi$ scattering, where each pair is in $S$-wave only, constrained by 45 energy levels (the colored and black levels, excluding the $\ccz$ bound state, shown in the $[000]A_1^+$ panel of Fig.~\ref{fig:specs}).
We have included the $\psi\phi$ channel for which constraint is provided by three levels close to $\psi\phi$ threshold dominated by a $\psi\phi$-like operator construction.

A small complication comes from the presence of a degenerate pair of $\psi_{[100]}\omega_{[100]}$ levels in the non-interacting limit. In order for there to be two such solutions to the quantization condition, the $t$-matrix must feature a $\psi\omega$ \mbox{$D$-wave} as well as $S$-wave, although the impact of what will be a very weak amplitude near threshold is just to supply an energy level lying very close to the non-interacting energy. The simplest option, which we will adopt, is to add a $J^{PC}=4^{++}$ amplitude, $\psi\omega\SLJc{5}{D}{4}\to\psi\omega\SLJc{5}{D}{4}$, parameterized with a $K$-matrix constant.

\begin{figure}
\includegraphics[width=0.82\columnwidth]{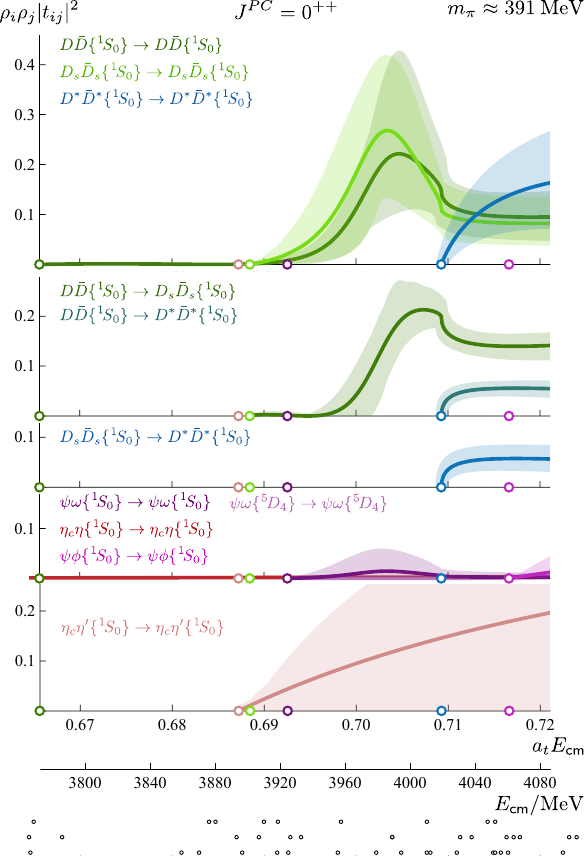}
\caption{Scattering amplitudes with $J^{PC}=0^{++}$ determined from the $[000]A_1^+$ irrep plotted as $\rho_i\rho_j |t_{ij}|^2$ which is limited to a maximum value of 1 by unitarity. Circles on the horizontal axes indicate kinematic thresholds. The open circles at the bottom show the locations of the energy levels providing constraint on the amplitudes. }
\label{fig:fit_A1pP}
\end{figure}

The $J^{PC}=0^{++}$ $t$-matrix is parameterized using a \mbox{$K$-matrix} with a single pole and a matrix of constants,
\begin{align}
K_{ij}=\frac{g_i \, g_j}{m^2-s} + \gamma_{ij} \, ,
\end{align}
and in practice the spectrum mainly lying on the non-interacting energies suggests that we can fix many of the free parameters to zero. In particular, many parameters that when allowed to vary resulted in a value consistent with zero, are then fixed to zero, and the minimization re-run. The Chew-Mandelstam phase-space subtracted at the $K$-matrix pole location is used. An example result is,

\begin{widetext}
\begin{center}
{\small
\setcounter{MaxMatrixCols}{20}
\begin{tabular}{rll}
$a_t m=$ & $(0.7047 \pm 0.0015 \pm 0.0004)$ & \multirow{11}{*}{ $\begin{bmatrix*}[r]   1.00 & \text{-}0.08 & \text{-}0.14 & \text{-}0.26 & \text{-}0.28 & \text{-}0.16 &   0.23 &   0.15 & \text{-}0.13 & \text{-}0.06 & \text{-}0.16\\
&  1.00 &   0.55 &   0.67 &   0.73 & \text{-}0.16 & \text{-}0.72 & \text{-}0.36 & \text{-}0.07 & \text{-}0.08 & \text{-}0.12\\
&&  1.00 &   0.54 &   0.66 & \text{-}0.06 & \text{-}0.21 & \text{-}0.12 & \text{-}0.02 &   0.03 & \text{-}0.03\\
&&&  1.00 &   0.57 & \text{-}0.07 & \text{-}0.51 & \text{-}0.38 & \text{-}0.03 &   0.05 & \text{-}0.10\\
&&&&  1.00 & \text{-}0.01 & \text{-}0.36 & \text{-}0.26 &   0.02 & \text{-}0.08 & \text{-}0.05\\
&&&&&  1.00 &   0.24 &   0.04 &   0.25 &   0.04 &   0.04\\
&&&&&&  1.00 &   0.31 &   0.13 &   0.09 &   0.13\\
&&&&&&&  1.00 &   0.00 &   0.02 &   0.05\\
&&&&&&&&  1.00 &   0.02 &   0.03\\
&&&&&&&&&  1.00 &   0.02\\
&&&&&&&&&&  1.00\end{bmatrix*}$ } \\ 
$a_t g_{\DD}=$     & $(0.150 \pm 0.036 \pm 0.022)$ & \\
$a_t g_{\DsDs}=$   & $(0.193 \pm 0.049 \pm 0.042)$ & \\
$a_t g_{\psiom}=$  & $(0.109 \pm 0.095 \pm 0.145)$ & \\
$a_t g_{\DstDst}=$ & $(0.368 \pm 0.110 \pm 0.066)$ & \\
$\gamma_{\eta_c\eta,\eta_c\eta}=$   & $(0.122 \pm 0.095 \pm 0.052)$ & \\
$\gamma_{\DD,\DD}=$                 & $(-0.476 \pm 0.301 \pm 0.103)$ & \\
$\gamma_{\DD,\DsDs}=$               & $(-1.23 \pm 0.29 \pm 0.12)$ & \\
$\gamma_{\eta_c\etap,\eta_c\etap}=$ & $(1.96 \pm 0.97 \pm 0.69)$ & \\
$\gamma_{\psi\phi,\psi\phi}=$       & $(0.99 \pm 0.90 \pm 0.15)$ & \\
$\gamma_{\psiom,\psiom\SLJc{5}{D}{4}}=$  & $(258 \pm 275 \pm 138) \cdot a_t^8$ & \\[1.3ex]
&\multicolumn{2}{l}{ $\chi^2/ N_\mathrm{dof} = \frac{56.4}{45-11} = 1.66$\,,}
\end{tabular}
\vspace{-0.8cm}
\begin{equation}\label{eq:7chan}\end{equation}
}
\end{center}
\end{widetext}
where all parameters not listed have been fixed to zero. The $K$-matrix pole couplings to open-charm channels appear to be significantly non-zero. Figure~\ref{fig:fit_A1pP} shows the corresponding amplitudes, where clear peaks are visible in the $\DD$ and $\DsDs$ amplitudes, along with a rapid turn-on at threshold of amplitudes leading to $\DstDst$. Examination of the complex energy-plane singularities of this amplitude, reported on later in the manuscript, will lead us to conclude that these effects are due to a single resonance. We will later show that the main features of this amplitude are robust when we vary the specific parameterization used, and when we add constraints from moving-frame irrep spectra.

The value of $\chi^2/N_\mathrm{dof}$ for this fit suggests that there is some mild tension between the computed spectrum and this amplitude, but correlations between computed energy levels play a significant role. All levels are described with a maximum deviation of 1.5$\sigma$, and the same amplitude in an \emph{uncorrelated} fit results in $\chi^2/N_\mathrm{dof}=\frac{34.0}{45 - 11} = 1.00$. The ``global'' systematic uncertainty on the input spectrum introduced in Section~\ref{sec:stable_hadrons} influences the $\chi^2$ and the associated parameter errors (removing it yields $\chi^2/N_\mathrm{dof}=\frac{74.4}{45 - 11} = 2.19$, and errors roughly half as large), but both the parameter central values and the qualitative behavior of the amplitudes remains unchanged. The result presented in Eq.~\ref{eq:7chan} should be viewed as being a conservative estimate of the amplitudes.

The inclusion of the $\eta_c\etap$ channel and the associated levels introduced the level-matching problem described in Section~\ref{sec:spectra}. The large uncertainties on these levels, coupled with the high density of zeroes of the quantization condition, makes many matching assignments plausible. Fortunately, the noisy levels appear to overlap only with the $\eta_c\etap$ operators, suggesting a decoupling that can be built into the amplitude. When $\bm{D}$ in Eq.~\ref{eq:det} was eigendecomposed, the zeroes found in the eigenvalue associated with the $\eta_c\etap\SLJc{1}{S}{0}$ channel could be matched with the levels that have large overlap with the $\eta_c\etap$ operators. The $\eta_c\eta$ and $\psi\phi$ channels, which also appear to be decoupled, were paired in the same way with the eigendecomposition solutions of the quantization condition. The remaining levels were matched by pairing levels working from the smallest energy difference first. 

In order to make use of moving-frame irrep spectra to provide additional constraint on the $0^{++}$ amplitude, we must first constrain the other $J^{PC}$ amplitudes which enter into these irreps by considering other rest-frame irreps in which they are leading. 

\subsection{$J^{PC}=3^{++}$ from rest-frame $A_2^{++}$ irrep}
\label{sec:3pP_scattering}

$3^{++}$ amplitudes appear in several irreps from which we wish to extract $0^{++}$ or $2^{++}$, but we can constrain their low-energy behavior using the spectrum in the $[000]A_2^{++}$ irrep (Fig.~\ref{fig:specs}) where it appears in relative isolation. In the energy region below $a_t E_\cm = 0.72$, where we wish to constrain the amplitudes for use in extracting $0^{++}$ and $2^{++}$, very few levels are present. The lowest level is dominated by overlap with a $\psi\omega$ operator, and on the $L/a_s=24$ volume this is located close to its non-interacting energy. At higher energies larger shifts can be seen and a narrow resonance may be present, as anticipated in Fig.~\ref{fig:spec_qqbar} where the use of only $q\bar{q}$-like operator constructions results in a level around $a_t E_\cm\approx 0.725$. 

Meson-meson channels contributing to $3^{++}$ are (in order of threshold opening): $\DDst\SLJc{3}{D}{3}$, $\psi\omega\SLJc{3,5}{D}{3}$, $\DsDsst\SLJc{3}{D}{3}$, $\DstDst\SLJc{5}{D}{3}$, $\psi\phi\SLJc{3,5}{D}{3}$, and $\DsstDsst\SLJc{5}{D}{3}$. Excluded from this list is $\eta_c \sigma\SLJc{1}{F}{3}$ which we expect to be heavily suppressed by the angular momentum barrier.\footnote{
We included in our basis a three-hadron $\eta_c \pi \pi \sim \eta_c \sigma$ operator constructed from an $\eta_c$ and the variational solution in the $\pi\pi$ system corresponding to the $\sigma$ which is a near-threshold $\pi\pi$ bound-state on these lattices~\cite{Briceno:2016mjc,Briceno:2017qmb}. We observe that a level with large overlap onto this operator is consistent with an $\eta_c$ combined with the lowest $\sigma$ level observed in Ref.~\cite{Briceno:2017qmb} with no clear ``additional interactions''. This level is located around $a_tE_\cm=0.74$ with a relatively large uncertainty.} 

\begin{figure}[h]
\includegraphics[width=0.9\columnwidth]{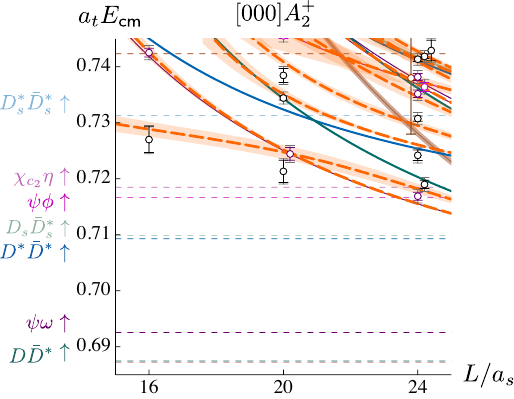}
\caption{As for the right panel of Fig.~\ref{fig:amp_spec_Swave_two_channel}, except in the $[000]A_2^{++}$ irrep with the solutions from the amplitude in Eq.~\ref{eq:fit_3pP}. Several channels such as $\etce$ and $\eta_c\pi\pi$ open below the plotted range.}
\label{fig:spec_fitted_A2pP}
\end{figure}

\begin{figure*}
\includegraphics[width=0.9\textwidth]{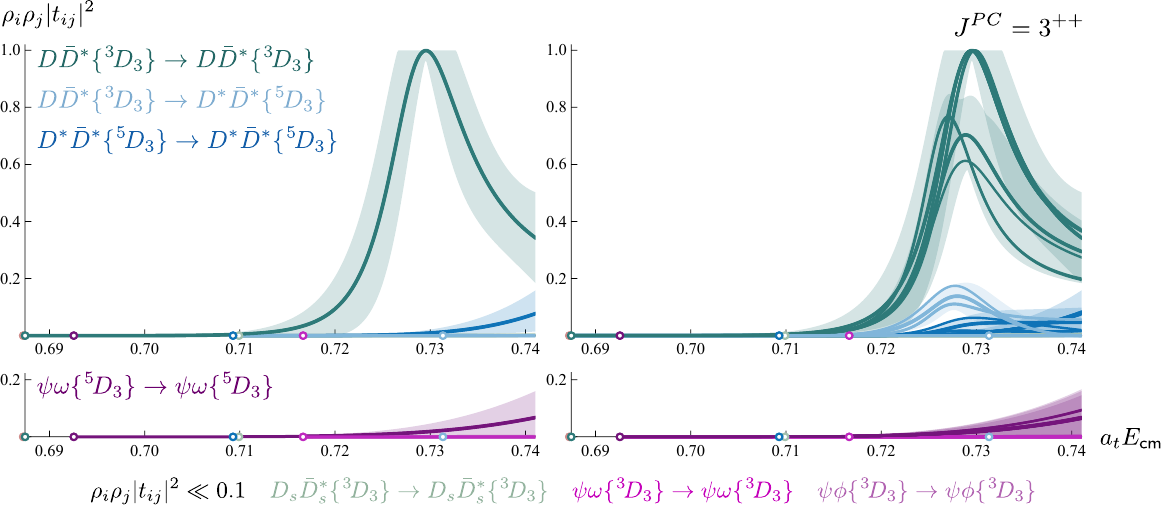}
\caption{$J^{PC}=3^{++}$ reference amplitude (left) as given in Eq.~\eqref{eq:fit_3pP} and other acceptable descriptions of the finite volume spectra obtained by parameterisation variations (right) as summarized in Table~\ref{tab:J3pP_par_var}. Several channels included, but found to be very small, are listed below the figure. $\DDst\SLJc{3}{D}{3}\to\DDst\SLJc{3}{D}{3}$ and in some cases $\DDst\SLJc{3}{D}{3}\to\DstDst\SLJc{5}{D}{3}$ have clear peaks indicative of a nearby resonance.
}
\label{fig:amp_F_par_var}
\end{figure*}

We consider 16 levels below $a_t E_\cm=0.743$ having excluded the single level with large overlap onto the $\eta_c \sigma$ operator. A description using a $K$-matrix pole and matrix of constants, with Chew-Mandelstam phase-space subtracted at the pole, is given by,
\begin{widetext}
\begin{center}
{\small
\begin{tabular}{rll}
$a_t m=$ & $(0.7295 \pm 0.0017 \pm 0.0002)$ & \multirow{10}{*}{ $\begin{bmatrix*}[r]   1.00 &   0.06 &  0.00 &   0.00 &   0.26 & \text{-}0.25 & \text{-}0.22 & \text{-}0.17 & \text{-}0.13 & \text{-}0.27\\
&  1.00 & \text{-}0.01 & \text{-}0.00 &   0.88 &   0.32 &   0.37 &   0.00 &   0.02 & \text{-}0.08\\
&&  1.00 & \text{-}0.00 & \text{-}0.01 & \text{-}0.00 &   0.00 &   0.00 &   0.00 &   0.00\\
&&&  1.00 & \text{-}0.00 & \text{-}0.00 &   0.01 &   0.00 & \text{-}0.00 & \text{-}0.00\\
&&&&  1.00 &   0.29 &   0.50 &   0.02 & \text{-}0.00 & \text{-}0.13\\
&&&&&  1.00 &   0.27 &   0.08 &   0.08 &   0.03\\
&&&&&&  1.00 &   0.23 &   0.06 &   0.08\\
&&&&&&&  1.00 &   0.02 &   0.14\\
&&&&&&&&  1.00 &   0.07\\
&&&&&&&&&  1.00\end{bmatrix*}$ } \\ 
$g_{\DDst \SLJc{3}{D}{3}}=$   & $(2.51 \pm 0.35 \pm 0.08) \cdot a_t$ & \\
$g_{\DstDst \SLJc{5}{D}{3}}=$ & $(0.00 \pm 1.38 \pm 0.12) \cdot a_t$ & \\
$g_{\DsDsst \SLJc{3}{D}{3}}=$ & $(0.00 \pm 0.69 \pm 0.07) \cdot a_t$ & \\
$\gamma_{\DDst \SLJc{3}{D}{3}   \to \DDst \SLJc{3}{D}{3}}=$      & $(53 \pm 153 \pm 40)\cdot a_t^4$ & \\
$\gamma_{\DstDst \SLJc{5}{D}{3} \to \DstDst \SLJc{5}{D}{3}}=$  & $(-462 \pm 122 \pm 105)\cdot a_t^4$ & \\
$\gamma_{\DsDsst \SLJc{3}{D}{3} \to \DsDsst \SLJc{3}{D}{3}}=$  & $(54 \pm 184 \pm 24)\cdot a_t^4$ & \\
$\gamma_{\psiom \SLJc{3}{D}{3}  \to \psiom \SLJc{3}{D}{3}}=$    & $(343 \pm 210 \pm 55)\cdot a_t^4$ & \\
$\gamma_{\psiom \SLJc{5}{D}{3}  \to \psiom \SLJc{5}{D}{3}}=$    & $(-26 \pm 40 \pm 13)\cdot a_t^4$ & \\
$\gamma_{\psiphi \SLJc{3}{D}{3} \to \psiphi \SLJc{3}{D}{3}}=$  & $(-19 \pm 628 \pm 75)\cdot a_t^4$ & \\[1.3ex]
&\multicolumn{2}{l}{ $\chi^2/ N_\mathrm{dof} = \frac{8.34}{16-10} = 1.39$\,,}
\end{tabular}
\vspace{-0.9cm}
\begin{equation}\label{eq:fit_3pP}\end{equation}
}
\end{center}
\end{widetext}
where a noticeable feature is a pole with a significant coupling to $D \bar{D}^*\{\SLJ{3}{D}{3}\}$. We will refer to this description as the ``reference amplitude''. The reproduction of the lattice QCD energy levels from the finite volume formalism is shown in Fig.~\ref{fig:spec_fitted_A2pP}, and the amplitudes in Eq.~\ref{eq:fit_3pP} appear in the left panel of Fig.~\ref{fig:amp_F_par_var}.

The amplitude above proves to not be a \emph{unique} description of the finite volume spectrum, with other parameterizations giving solutions which have a significant $D \bar{D}^\ast, D^\ast \bar{D}^\ast$ cross-term, as shown in the right panel of Fig.~\ref{fig:amp_F_par_var} and summarised in Table~\ref{tab:J3pP_par_var} in Appendix \ref{app:Fwave_par_vars}.
For the purposes of serving as a ``background wave'' in irreps where we seek $0^{++}$ and $2^{++}$ amplitudes, we only require the $3^{++}$ amplitude below $a_t E_\cm \approx 0.72$, and there the various amplitude descriptions all broadly agree. We will use the reference amplitude presented above for this purpose.

\clearpage

\subsection{$J^{PC}=\{1,2,3\}^{-+}$ from rest-frame irreps}
\label{sec:amps_2mP_3mP}
In order to use moving-frame irreps to constrain $0^{++}$ and $2^{++}$ partial-waves, we must also consider negative parities, and these are most directly extracted from the at-rest irreps presented in Fig.~\ref{fig:specs_mP}. The spectra indicate that the interactions are relatively weak, with the only non-trivial feature being an ``extra'' level located around $a_tE_\cm=0.68$ in the $E_2^-$ and $T_2^-$ irreps. The mild volume-dependence and large overlaps of this level with single-hadron-like operators subduced from $2^{-+}$ suggest that this is a stable $\eta_{c2}$ state. In Ref.~\cite{Liu:2012ze}, which used only single-hadron-like operators, the pattern of states extracted in this energy region resembled a quark model $1D$ multiplet, with this state being the $q\bar{q}(\SLJ{1}{D}{2})$ member~\cite{Godfrey:1985xj}. The state lies below all open-charm decay thresholds, but slightly above $\eta_c\pi\pi$. For the same reasons as in Section \ref{sec:3pP_scattering}, we expect only a very weak coupling to $\eta_c\pi\pi$ and do not include this three-meson channel in our amplitude analysis.

Considering $\{1,2,3\}^{-+}$, the meson-meson partial-waves that contribute are given in Table~\ref{tab:pwa_chans}. We choose to first determine $J^{PC}=2^{-+}$ and exotic $3^{-+}$ using energy levels in the $E^{-}$, $T_2^{-}$ and $A_2^{-}$ irreps. Considering all levels below $a_t E_\cm = 0.73$ except those with dominant overlap with an $\eta_c\pi\pi$ operator (which are decoupled from other operators), we have 41 energies. 
Using a pole plus constant $K$-matrix for $2^{-+}$ and a matrix of constants for $3^{-+}$, the following parameters provide a good description of the finite-volume spectra:

\begin{widetext}
\begin{center}
{
\small

\begin{tabular}{rll}
$a_t m_{2^-} =$ & $(0.67538 \pm 0.00063 \pm 0.00025)$ & \multirow{9}{*}{ $\begin{bmatrix*}[r]   1.00 &   0.06 & \text{-}0.08 & \text{-}0.05 & \text{-}0.06 & \text{-}0.28 & \text{-}0.11 & \text{-}0.13 & \text{-}0.24\\
&  1.00 & \text{-}0.99 &   0.32 &   0.18 &   0.25 &   0.13 &   0.07 &   0.20\\
&&  1.00 & \text{-}0.37 & \text{-}0.17 & \text{-}0.20 & \text{-}0.11 & \text{-}0.05 & \text{-}0.15\\
&&&  1.00 &   0.21 &   0.18 &   0.10 &   0.07 &   0.52\\
&&&&  1.00 &   0.18 &   0.01 &   0.00 &   0.21\\
&&&&&  1.00 &   0.17 &   0.14 &   0.34\\
&&&&&&  1.00 &   0.45 &   0.20\\
&&&&&&&  1.00 &   0.16\\
&&&&&&&&  1.00\end{bmatrix*}$ } \\ 
$g_{\DDst\SLJc{3}{P}{2}} = $ & $(-1.73 \pm 0.64 \pm 0.25)$ & \\ 
$\gamma_{\DDst\SLJc{3}{P}{2}\to\DDst\SLJc{3}{P}{2}} =$  & $(43.1 \pm 35.6 \pm 20.3) \cdot a_t^2$ & \\
$\gamma_{\DDst\SLJc{3}{F}{2}\to\DDst\SLJc{3}{F}{2}} =$ & $(3676 \pm 2118 \pm 1640) \cdot a_t^6$ & \\
$\gamma_{\DsDsst\SLJc{3}{P}{2}\to\DsDsst\SLJc{3}{P}{2}} = $  & $(21.9 \pm 27.3 \pm 8.8) \cdot a_t^2$ & \\
$\gamma_{\DstDst\SLJc{3}{P}{2}\to\DstDst\SLJc{3}{P}{2}} = $  & $(-4.9 \pm 24.2 \pm 62.2) \cdot a_t^2$ & \\
$\gamma_{\psiom\SLJc{3}{P}{2}\to\psiom\SLJc{3}{P}{2}} = $ & $(-6.32 \pm 11.4 \pm 68.3) \cdot a_t^2$ & \\
$\gamma_{\psiom\SLJc{5}{P}{2}\to\psiom\SLJc{5}{P}{2}} = $ & $(-6.63 \pm 12.0 \pm 59.7) \cdot a_t^2$ & \\
$\gamma_{\cct\eta\SLJc{5}{S}{2}\to\cct\eta\SLJc{5}{S}{2}} = $ & $(-0.42 \pm 0.91 \pm 1.56)$ & \\[1.3ex] 
\end{tabular}
\begin{tabular}{rll}
$\gamma_{\eta_c\eta\SLJc{1}{F}{3}\to \eta_c\eta\SLJc{1}{F}{3}} =$   & $(243 \pm 295 \pm 49) \cdot a_t^6$ & \multirow{3}{*}{ $\begin{bmatrix*}[r]   1.00 &   0.04 &   0.05\\
&  1.00 &   0.09\\
&&  1.00\end{bmatrix*}$ } \\ 
$\gamma_{\DDst\SLJc{3}{F}{3}\to\DDst\SLJc{3}{F}{3}} = $  & $(-13 \pm 1758 \pm 460) \cdot a_t^6$ & \\
$\gamma_{\psiom\SLJc{5}{P}{3}\to \psiom\SLJc{5}{P}{3}} =$  & $(-14.4 \pm 9.9 \pm 51.3) \cdot a_t^2$ & \\[1.3ex]
&\multicolumn{2}{l}{ $\chi^2/ N_\mathrm{dof} = \frac{24.9}{41-12} = 0.86$\,,}
\end{tabular}
}

\vspace{-0.5cm}
\begin{equation}\label{eq:fit_2mP_3mP}\end{equation}
\end{center}
\end{widetext}
The $K$-matrix pole in $2^{-+}$ is allowed a coupling only to the lowest-lying open-charm partial-wave, $\DDst\SLJc{3}{P}{2}$; all other couplings are fixed to zero.

\vspace{3mm}

An independent description of energy levels in the $T_1^{-}$ and $A_2^{-}$ irreps yields $1^{-+}$ and $3^{-+}$ amplitudes. While a hybrid meson is expected in $1^{-+}$, it will lie above the considered energy region, and well above open-charm three-meson thresholds. Using 20 levels below $a_t E_\cm=0.721$, the small energy shifts can be described by constant \mbox{$K$-matrices},

\begin{widetext}
\begin{center}

\begin{tabular}{rll}
$\gamma_{\DDst\SLJc{3}{P}{1}\to\DDst\SLJc{3}{P}{1}} =$ & $(14.0 \pm 7.6 \pm 8.1) \cdot a_t^2$ & \multirow{6}{*}{ $\begin{bmatrix*}[r]   1.00 &   0.18 &  -0.00 &  -0.02 &   0.01 &   0.09\\
&  1.00 &   0.00 &   0.01 &   0.02 &   0.10\\
&&  1.00 &   0.15 &   0.16 &   0.03\\
&&&  1.00 &   0.65 &   0.02\\
&&&&  1.00 &   0.03\\
&&&&&  1.00\end{bmatrix*}$ } \\ 
$\gamma_{\eta_c\eta\SLJc{1}{P}{1}\to\eta_c\eta\SLJc{1}{P}{1}} =$ & $(1.73 \pm 1.51 \pm 0.63) \cdot a_t^2$ & \\
$\gamma_{\psiom\SLJc{1}{P}{1}\to\psiom\SLJc{1}{P}{1}}=$          & $(-36.4 \pm 22.8 \pm 6.14)\cdot a_t^2$ & \\
$\gamma_{\psiom\SLJc{3}{P}{1}\to\psiom\SLJc{3}{P}{1}}=$          & $(-1.61 \pm 26.64 \pm 5.89)\cdot a_t^2$ & \\
$\gamma_{\psiom\SLJc{5}{P}{1}\to\psiom\SLJc{5}{P}{1}}=$          & $(23.33 \pm 27.14 \pm 6.01)\cdot a_t^2$ & \\
$\gamma_{\cco\eta\SLJc{3}{S}{1}\to\cco\eta\SLJc{3}{S}{1}}=$      & $(0.65 \pm 1.06 \pm 0.99)$ & \\

\end{tabular}
\begin{tabular}{rll}
$\gamma_{\eta_c\eta\SLJc{1}{F}{3}\to\eta_c\eta\SLJc{1}{F}{3}} =$ & $(-282 \pm 201 \pm 61) \cdot a_t^6$ & \multirow{3}{*}{ $\begin{bmatrix*}[r]   1.00 &  -0.06 &   0.15\\
&  1.00 &   0.02\\
&&  1.00\end{bmatrix*}$ } \\ 
$\gamma_{\DDst\SLJc{3}{F}{3}\to\DDst\SLJc{3}{F}{3}} = $ & $(1213 \pm 1286 \pm 1268) \cdot a_t^6$ & \\
$\gamma_{\psiom\SLJc{5}{P}{3}\to\psiom\SLJc{5}{P}{3}} =$ & $(-24.4 \pm 20.6 \pm 5.0) \cdot a_t^2$ & \\[1.3ex]
&\multicolumn{2}{l}{ $\chi^2/ N_\mathrm{dof} = \frac{7.11}{20-9} = 0.65$\,.}
\end{tabular}

\vspace{-0.5cm}
\begin{equation}\label{eq:fit_1mP_3mP}\end{equation}
\end{center}
\end{widetext}
where the $3^{-+}$ amplitude so obtained is statistically compatible with that extracted previously in Eq.~\ref{eq:fit_2mP_3mP}.

\begin{figure}[h]
\includegraphics[width=0.90\columnwidth]{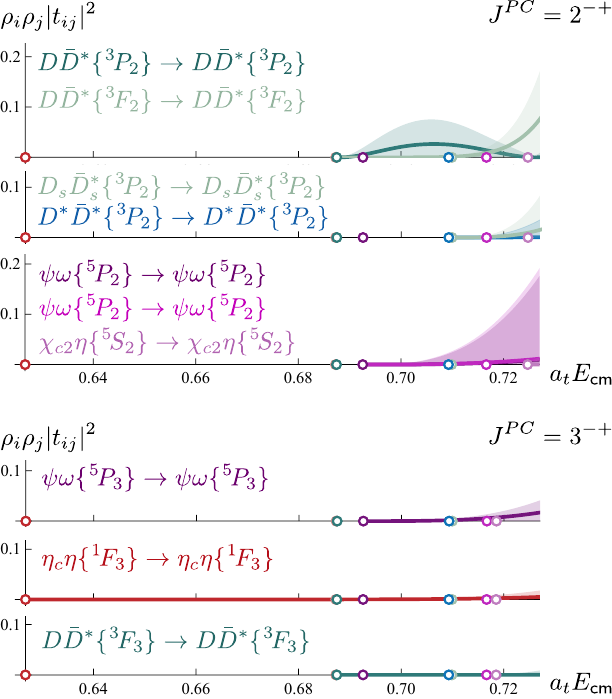}
\caption{Scattering amplitudes determined in $J^{PC}=2^{-+}$ and $3^{-+}$ presented in Eq.~\ref{eq:fit_2mP_3mP}. The bound-state present in $2^{-+}$, while coupled to $\DDst$, does not produce significant scattering above threshold.}
\label{fig:amp_J2mP_J3mP}
\end{figure}

\begin{figure}[h]
\vspace{0.3cm}
\includegraphics[width=0.90\columnwidth]{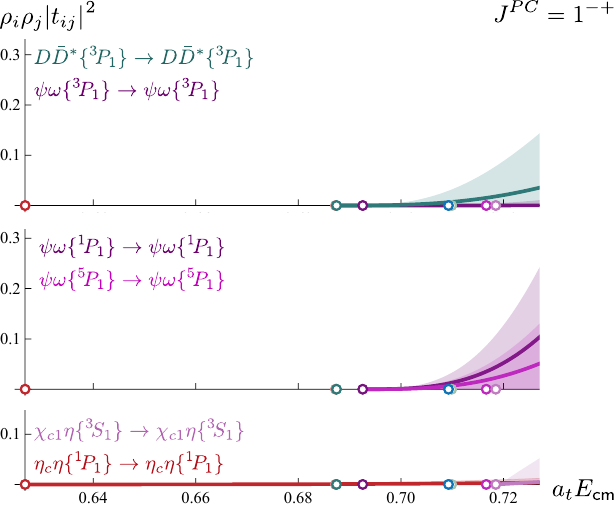}
\caption{As Fig.~\ref{fig:amp_J2mP_J3mP} but for the $J^{PC}=1^{-+}$ amplitude presented in Eq.~\ref{eq:fit_1mP_3mP}.} 
\label{fig:amp_ref_J1mP}
\end{figure}

The $J^{-+}$ amplitudes are shown in Figs.~\ref{fig:amp_J2mP_J3mP} and \ref{fig:amp_ref_J1mP}.	It is clear that the negative parity waves are weak at low energies, and we will later find they have only modest influence in moving-frame irreps.

\pagebreak
\subsection{$J^{PC}=2^{++}$ from rest frame irreps}
\label{sec:2pP_rest}

We now turn to the $J^{PC}=2^{++}$ partial wave, investigating the energy region up to just above the $\psi\phi$ threshold where, based on Fig.~\ref{fig:spec_qqbar}, we expect to reveal the lowest resonance in this channel. Several meson-meson channels contribute, and the large energy shifts seen in Fig.~\ref{fig:specs} indicate significant interactions. 

We begin by determining amplitudes using energies in the $E^{+}$ and $T_2^{+}$ irreps, where the $3^{++}$ contributions to $T_2^{+}$ are fixed using the reference amplitude determined in section~\ref{sec:3pP_scattering} -- these amplitudes are quite small in this energy region and the precise details do not significantly impact the results for $2^{++}$.
We exclude the $\eta_c\etap$ channel and a single energy level that arises in the $E^+$ irrep on the $L/a_s=24$ volume dominated by overlap onto an \mbox{$\eta_c\etap$-like} operator. We also exclude one level having dominant overlap onto an $\eta_c\pi\pi$-like operator in the $T_2^+$ irrep on the $L/a_s=24$ volume. In total there are 47 levels that can be used to constrain the amplitudes.

\pagebreak
Considering all $S$-wave channels open below ${a_t E_\cm=0.717}$, and those $D$-wave channels opening at lower energies (except for $\eta_c\etap$), we will describe the $2^{++}$ sector as a coupled $\eta_c\eta\SLJc{1}{D}{2}$, $\DD\SLJc{1}{D}{2}$, $\DDst\SLJc{3}{D}{2}$, $\DsDs\SLJc{1}{D}{2}$, $\psi\omega\SLJc{5}{S}{2}$, $\DstDst\SLJc{5}{S}{2}$ and $\psi\phi\SLJc{5}{S}{2}$ system.\footnote{
  Similar to section~\ref{sec:Swave_scattering_coupled_rest}, some $\psi\omega$ $D$-waves are required to produce a sufficient number of solutions. There proves to be insufficient constraint to uniquely determine all amplitudes featuring $\psi\omega$ in $D$-wave. 
The required number of levels in the considered energy region is obtained from the finite-volume determinant condition provided $\SLJ{5}{S}{2}$, $\SLJ{3}{D}{3}$ and $\SLJ{5}{D}{3}$ waves are included (the latter two from $3^{++}$).
}

\begin{figure}
\includegraphics[width=0.99\columnwidth]{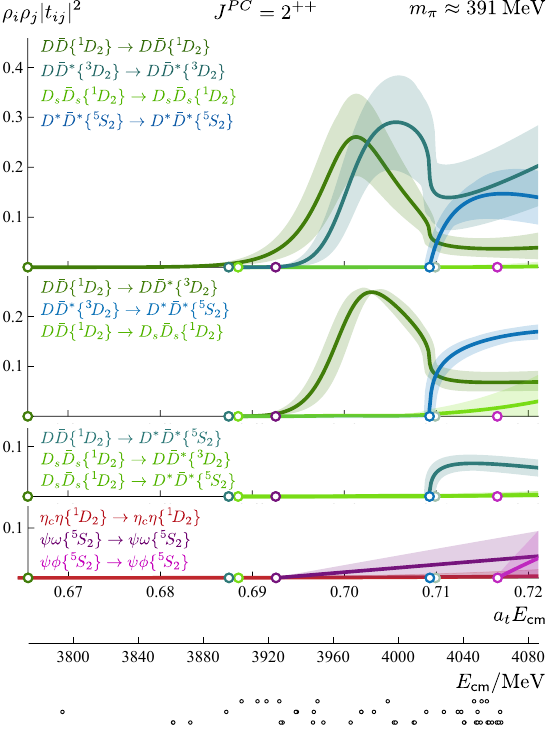}
\caption{
  As Fig.~\ref{fig:fit_A1pP}, but for amplitudes with $J^{PC}=2^{++}$ in Eq.~\ref{eq:fit_2pP_rest} determined from the rest frame $[000]E^+$ and $[000]T_2^+$ irreps.
  }
\label{fig:amps_2pP_rest}
\end{figure}

A $K$-matrix of the form in Eq.~\ref{eq:kmat_poles_poly}, which includes the appropriate $k_i^\ell$ threshold factors, is capable of describing the energy levels. One suitable example in which the \mbox{$K$-matrix} pole has couplings only to open-charm channels, and a Chew-Mandelstam phase-space subtracted at the $K$-matrix pole location, is,

\begin{widetext}
\begin{center}
{\small
\begin{tabular}{rll}
$a_tm = $ & $(0.7030 \pm 0.0010 \pm 0.0002)$ & \multirow{8}{*}{ $\begin{bmatrix*}[r]   1.00 &  -0.05 &  -0.00 &  -0.18 &  -0.13 &  -0.06 &  -0.08 &  -0.01\\
&  1.00 &  -0.08 &  -0.42 &   0.15 &   0.05 &  -0.05 &  -0.13\\
&&  1.00 &   0.08 &   0.01 &  -0.05 &  -0.01 &  -0.00\\
&&&  1.00 &   0.02 &   0.02 &   0.08 &  -0.03\\
&&&&  1.00 &   0.04 &   0.02 &  -0.01\\
&&&&&  1.00 &   0.06 &   0.06\\
&&&&&&  1.00 &   0.22\\
&&&&&&&  1.00\end{bmatrix*}$ } \\ 
$g_{\DDst\SLJc{3}{D}{2}} = $ & $(-30.1 \pm 4.5 \pm 0.8) \cdot a_t$ & \\
$g_{\DsDs\SLJc{1}{D}{2}} = $  & $(1.53 \pm 2.17 \pm 0.40) \cdot a_t$ & \\
$g_{\DstDst\SLJc{5}{S}{2}} = $  & $(1.67 \pm 0.18 \pm 0.13) \cdot a_t^{-1}$ & \\
$\gamma_{\eta_c\eta\SLJc{1}{D}{2},\eta_c\eta\SLJc{1}{D}{2}} = $ & $(20.4 \pm 23.9 \pm 8.17) \cdot a_t^4$ & \\
$\gamma_{\DD\SLJc{1}{D}{2},\DsDs\SLJc{1}{D}{2}} = $ & $(182 \pm 138 \pm 18) \cdot a_t^4$ & \\
$\gamma_{\psiom\SLJc{5}{S}{2},\psiom\SLJc{5}{S}{2}} = $  & $(-0.884 \pm 0.449\pm 0.057)$ & \\
$\gamma_{\psi\phi\SLJc{5}{S}{2},\psi\phi\SLJc{5}{S}{2}} = $  & $(1.61 \pm 0.77 \pm 0.04)$ & \\
$g_{\DD\SLJc{1}{D}{2}}=$ & $10\cdot a_t$ {\bf (fixed)}\\[1.3ex]
&\multicolumn{2}{l}{ $\chi^2/ N_\mathrm{dof} = \frac{48.0}{47-8} = 1.23$\,,}
\end{tabular}
}
\vspace{-0.5cm}
\begin{equation}\label{eq:fit_2pP_rest}\end{equation}
\end{center}
\end{widetext}
where the resulting amplitude is plotted in Fig.~\ref{fig:amps_2pP_rest}. A clear resonance-like bump is observed in $D\bar{D}$ and $D\bar{D}^*$, along with a rapid turn-on of $D^* \bar{D}^*$ at threshold.

The amplitude presented in Eq.~\ref{eq:fit_2pP_rest} has the unusual feature that the $K$-matrix pole coupling to $D\bar{D}$ is \emph{fixed} to an apparently arbitrary value. The origin of this is an empirical observation that when describing the finite-volume spectra, there proves to be essentially no sensitivity to the \emph{absolute scale} of the couplings $g$, but only to their \emph{ratios}. This is a novel finding, so far unique to this case, but one which seems to have an explanation in terms of there being a $\chi_{c2}$ resonance having a large coupling to the \emph{kinematically closed} $S$-wave $D^* \bar{D}^*$ channel.

The coupling-ratio phenomenon can be illustrated using a simple two-channel Flatt\'e amplitude specialized to describe a resonance lying above threshold for channel 1, and below threshold for channel 2,\footnote{For simplicity we will put both channels in $S$-wave, although the logic requires only the higher channel to be in $S$-wave.}
\begin{equation}
t_{ij}(s) = \frac{g_i \, g_j}{m_0^2 -s -i \sum_{k=1}^2 g_k^2 \, \rho_k(s)} \, ,
\label{eq:flatte}
\end{equation}
where it is convenient to remove the channel 2 ``self-energy'' contribution to the resonance mass by defining an $m$ such that $m_0^2 = m^2 - g_2^2 | \rho_2(m^2) |$, so that the Flatt\'e denominator takes the form
\begin{equation*}
D(s) = m^2 -s - i g_1^2\,  \rho_1(s) - i g_2^2\,  \big( \rho_2(s) - \rho_2(m^2) \big) \, .
\end{equation*}
If we restrict to the region around $s = m^2$, taken to be below the threshold for channel 2, we can approximate
\begin{align*}
D(s \approx m^2) &= \\ - g_2^2 \tfrac{\beta}{m} &\left( 1 + \tfrac{2 m^2}{g_2^2  \beta} \right) \left[ \sqrt{s} - m + i \big(\tfrac{g_1}{g_2} \big)^2 \frac{m/\beta}{ 1 + \tfrac{2 m^2}{g_2^2  \beta} } \rho_1(s) \right] \, ,
\end{align*}
where $\beta = \tfrac{4m_2^2}{m^2} \tfrac{1}{|\rho_2(m^2)|}$. This indicates an amplitude that depends only on the \emph{ratio} $g_1/g_2$ in the limit that $g_2 \gg \sqrt{\tfrac{2m^2}{\beta}}$. Some consequences of this property are investigated in Appendices~\ref{app:toy_D} and \ref{app:extra_poles}.

Interpretations of the amplitude given in Eq.~\ref{eq:fit_2pP_rest} in terms of resonant content and channel couplings will be presented in Section \ref{sec:poles} by considering the rigorously defined complex-energy pole content of the $t$-matrix.

\pagebreak
\subsection{$J^{PC}=2^{++}$ from rest and moving-frame irreps}
\label{sec:2pP_rest_and_mov}

Additional constraint on the $2^{++}$ amplitude comes from moving-frame irreps. In order to use energy levels in the $[001],[002]B_{1,2}$ irreps to further constrain the $2^{++}$ amplitude, we supply our previously determined $2^{-+}$ and $3^{\pm +}$ amplitudes, fixing them to the central values found. The determinant condition one is working with here is of unprecedented size, featuring seven $2^{++}$ channels, seven $2^{-+}$ channels, three $3^{-+}$ channels and six $3^{++}$ channels. The techniques presented in Ref.~\cite{Woss:2020cmp} are invaluable in handling such a large dimensional problem.

In practice we choose to reduce the complexity of the minimization problem by adding only the energy levels from the irreps $[001]B_1$ and $[002]B_1$ on the $L/a_s=20,\,24$ volumes, leading to a total of 86 levels to constrain the $2^{++}$ interactions. We checked that the resulting amplitudes also give a reasonable description of the computed finite-volume spectra in the $[001]B_2$ and $[002]B_2$ irreps.

A challenge associated with using these 86 levels in a minimization is that there are considerable data correlations between energy levels computed on the same lattice volume. Upon eigendecomposition, the data correlation matrix is found to have a relatively small number of large eigenvalues which are likely to be reliably determined, and many more much smaller eigenvalues that may not be well determined on a limited number of gauge configurations. In an earlier study, various approaches to deal with such correlations were explored, such as uncorrelated fits, fitting to subsets of the data, and removing the smallest eigenvalues by a singular value decomposition (SVD)~\cite{Cheung:2020mql}. Alternative strategies and summaries of the issue can be found in Refs.~\cite{Michael:1993yj,Michael:1994sz,Dowdall:2019bea,Bruno:2022mfy}. Here we opt to remove eigenmodes with the smallest eigenvalues when inverting the covariance matrix, associating the cut with a reduction in the number of degrees of freedom by which we judge the $\chi^2$. We find that retaining all eigenvalues $\lambda_i$ where $\Lambda=\lambda_i/\max(\lambda)>0.02$ results in a reasonable description of the data. This cut leads to removal of 4 and 19 eigenmodes from the $L/a_s=20$ and $24$ spectra respectively. A detailed discussion is presented in Appendix~\ref{app:svd}.

Using the same $J^{PC}=2^{++}$ parameterization as in Eq.~\ref{eq:fit_2pP_rest}, with the additional moving-frame energy levels included, we obtain,

\begin{widetext}
\begin{center}
\small
\begin{tabular}{rll}
$a_tm = $ & $(0.7025 \pm 0.0012 \pm 0.0007)$ & \multirow{8}{*}{ $\begin{bmatrix*}[r]   1.00 &  -0.04 &  -0.00 &  -0.23 &  -0.14 &  -0.03 &  -0.02 &  -0.03\\
&  1.00 &  -0.08 &  -0.28 &   0.13 &   0.10 &  -0.07 &  -0.12\\
&&  1.00 &   0.00 &  -0.02 &  -0.51 &   0.25 &   0.08\\
&&&  1.00 &  -0.02 &  -0.15 &  -0.03 &  -0.06\\
&&&&  1.00 &   0.03 &   0.06 &   0.02\\
&&&&&  1.00 &  -0.23 &  -0.01\\
&&&&&&  1.00 &   0.27\\
&&&&&&&  1.00\end{bmatrix*}$ } \\ 
$g_{\DDst\SLJc{3}{D}{2}} = $  & $(-37.9 \pm 5.0 \pm 3.94) \cdot a_t$ & \\
$g_{\DsDs\SLJc{1}{D}{2}} = $  & $(-3.3 \pm 4.3 \pm 2.5) \cdot a_t$ & \\
$g_{\DstDst\SLJc{1}{S}{2}} = $ & $(1.58 \pm 0.15 \pm 0.22) \cdot a_t^{-1}$ & \\
$\gamma_{\eta_c\eta\SLJc{1}{D}{2}\to\eta_c\eta\SLJc{1}{D}{2}} = $ & $(16.3 \pm 23.1 \pm 7.5) \cdot a_t^4$ & \\
$\gamma_{\DD\SLJc{1}{D}{2}\to\DsDs\SLJc{1}{D}{2}} = $  & $(-81 \pm 129 \pm 100) \cdot a_t^4$ & \\
$\gamma_{\psiom\SLJc{5}{S}{2}\to\psiom\SLJc{5}{S}{2}} = $ & $(0.55 \pm 0.72 \pm 0.81)$ & \\
$\gamma_{\psi\phi\SLJc{5}{S}{2}\to\psi\phi\SLJc{5}{S}{2}} = $  & $(2.19 \pm 0.77 \pm 0.11)$ & \\
$g_{\DD\SLJc{1}{D}{2}}=$ & $10\cdot a_t$ (fixed)\\[1.3ex]
&\multicolumn{2}{l}{ $\chi^2/ N_\mathrm{dof} = \frac{62.8}{86 - 8 - 23} = 1.14$\,,}
\end{tabular}

\vspace{-0.5cm}
\begin{equation}\label{eq:fit_2pP_rest+flight}\end{equation}
\end{center}
\end{widetext}
which agrees within uncertainties with the amplitude determined from the $[000]E^{++}$ and $[000]T_2^{++}$ energies alone.\footnote{The number of degrees of freedom is taken to be ${N_\mathrm{dof} = N_\mathrm{levels}-N_\mathrm{pars.}-N_\mathrm{reset}}$, where $N_\mathrm{levels}$ is the number of energies, $N_\mathrm{pars.}$ is the number of free parameters, and $N_\mathrm{reset}$ is the number of eigenmodes removed from the covariance matrix by the cut on eigenvalues.} This parameterization's description of the rest frame energy levels can be seen in the middle and right panels of Fig.~\ref{fig:rest_irreps_fvs}.

\begin{figure}
\includegraphics[width=0.84\columnwidth]{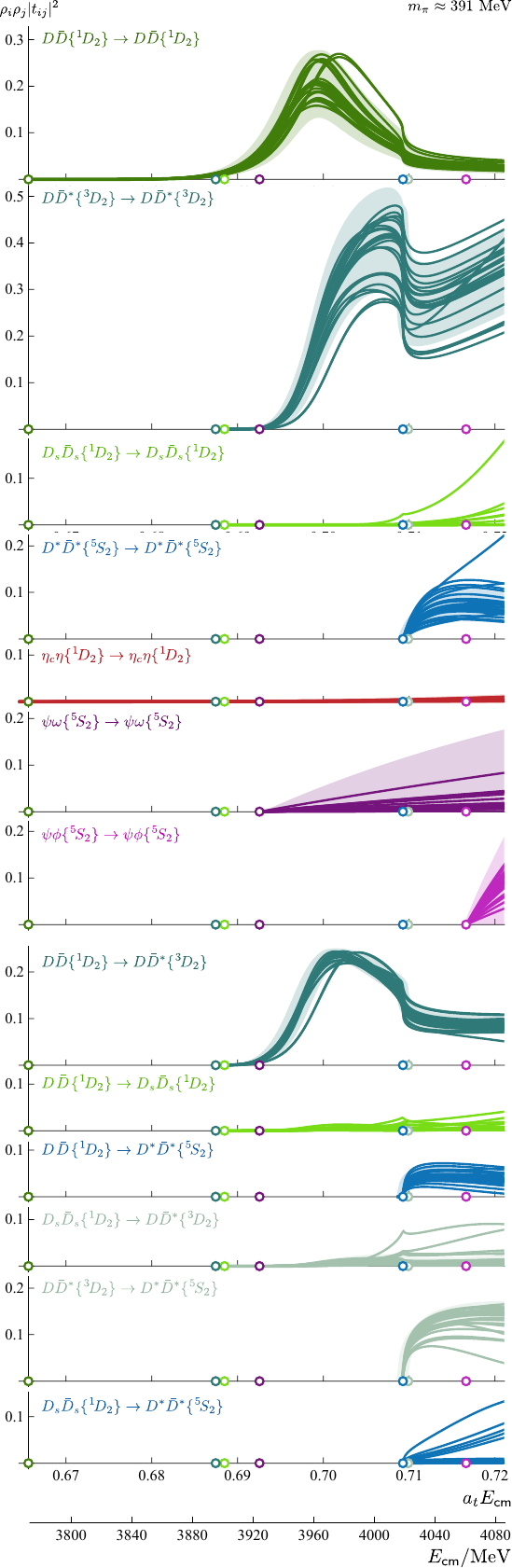}
\caption{
As Fig.~\ref{fig:amps_2pP_rest}, but for amplitudes with $J^{PC}=2^{++}$ in Eq.~\eqref{eq:fit_2pP_rest+flight} determined from the rest and moving frame irreps. Solid curves show the central values from the parameterization variations summarized in Table~\ref{tab:J2pP_par_var}.
}
\label{fig:Dwave_amp_par_var}
\end{figure}

The choice of parameterization form is varied to investigate bias associated with choosing any specific form. We take the form implied in Eq.~\ref{eq:fit_2pP_rest+flight} and vary which constants and couplings are present or set to zero compared to Eq.~\ref{eq:fit_2pP_rest+flight}, we vary the Chew-Mandelstam subtraction point, and also replace it with the simple phase-space. In addition, the choice of which $K$-matrix pole coupling is fixed is varied, choosing $g_{\DsDs\SLJc{1}{D}{2}}$ or $g_{\DDst\SLJc{1}{D}{2}}$ rather than $g_{\DD\SLJc{1}{D}{2}}$, although this is found to have a negligible effect. We also explore the sensitivity to the data correlation eigenvalue cutoff. Overall, we consider 24 parameterizations that give a reasonable description of the finite-volume spectra and these are summarised in Table \ref{tab:J2pP_par_var} in Appendix~\ref{app:Dwave_par_vars}.
The central values of these parameterizations are compared with the reference parameterization from Eq.~\ref{eq:fit_2pP_rest+flight} in Fig.~\ref{fig:Dwave_amp_par_var}, and we find that the central values of the majority of parameterizations fall within the error bands obtained from Eq.~\ref{eq:fit_2pP_rest+flight}.
\footnote{One obvious weakness is that there are relatively few parameterizations with couplings between open-charm channels and $\psi\omega\SLJc{5}{S}{2}$. Although no good $\chi^2$ minima were found with the $g_{\psiom\SLJc{5}{S}{2}}$ parameter allowed to vary, many were attempted and all of these appeared to produce a small $\psi\omega\SLJc{5}{S}{2}$ amplitude. This parameter was freed in some amplitude determinations of the rest-frame energies, only one of which is given in Eq.~\ref{eq:fit_2pP_rest_fudged}, and in that case it was found to be consistent with zero. A $\gamma_{\DD\SLJC{1}{D}{2}\to\psiom\SLJC{5}{S}{2}}$ term was included in one parameterization. Considering the relevant spectra in Fig.~\ref{fig:specs}, there are no clear large shifts involving the $\psi\omega$ levels, and so perhaps it is a reasonable conclusion that these amplitudes are small.}

The amplitudes in Fig.~\ref{fig:Dwave_amp_par_var} that are not small show very similar features to those in Fig.~\ref{fig:amps_2pP_rest}. There is a clear resonance-like bump in $D\bar{D}$ and $D\bar{D}^\ast$, and a rapid turn-on of $D^\ast \bar{D}^\ast$ at threshold. A small number of parameterizations appear to have some large $D_s\bar{D}_s$ amplitudes at high energies, although there is relatively little constraint in this region. The other amplitudes, including all closed-charm channels, are consistent with being small. We will explore the singularity content of these amplitudes in Section~\ref{sec:poles}.

\pagebreak
\subsection{$J^{PC}=0^{++}$ below $\etcep$ and $\DsDs$ thresholds including moving frame energies}
\label{sec:Swave_threshold_mov}

\begin{figure*}
\includegraphics[width=1.0\textwidth]{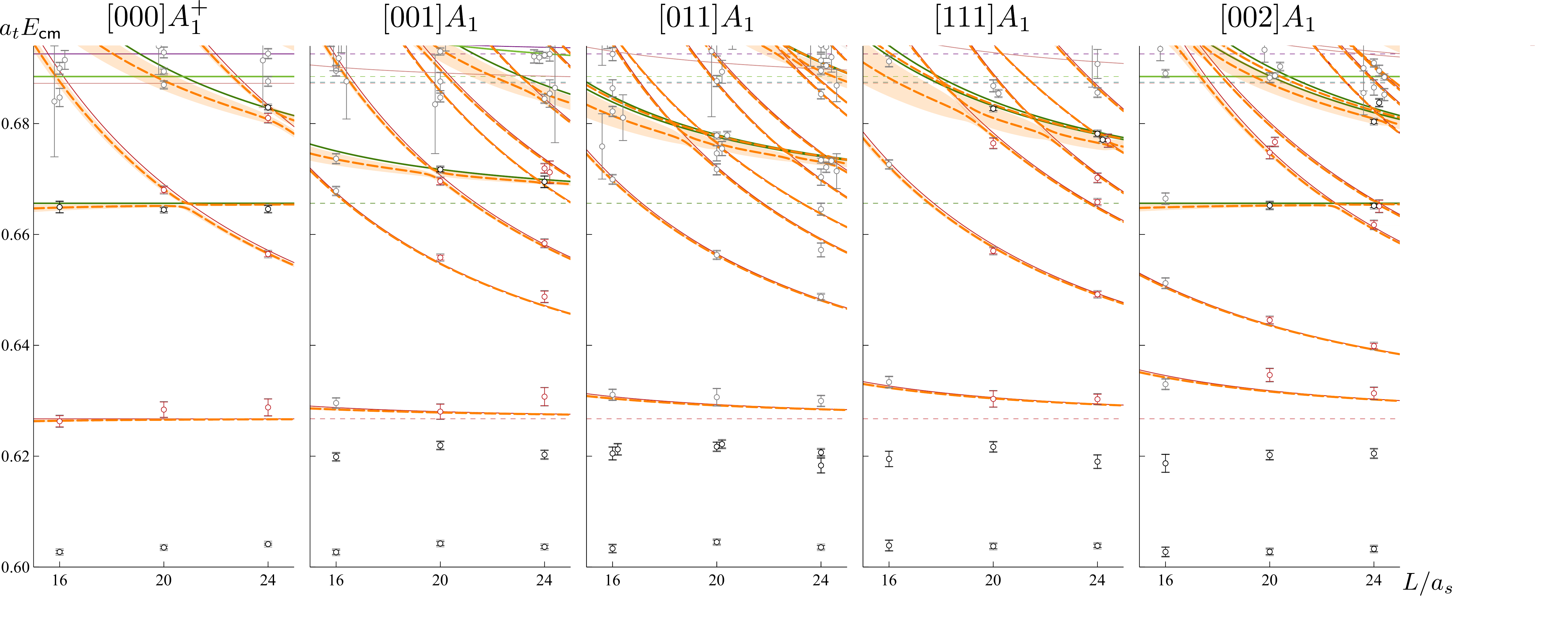}
\caption{Energy levels from $[000]A_1^{+}$ and moving frame $[ijk]A_1$ irreps as in Fig.~\ref{fig:specs} (points) compared with the spectra from the coupled-channel $\eta_c\eta\SLJc{1}{S}{0}-\DD\SLJc{1}{S}{0}$ amplitude in Eq.~\ref{eq:fit_2chan_mov} using the finite-volume quantization condition Eq.~\ref{eq:det} (dashed orange curves and bands). 
Energies plotted in gray were not used in this amplitude determination, neither are the bound state levels around $a_tE_\cm=0.62$ and below corresponding to the stable $\chi_{c0,2}(1P)$.}
\label{fig:specs_A1s_fitted}
\end{figure*}

The region around the $D\bar{D}$ threshold, previously constrained using only rest-frame irrep energy levels, can be reconsidered including the additional constraint from moving-frame irreps. This analysis further confirms the previous conclusion that there is no near-threshold scalar bound-state in this system.
In addition to $[000]A_1^{+}$, constraint comes from energy levels in the $[001]A_1$, $[111]A_1$ and $[002]A_1$ irreps, with a total of 43 energy levels below $\eta_c\eta^\prime$ threshold. For this selection of irreps, in this energy region, all higher partial waves can be neglected.\footnote{We choose to exclude energy levels in the $[011]A_1$ irrep, which receive contributions from $2^{-+}$ that may not be negligible due to a $2^{-+}$ bound state. We later show in Fig.~\ref{fig:specs_A1s_fitted} that the levels in this irrep are in fact in good agreement in this energy region.}

\begin{figure}
\includegraphics[width=0.87\columnwidth]{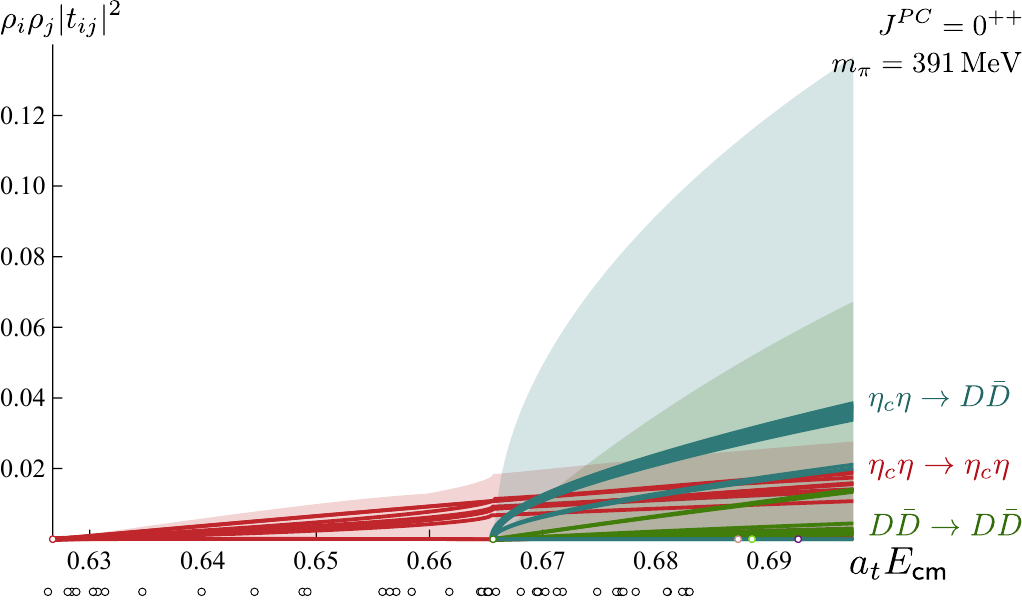}
\caption{As Fig.~\ref{fig:amps_2pP_rest}, but for coupled-channel $\eta_c\eta\SLJc{1}{S}{0}-\DD\SLJc{1}{S}{0}$ amplitudes with $J^{PC}=0^{++}$ determined from $[000]A_1^{+}$ and moving frame energies. The individual curves correspond to the central values of the parameterization variations listed in Table~\ref{tab:Swave_threshold_par_var}. The bands show the extent of the uncertainties from the amplitude in Eq.~\ref{eq:fit_2chan_mov}.}
\label{fig:amp_spec_Swave_two_channel_mov}
\end{figure}

For these levels, a reasonable description using a constant $K$-matrix is found,
%
\begin{tabular}{rll}
$\gamma_{\eta_c\eta\to\eta_c\eta}=$ & $0.369 \pm 0.145 \pm 0.047$ & \multirow{3}{*}{ $\begin{bmatrix*}[r]   1.00 &  \text{-}0.37 &   0.06\\
&  1.00 &  \text{-}0.31\\
&&  1.00\end{bmatrix*}$ } \\ 
$\gamma_{\eta_c\eta\to \DD} =$      & -$0.638 \pm 0.157 \pm 0.988$ & \\
$\gamma_{\DD\to\DD} =$              & $0.172 \pm 0.324 \pm 2.162$ & \\
\end{tabular}
\vspace*{-3mm}
\begin{equation}
\chi^2/ N_\mathrm{dof} = \tfrac{40.5}{43-3-5} = 1.16\,,
\label{eq:fit_2chan_mov}
\end{equation}
that is in qualitative agreement with the amplitude found earlier. The description of the finite-volume spectra is shown in Figure~\ref{fig:specs_A1s_fitted}. This parameterization and 9 other variations\footnote{Details of the parameterization variations are provided in Appendix~\ref{app:Swave_par_vars_two_chan}.} are plotted in Fig.~\ref{fig:amp_spec_Swave_two_channel_mov}, where we again observe no signal indicating strong interactions near $D\bar{D}$ threshold.

\subsection{$J^{PC}=0^{++}$ up to and including $\psi\phi$ threshold}
\label{sec:Swave_scattering_final}

\begin{figure*}
\includegraphics[width=0.85\textwidth]{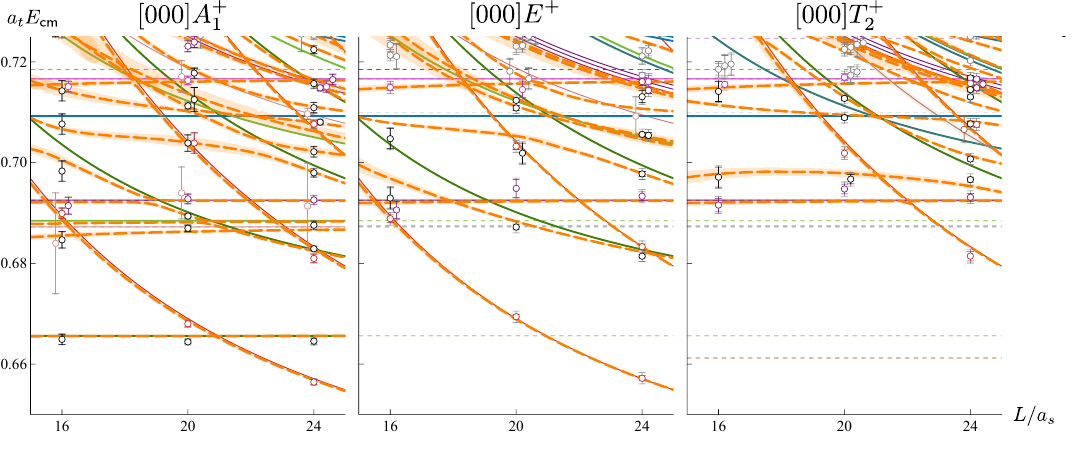}
\caption{As Fig.~\ref{fig:specs_A1s_fitted} but for spectra in $[000]A_1^+$, $[000]E^+$ and $[000]T_2^+$ irreps compared with solutions from the amplitudes in Eqs.~\ref{eq:fit_0pP_rest+flight} (left panel) and \ref{eq:fit_2pP_rest+flight} (middle and right panels).}
\label{fig:rest_irreps_fvs}
\end{figure*}

Our most highly constrained $0^{++}$ amplitude comes from simultaneously describing energy levels in the $[000]\, A_1^+$, $[001]\, A_1$, $[111]\, A_1$ and $[002]\, A_1$ irreps up to $a_tE_\cm=0.724$ at rest (just above $\psi\phi$ threshold) and up to $a_t E_\cm=0.69$ in moving frame irreps  (just above $\DsDs$ threshold). The 90 energy levels are subject to a significant degree of data correlation which we mollify by removing small eigenmodes below a cutoff $\Lambda = 0.02$ as described in Section \ref{sec:2pP_rest_and_mov}.\footnote{This results in the removal of 16 eigenmodes. A range of values are used when we vary the parameterization, including neglecting the correlations entirely.}

An amplitude of the form used in Section~\ref{sec:Swave_scattering_coupled_rest} describes the spectra with parameter values,
\begin{widetext}
\vspace*{-3mm}
\small
\begin{align}
\begin{tabular}{rll}
$a_t m = $ & $(0.7065 \pm 0.0015 \pm 0.0004)$ & \multirow{10}{*}{ $\begin{bmatrix*}[r]   1.00 & \text{-}0.05 & \text{-}0.17 &   0.02 & \text{-}0.33 & \text{-}0.26 &   0.23 & \text{-}0.06 & \text{-}0.03 & \text{-}0.23\\
&  1.00 &   0.52 & \text{-}0.44 &   0.64 &   0.08 & \text{-}0.07 &   0.04 & \text{-}0.04 &   0.01\\
&&  1.00 & \text{-}0.55 &   0.69 &   0.05 & \text{-}0.23 & \text{-}0.03 &   0.00 &   0.02\\
&&&  1.00 & \text{-}0.54 &   0.05 &   0.33 &   0.02 & \text{-}0.03 &   0.05\\
&&&&  1.00 &   0.09 & \text{-}0.24 &   0.02 & \text{-}0.08 &   0.03\\
&&&&&  1.00 & \text{-}0.07 &   0.22 &   0.01 &   0.08\\
&&&&&&  1.00 &   0.05 & \text{-}0.03 & \text{-}0.04\\
&&&&&&&  1.00 & \text{-}0.00 &   0.03\\
&&&&&&&&  1.00 & \text{-}0.01\\
&&&&&&&&&  1.00\end{bmatrix*}$ } \\ 
$a_t g_{\DD\SLJc{1}{S}{0}}=$      & $(0.1174 \pm 0.0226 \pm 0.0039)$ & \\
$a_t g_{\DsDs\SLJc{1}{S}{0}}=$    & $(0.189 \pm 0.046 \pm 0.026)$ & \\
$a_t g_{\psiom\SLJc{1}{S}{0}}=$   & $(-0.127 \pm 0.069\pm 0.230)$ & \\
$a_t g_{\DstDst\SLJc{1}{S}{0}}=$  & $(0.330 \pm 0.095 \pm 0.023)$ & \\
$\gamma_{\eta_c\eta\SLJc{1}{S}{0}\to\eta_c\eta\SLJc{1}{S}{0}}=$     & $(0.144 \pm 0.097 \pm 0.038)$ & \\
$\gamma_{\DD\SLJc{1}{S}{0}\to\DsDs\SLJc{1}{S}{0}}=$                 & $(-0.974 \pm 0.301 \pm 0.027)$ & \\
$\gamma_{\eta_c\etap\SLJc{1}{S}{0}\to\eta_c\etap\SLJc{1}{S}{0}}=$   & $(2.55 \pm 1.03 \pm 0.73)$ & \\
$\gamma_{\psi\phi\SLJc{1}{S}{0}\to\psi\phi\SLJc{1}{S}{0}}=$         & $(1.36 \pm 0.90 \pm 0.26)$ & \\
$\gamma_{\psiom\SLJc{5}{D}{4}\to\psiom\SLJc{5}{D}{4}}=$    & $(162 \pm 254 \pm 43) \cdot a_t^8$ & \\[1.3ex]
&\multicolumn{2}{l}{ $\chi^2/ N_\mathrm{dof} = \frac{91.0}{90 - 10 - 16} =  1.42$\,,}
\end{tabular}
\label{eq:fit_0pP_rest+flight}
\end{align}
\end{widetext}
where the description of the rest frame energy levels can be seen in the leftmost panel of Fig.~\ref{fig:rest_irreps_fvs}.
We have fixed the $J^{PC}=1^{-+},3^{-+}$ amplitudes to the results in Eq.~\ref{eq:fit_1mP_3mP}, the $2^{++}$ amplitude is fixed to the result in Eq.~\ref{eq:fit_2pP_rest+flight}, and the $3^{++}$ amplitude is fixed to the result in Eq.~\ref{eq:fit_3pP}.

Exploring a range of $K$-matrix parameterizations, we find certain features that must be present to successfully describe the lattice QCD spectra. Coupling a $K$-matrix pole to the open-charm channels ($D\bar{D}$, $D_s \bar{D}_s$, $D^* \bar{D}^*$) appears to be required, and these couplings are always significantly non-zero, while the $K$-matrix entries corresponding to the $\psi\omega$ channels are always small and typically consistent with zero.

\begin{figure}
\includegraphics[width=0.86\columnwidth]{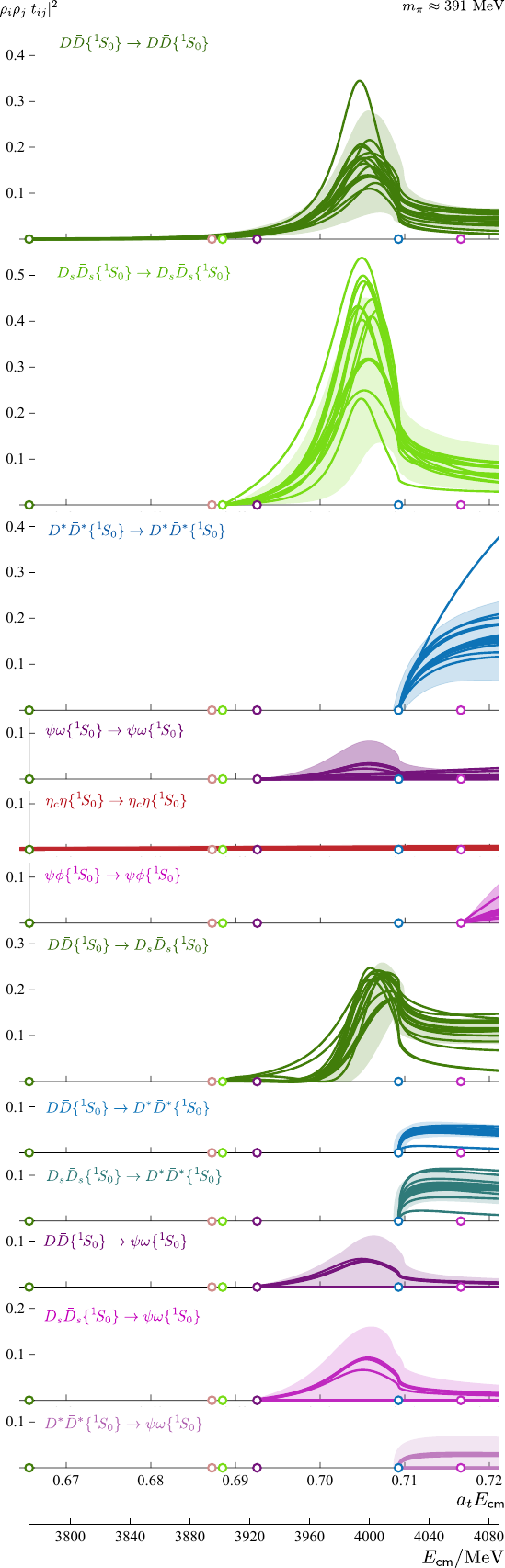}
\caption{
As Fig.~\ref{fig:Dwave_amp_par_var} but for $J^{PC}=0^{++}$ amplitudes with the band showing the amplitude in Eq.~\ref{eq:fit_0pP_rest+flight}. Parameterization variations are summarized in Table~\ref{tab:Swave_mov_par_var}. }
\label{fig:Swave_amp_par_var}
\end{figure}

The results of describing the finite-volume spectra with a range of parameterization choices (listed in Table~\ref{tab:Swave_mov_par_var}) are presented in Fig.~\ref{fig:Swave_amp_par_var}, where we see that they are qualitatively similar with a single large enhancement around $a_t E_\cm\approx 0.705$. The $\DstDst\SLJc{1}{S}{0}$ channel opens rapidly at threshold, a phenomenon we will later associate with a large resonance coupling to $\DstDst\SLJc{1}{S}{0}$. 

\vspace{3mm}
With a set of well-constrained $0^{++}$ and $2^{++}$ scattering amplitudes in hand, in the following sections we will determine their pole singularities and present a corresponding interpretation in terms of resonances.

\section{Resonance poles}                   

\label{sec:poles}

Scattering amplitudes, considered as a function of \emph{complex} values of the scattering energy squared can have only certain features due to analyticity. As well as the branch cuts required by unitarity, \emph{pole singularities} can be present, having an interpretation as the bound-states and resonances of the scattering system. 

A new branch cut for each channel, opening at the kinematical threshold, defines a Riemann sheet structure, with the \emph{physical sheet}, where scattering occurs for real energies, having $\mathrm{Im} \, k_i > 0$ for all channels, $i$. For a given energy, the unphysical sheet reached by moving down through the cut, known as the \emph{proximal sheet}, has ${\mathrm{Im} \, k_i < 0}$ for all kinematically open channels and ${\mathrm{Im} \, k_i > 0}$ for all closed channels.

Close to a pole,
\begin{equation}
t_{ij}(s \approx s_0) = \frac{c_i \,c_j}{s_0 - s} \, ,
\end{equation}
and a nearby pole on the proximal sheet will generate rapid energy dependence on the real energy axis, typically taking the form of a peak, the canonical resonance lineshape. The pole location in the complex energy plane has an interpretation in terms of the resonance mass and width, ${\sqrt{s_0} = m \pm \tfrac{i}{2} \Gamma}$, while the factorized pole residues give the channel couplings, $c_i$. Except when they lie close to thresholds, poles on \emph{other} unphysical sheets typically have only a weak influence on physical scattering. 

The $K$-matrix parameterizations we have explored in this paper have good analytic properties, such that they can be continued into the complex energy plane without difficulty. We will explore to what extent resonance pole locations and channel couplings are independent of the details of the specific parameterization chosen. Experience shows that for narrow resonances, where the pole is close to physical scattering (such as the $\rho$~\cite{Dudek:2012xn,Wilson:2015dqa}), very little variation under changes in parameterization is seen, while for broad resonances, lying far into the complex plane (such as the $\sigma$~\cite{Pelaez:2015qba,Briceno:2016mjc,Rodas:2023twk} or the $D_0^\ast$~\cite{Gayer:2021xzv}), a much more significant scatter over parameterizations can be observed.

While factorized pole-residue couplings, $c_i$, are the most rigorous way to quantify the coupling of a resonance to a channel, it is also common to use partial-widths, $\Gamma_i$, or branching ratios, $\mathrm{Br}_i$, to describe decay rates to \emph{open} channels.
A prescription relating couplings to partial-widths, expected to be reasonable for narrow resonances, has been provided by the PDG~\cite{ParticleDataGroup:2022pth},
\begin{align}
\mathrm{Br}_i=\frac{\Gamma_i}{\Gamma}=|c_i|^2 \frac{\rho_i(\mathrm{Re}\:s_0)}{\sqrt{(\mathrm{Re}\:s_0)}}\,.
\label{eq:partial}
\end{align}

\pagebreak
\subsection{Scalar resonance}\label{sec:poles:scalar}
\begin{figure*}
\includegraphics[width=0.9\textwidth]{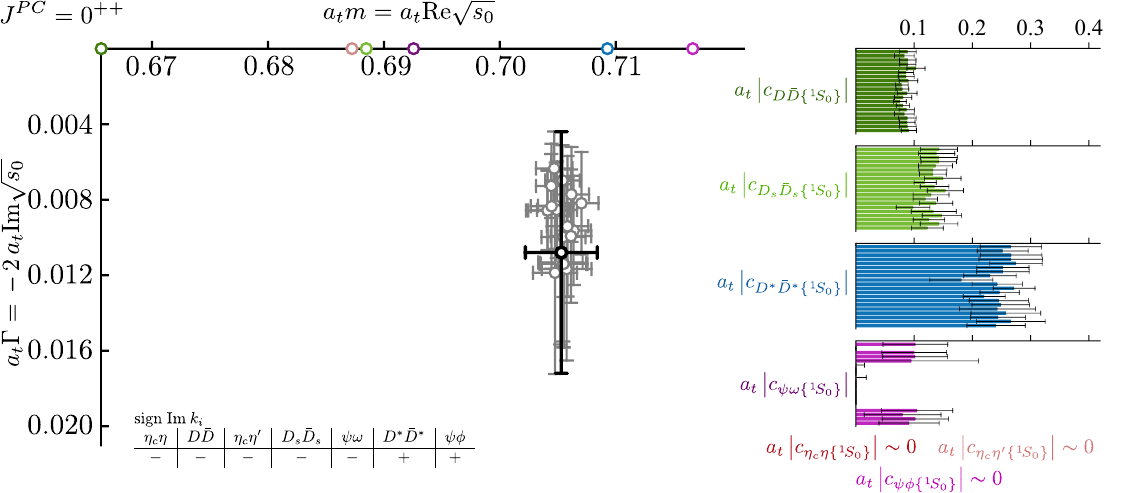}
\caption{The $J^{PC}=0^{++}$ pole and couplings found on the ``proximal'' sheet between $\psi\phi$ and $\DstDst$ thresholds. 
In the left panel grey points indicate the pole position on each successful amplitude parameterization and the black point shows the final quoted pole position and uncertainty as described in the text. In the right panel each histogram bar represents the value of that coupling found in one successful parameterization.}
\label{fig:Swave_pole_par_var}
\end{figure*}

For $J^{PC}= 0^{++}$, considering the analysis in section~\ref{sec:Swave_scattering_final} with variation of parameterizations summarized in appendix~\ref{app:Swave_par_vars}, we consistently find a pole on the proximal sheet between $\psiom$ and $\DstDst$ thresholds. We denote the relevant Riemann sheet using the notation 
\begin{align}
&\mathrm{sign}(\mathrm{Im}(k_{\etce},k_{\DD},k_{\etcep},k_{\DsDs},k_{\psiom},k_{\DstDst},k_{\psi\phi}))\nonumber\\ 
&=(-,-,-,-,-,+,+)\nonumber\\ 
&={\scriptstyle(\etce{[-]},\DD{[-]},\etcep{[-]},\DsDs{[-]},\psiom{[-]},\DstDst{[+]},\psi\phi{[+]})}.\nonumber
\end{align}
We always order the channels by their threshold energies, so that a proximal sheet can be identified by a sequence of ``$-$" followed by a sequence of ``$+$". 
Thus at these energies $\scriptstyle(\etce{[-]},\DD{[-]},\etcep{[-]},\DsDs{[-]},\psiom{[-]},\DstDst{[+]},\psi\phi{[+]})$ is the proximal sheet. The pole on this sheet found when varying the parameterization is shown in Figure~\ref{fig:Swave_pole_par_var}.
The pole is located at
\begin{align}
a_t\sqrt{s_0}  =    &        (0.7050 \pm 0.0025) - \tfrac{i}{2}(0.0120 \pm 0.0070) \nonumber\\
\sqrt{s_0}  \approx &  \, 3995 \pm 14 - \tfrac{i}{2}(67 \pm 38) \,\mathrm{MeV} \nonumber\,,
\end{align}
where the quoted uncertainties are conservatively taken as the envelope of the individual uncertainties from each parameterization, and the quoted central values are taken as the centre of the envelope in complex-$E_\cm$.

The pole residue factorizes into channel couplings,
\begin{align}
a_t |c_{\etce\SLJc{1}{S}{0}}|   \approx \; & 0\nonumber\\
a_t |c_{\DD\SLJc{1}{S}{0}}|     =       \; & 0.093(28) \nonumber\\
a_t |c_{\etcep\SLJc{1}{S}{0}}|  \approx \; & 0\nonumber\\
a_t |c_{\DsDs\SLJc{1}{S}{0}}|   =       \; & 0.128(56) \nonumber\\
a_t |c_{\psiom\SLJc{1}{S}{0}}|  =       \; & 0.083(83) \nonumber\\
a_t |c_{\DstDst\SLJc{1}{S}{0}}| =       \; & 0.227(97)\nonumber\\
a_t |c_{\psiphi\SLJc{1}{S}{0}}| \approx \; & 0  \, ,
\end{align}
where the uncertainties quoted again reflect the envelope over all of the individual parameterizations. We find no evidence for significant couplings to channels with a charmonium and light meson.

\begin{figure*}
\includegraphics[width=0.9\textwidth]{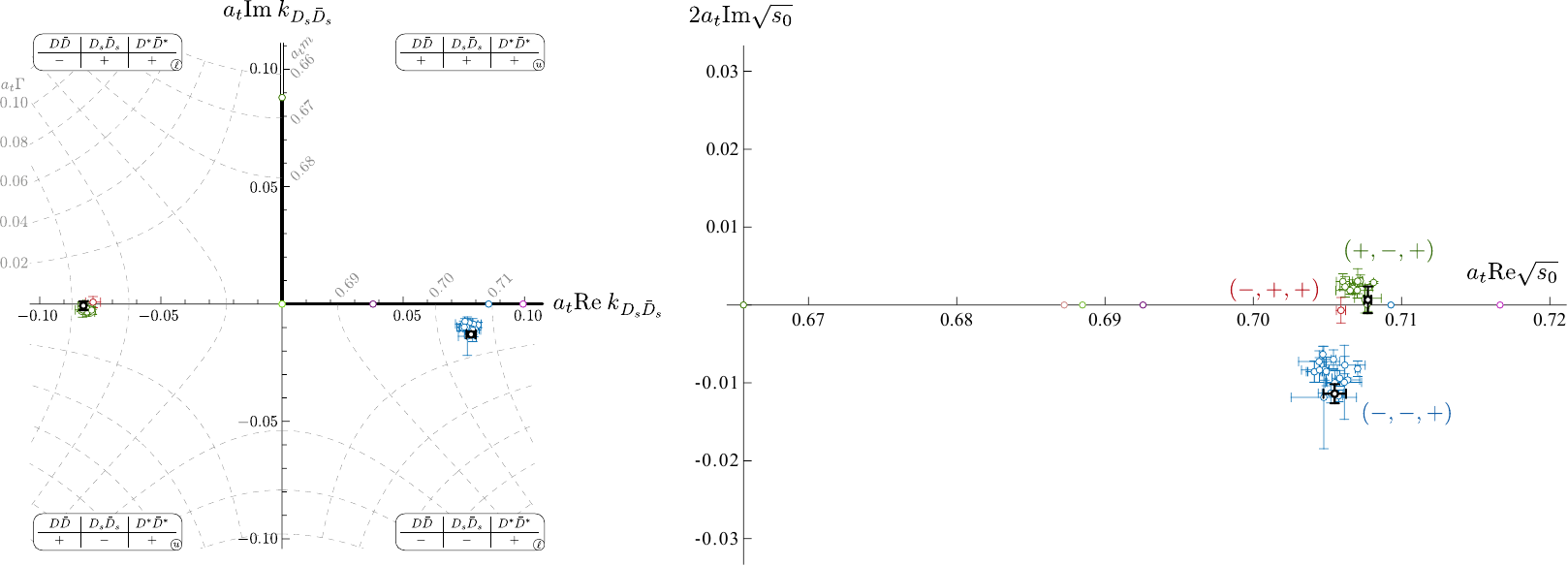}
\caption{Poles found in the $J^{PC}=0^{++}$ amplitudes plotted in the complex $k_{\DsDs}$ and $\sqrt{s}$ planes. The poles of the reference parameterization Eq.~\ref{eq:fit_0pP_rest+flight} are identified in black. The resonance pole on the proximal sheet is plotted in blue, and additional poles on other sheets are plotted in red and green. In one parameterization a ``mirror'' pole is found on the $(-,+,+)$ sheet, in all others it is found on the $(+,-,+)$ sheet.}
  \label{fig:Swave_extra_pole}
\end{figure*}

The corresponding partial widths are,
\begin{align}
\Gamma(\DD\SLJc{1}{S}{0}) & = 0.0040(23) \, a_t^{-1} \approx 23(13) \,\mathrm{MeV}\nonumber\\
\Gamma(\DsDs\SLJc{1}{S}{0}) & = 0.0049(46)\, a_t^{-1} \approx 28(26) \, \mathrm{MeV}\nonumber\\
\Gamma(\psiom\SLJc{1}{S}{0}) & = 0.016(^{+31}_{-16}) \, a_t^{-1} \approx 9 ^{+18}_{\:-9} \, \mathrm{MeV}\,,
\end{align}
and summing these we obtain a value in good agreement with the total width obtained from the pole location: 60(34) MeV compared with 67(38) MeV from $2\;\mathrm{Im}\sqrt{s_0}$.

In all cases, the scattering amplitudes contain several additional poles.
When a pole has a non-zero imginary part in $s$, a complex-conjugate pair of poles $(s_0,s_0^\ast)$ must arise on a common Riemann sheet. These are exact complex conjugates and so are easily identifiable. However, the distance of the relevant \emph{half} $s$-plane should be considered. Since physical scattering occurs at $s+i\epsilon$ on the \emph{upper} half $s$-plane of the physical sheet, then the part of the closest unphysical sheet (the proximal sheet) that is nearby is its \emph{lower} half $s$-plane. Other relevant unphysical sheets can be connected to these via their \emph{upper} half $s$-planes as we shall see below. 

Due to the presence of a large number of channels, it is inevitable that additional poles on other Riemann sheets are present in all cases. Many of these can be considered ``trivial copies'' of the resonance pole identified above, others are far from the region where energy levels are present and so cannot be reliably claimed. Very distant poles, far from any (real-valued) energy levels, typically vary between parameterizations or are sometimes absent entirely, and can thus be considered inessential to describe the physics present.\footnote{Occasionally distant poles occur on the physical sheet. In the scalar channel, these are typically a GeV or more in their imaginary parts and so are not considered relevant.} 

One family of poles that can be dismissed as ``trivial copies'' of the resonance pole are found on the sheets where the sign of $\mathrm{Im} \, k_i$ for a decoupled (or very weakly coupled) channel is flipped. For example, since $\eta_c\eta$ is decoupled, there is no sensitivity to the sign of $\mathrm{Im}\, {k_{\etce}}$. This can be seen from simple Flatt\'e-like amplitudes which have a denominator like
\begin{align}
D= m^2-s-ig_1^2\rho_1-ig^2_2\rho_2\,.
\end{align}
It is the zeros of $D$ that are the poles of the amplitude. If any $g_i$ tends to zero then the dependence on the choice of sheet for channel $i$ drops out since $\rho_i=2k_i/\sqrt{s}$ and a pole will be present for both signs of $\mathrm{Im} \, k_i$. We thus \emph{expect} there to be trivial copies due to the possible signs of $\mathrm{Im}\, k_i$ for $\etce$, $\etcep$, $\psiom$ and $\psi\phi$, which are typically observed to have zero or small couplings. For the remainder of this subsection, we do not consider these trivial copies, and focus only on the sheets defined by the signs of $\mathrm{Im}\, k_i$ for $\DD$, $\DsDs$ and $\DstDst$.

Considering the 8 possibilities for $\DD$, $\DsDs$ and $\DstDst$, for any given real scattering energy only a few sheets are relevant. 
Aside from the physical sheet and the proximal sheet, further ``hidden'' sheets may also be important. These are sheets that are not continuously connected to the real scattering line away from thresholds. 
Thus poles on such sheets can only exert their influence close to the relevant threshold where the distance in the complex plane to the physical scattering axis is short.

One relevant sheet for $0^{++}$ where we find an additional pole is $(\DD[+],\DsDs[-],\DstDst[+])$. This is sometimes referred to as the ``4th'' sheet, and its upper half $s$-plane is continuously connected to the lower half $s$-planes of the $(\DD[-],\DsDs[+],\DstDst[+])$ sheet above $\DsDs$ threshold, and the $(\DD[-],\DsDs[-],\DstDst[+])$ sheet below $\DsDs$ threshold.\footnote{For an earlier example of a 4th sheet pole, see Ref.~\cite{Dudek:2016cru}.} This is conveniently illustrated through a plot of the complex $k_{\DsDs}$ plane, which opens out the Riemann surface in $s$ into a single connected plane for the sheets nearest to $\DsDs$ threshold.
In Fig.~\ref{fig:Swave_extra_pole} we show the position of this additional pole, along with the pole on the proximal sheet, in both the complex $k_{\DsDs}$ and complex $\sqrt{s}$ planes.\footnote{In just one parameterization (the $\Lambda=0.032$ entry in table~\ref{tab:Swave_mov_par_var}) this pole is located on the $\scriptstyle{(\DD[-],\DsDs[+],\DstDst[+])}$ sheet instead of the $\scriptstyle{(\DD[+],\DsDs[-],\DstDst[+])}$ sheet.}

It is well-known that narrow resonances in coupled-channel systems often produce such a second pole, sometimes called a ``mirror'' pole.\footnote{In a two-channel Flatt\'e amplitude, the $(-,+)$ (or $(+,-)$) pole is relevant for resonances coupled to both channels but found below the second threshold, while the $(-,-)$ pole becomes important for narrow resonances coupled to both channels above both thresholds.} Suggestions have been made that further information may be inferred from the arrangement of poles~\cite{Morgan:1992ge,Morgan:1993td}, and in that interpretation, the arrangement in Fig.~\ref{fig:Swave_extra_pole} corresponds to an ``ordinary'' narrow resonance, as opposed to a state present due to strong attraction at threshold.

\begin{figure*}
\includegraphics[width=0.9\textwidth]{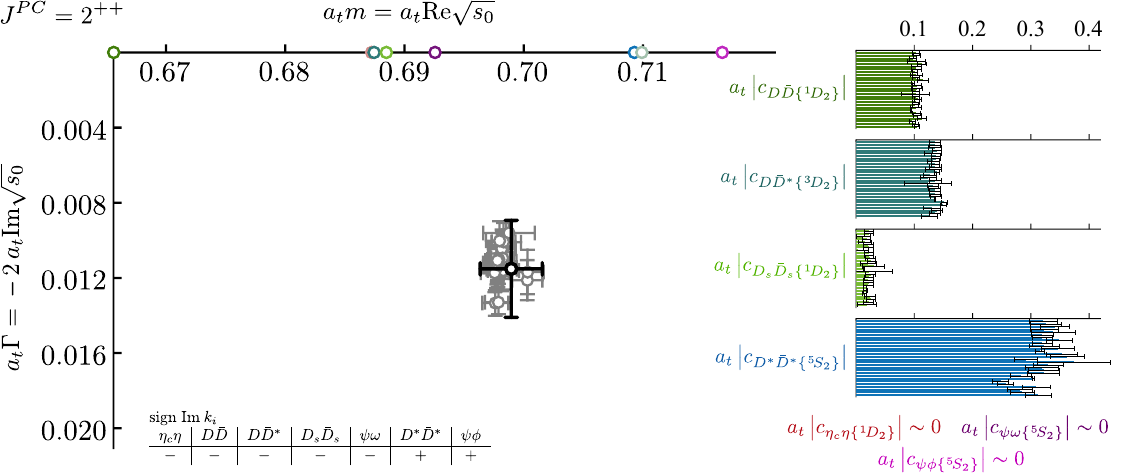}
\caption{The $J^{PC}=2^{++}$  pole and couplings found on the ``proximal'' sheet between $\psiom$ and $\DstDst$ thresholds for a set of successful amplitude parameterizations. The $\eta_c\etap$ channel is not included in the parameterizations. Couplings to the $\eta_c\eta$, $\psi\omega$ and $\psi\phi$ channels are found to be small, but only limited freedom is present in the parameterizations used.}
\label{fig:Dwave_pole_par_var}
\end{figure*}

\subsection{Tensor resonance}
The amplitudes in Section~\ref{sec:2pP_rest_and_mov} describing $J^{PC}=2^{++}$ are found to consistently feature a pole, shown in Figure~\ref{fig:Dwave_pole_par_var}, located at 
\begin{align}
a_t\sqrt{s_0}  =    &        (0.6990 \pm 0.0026) - \tfrac{i}{2}(0.0115 \pm 0.0026) \nonumber\\
\sqrt{s_0}  \approx & \, 3961 \pm 15 - \tfrac{i}{2}(65 \pm 15) \,\mathrm{MeV} \nonumber \, ,
\end{align}
on the proximal sheet between $\psiom$ and $\DstDst$ thresholds,
$({\scriptstyle \etce{[-]}, \DD{[-]}, \DDst{[-]}, \DsDs{[-]}, \psiom{[-]}, \DstDst{[+]}, \psi\phi{[+]} })$.

The couplings of this pole are determined to be
\begin{align}
a_t c_{\etce\SLJc{1}{D}{2}}   \approx \; & 0 \nonumber\\
a_t c_{\DD\SLJc{1}{D}{2}}     =       \; & 0.103(25) \nonumber\\
a_t c_{\DDst\SLJc{3}{D}{2}}   =       \; & 0.123(39) \nonumber\\
a_t c_{\DsDs\SLJc{1}{D}{2}}   =       \; & 0.032(32) \nonumber\\
a_t c_{\psiom\SLJc{5}{S}{2}}  \approx \; & 0 \nonumber\\
a_t c_{\DstDst\SLJc{5}{S}{2}} =       \; & 0.336(99) \nonumber\\
a_t c_{\psiphi\SLJc{5}{S}{2}} \approx \; & 0\,.
\end{align}
As in the scalar case, no significant coupling to closed-charm channels is observed. The relatively large coupling to $\DstDst$, with the resonance lying someway below threshold for decay into this channel, explains the rapid turn-on of $\DstDst$ at threshold. Note that the peculiar dependence only upon the ratio of $K$-matrix couplings $g_i$, and not the absolute scale, discussed in Section~\ref{sec:2pP_rest_and_mov}, is a property only of the parameterization, and not of the rigorously defined $t$-matrix pole couplings $c_i$, which take very similar values regardless of the choice of fixed ``$g$'' coupling.

The corresponding partial widths are,
\begin{align}
\Gamma(\DD\SLJc{1}{D}{2}) & = 0.0046(22) \, a_t^{-1} \approx 26(12) \, \mathrm{MeV}\nonumber\\
\Gamma(\DDst\SLJc{3}{D}{2}) & = 0.0039(25)\, a_t^{-1} \approx 22(14) \, \mathrm{MeV}\nonumber\\
\Gamma(\DsDs\SLJc{1}{D}{2}) & = 0.0003(^{5}_{3}) \, a_t^{-1} \approx  2^{+3}_{-2} \, \mathrm{MeV} \, ,  
\end{align}
and summing these produces 50(17) MeV, compared with 65(15) MeV obtained from the pole location. The large coupling to the closed $\DstDst$ channel is not accounted for in this prescription, which may explain the slight difference.

As was the case in $0^{++}$, additional poles are present for $2^{++}$, and they warrant further attention. Given the approximate decoupling observed to closed-charm final states, it is convenient to label sheets considering only $\DD$, $\DDst$, $\DsDs$ and $\DstDst$ channels. Additional poles on hidden sheets are present, and are presented in Appendix~\ref{app:extra_poles}.
On the \emph{physical sheet}, poles are observed for all parameterizations, and their presence is a concern given that it signals a violation of causality in the amplitude description. These poles are discussed in detail in Appendix~\ref{app:extra_poles} where they are found to be related to the \mbox{$D$-wave} barrier factor associated with the $\DDst\SLJC{3}{D}{2}$ channel, and upon modification of this factor, they disappear without the resonance pole being changed significantly. 

\pagebreak
\subsection{States in $J^{PC}=3^{++}$ and $2^{-+}$}

\begin{figure}
\includegraphics[width=0.99\columnwidth]{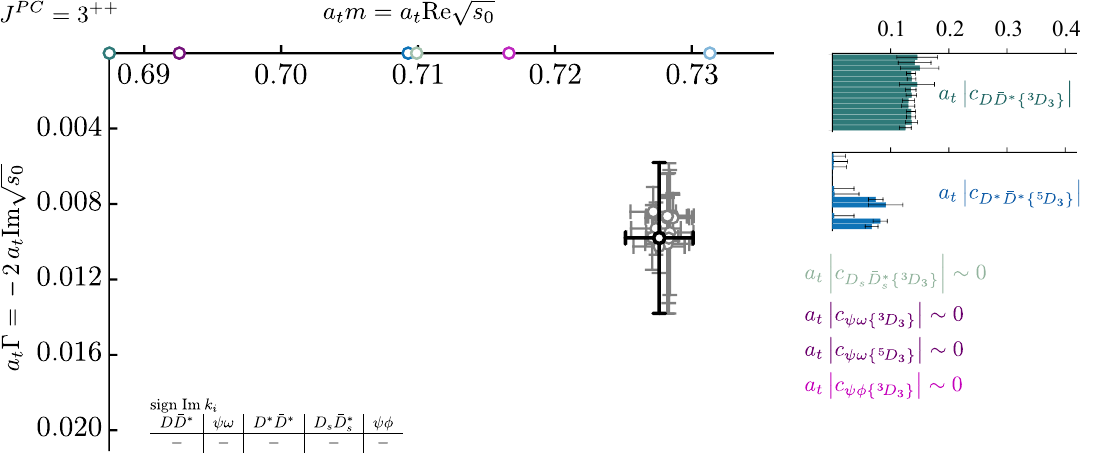}
\caption{The $J^{PC}=3^{++}$ pole and couplings found on the ``proximal'' sheet between $\psi\phi$ and $\DsstDsst$ thresholds for a set of successful amplitude parameterizations.}
\label{fig:Fwave_pole_par_var}
\end{figure}

We determined $J^{PC}=3^{++}$ amplitudes, primarily to constrain them as ``background'' waves in our determination of $2^{++}$. Successful descriptions of the finite-volume spectra include a $3^{++}$ resonance pole coupled to $\DDst\SLJc{3}{D}{3}$ and $\DstDst\SLJc{5}{D}{3}$. Several caveats apply to this result, as described in section~\ref{sec:3pP_scattering}. As shown in Fig.~\ref{fig:Fwave_pole_par_var}, considering multiple parameterizations, a pole is consistently found with
\begin{align}
a_t\sqrt{s_0}  =    &        (0.7276 \pm 0.0025) - \tfrac{i}{2}(0.0098 \pm 0.0040) \nonumber\\
\sqrt{s_0}  \approx & \, 4123\pm 14 - \tfrac{i}{2}(56 \pm 23) \,\mathrm{MeV} \nonumber
\end{align}
and couplings
\begin{align}
a_t c_{\DDst\SLJc{3}{D}{3}}   =       \; & 0.148(37) \nonumber\\
a_t c_{\psiom\SLJc{3}{D}{3}}  \approx \; & 0 \nonumber\\
a_t c_{\psiom\SLJc{5}{D}{3}}  \approx \; & 0 \nonumber\\
a_t c_{\DstDst\SLJc{5}{D}{3}} =       \; & 0.061(61) \nonumber\\
a_t c_{\DsDsst\SLJc{3}{D}{3}} \approx \; & 0 \nonumber\\
a_t c_{\psiphi\SLJc{5}{D}{3}} \approx \; & 0\,,
\end{align}
showing that, again, coupling to closed-charm channels is not significant. The partial widths are
\begin{align}
\Gamma(\DDst\SLJc{3}{D}{3}) & = 0.0098(50)\, a_t^{-1} \approx 55(38) \, \mathrm{MeV}\nonumber\\
\Gamma(\DstDst\SLJc{5}{D}{3}) & = 0.0011^{+22}_{-11} \, a_t^{-1} \approx  6^{+13}_{-6} \, \mathrm{MeV}\,.
\end{align}

In determining $J^{PC}=2^{-+}$, a stable bound--state pole coupled to $\DDst\SLJc{3}{P}{2}$ was found at $a_t\sqrt{s_0}=0.67538(68)$ in the reference parameterization given in Eq.~\ref{eq:fit_2mP_3mP}. This corresponds to a bound-state $\eta_{c2}$ pole with $\sqrt{s_0} \approx 3827(4)$ MeV. A coupling $a_tc_{\DDst\SLJc{3}{P}{2}}=25(15)i$ was also determined.

\subsection{Other possible singularities}

The amplitude parameterizations we have used have the advantage of exactly implementing coupled-channel unitarity in the physical $s$--channel scattering region where we have constraint from the finite-volume spectrum.  They have the cuts implied by $s$--channel unitarity, and are flexible enough to describe pole singularities corresponding to resonances, bound states and virtual bound states. What they do not contain is the physics of ``left-hand cuts'', i.e. the projection into $s$--channel partial-waves of scattering processes in the $t$-- and $u$--channels. In many simple cases these cuts appear far from the physical $s$--channel region, and are of limited relevance, but in certain circumstances they can enter in a way that may have a significant impact.

The closest such cuts relevant to the current study are due to $t,u$--channel pion exchanges, that arise when at least one of the scattering hadrons has nonzero intrinsic spin, leading to the analogue of the ``short nucleon cut''~\cite{HAMILTON1961172,PhysRev.123.692}. Such cuts open only a few tens of MeV below the physical $s$-channel threshold and thus may be of concern. Since the cut will generate an imaginary part in partial-wave amplitudes that is not accounted for in the derivation of the L\"uscher formalism, dealing with it correctly may require a modification of the finite-volume formalism~\cite{Raposo:2023nex}. A recent example considering the closely related case of doubly-charmed $I=0$ $DD^\ast$ scattering can be found in Ref. \cite{Du:2023hlu} which discusses the lattice calculation presented in Ref.~\cite{Padmanath:2022cvl}.

We leave the issue of explicitly accounting for ``left-hand cuts'' as a problem for future studies. Given that no internal inconsistencies have been observed in this calculation, with finite-volume spectra described perfectly well by amplitudes lacking explicit left-hand cut structures, it is possible that this effect is largely negligible. Indeed, if these effects \emph{are} large, then the issue of these cuts is likely to be of concern in \emph{all} studies of unstable charmonia.

\section{Interpretation and comparisons}    

\label{sec:interp}

\begin{table*}
\begin{tabular}{r|ccc|ccc|cccc}
Meson mass / MeV  & $\pi$ & $K$ & $\eta$ & $D$ & $D_s$ & $D^\ast$ & $\eta_c$      & $J/\psi$ & $\chi_{c0}$ & $\chi_{c2}$ \\
\hline
this calc.              & 391   & 550 & 587    & 1886 & 1951 & 2010     & 2965         &  3044     & 3423(3)     & 3519(2) \\[-0.3ex]
expt.                   & 140   & 494 & 548    & 1865 & 1969 & 2007     & {\it 2984}   &  3097     & {\it 3415}  & 3556
\end{tabular}
\caption{Comparing stable meson masses determined on this lattice, with the scale fixed using the physical $\Omega$--baryon mass, to their values in experiment (where states with significant decay widths have their masses shown in italics)~\cite{ParticleDataGroup:2022pth}. (Statistical uncertainties less than 0.5 MeV on lattice masses are not shown.)}
\label{tab:stable_MeV}
\end{table*}

Our key finding in this work is that, for ${m_\pi \approx 391 \, \mathrm{MeV}}$, the $0^{++}$ and $2^{++}$ charmonium sectors contain \emph{only} a single narrow resonance each, lying above the $D_s \bar{D}_s$ threshold, but slightly below the $D^* \bar{D}^*$ threshold. The scalar resonance has significant couplings to all open-charm decay channels, and the tensor to all open-charm except $D_s \bar{D}_s$. Neither resonance has any significant coupling to closed-charm channels. There are also bound states well below threshold corresponding to the $\chi_{c0}(1P)$ and the $\chi_{c2}(1P)$. There is no indication of any further states in the energy region considered. In particular, there is no sign of a scalar bound-state lying just below the $D\bar{D}$ threshold, where no significant attraction is observed. The results also suggest the existence of a narrow $3^{++}$ resonance and a $2^{-+}$ bound state. 

Throughout the course of this calculation, we considered several $S$-wave channels involving a closed-charm meson and a light meson: $\etce\SLJc{1}{S}{0}$, $\etcep\SLJc{1}{S}{0}$, $\psiom\SLJc{1}{S}{0}$,  $\psiom\SLJc{5}{S}{2}$, $\psi\phi\SLJc{1}{S}{0}$, $\psi\phi\SLJc{5}{S}{2}$, $\cco\eta\SLJc{3}{S}{1}$, and $\cct\eta\SLJc{5}{S}{2}$. 
None were found to have large scattering amplitudes, and no near-threshold singularities were identified associated with these channels.

The calculation was performed on three lattice volumes, but only a single lattice spacing, and a single choice of the degenerate light quark, strange quark and charm quark masses, with the light quarks being unphysically heavy. As indicated in Table~\ref{tab:stable_MeV}, which shows stable hadron masses, there is evidence that the charm-quark mass, and perhaps the strange-quark mass, may have been tuned to be slightly smaller than their physical values. Any phenomena that are sensitive to the mass difference between the up and down quarks, or QED effects, will not be correctly captured in this calculation. Discretization effects, while likely small for light mesons, can be larger for charmed and charmonium systems. For example, the $J/\psi$--$\eta_c$ hyperfine splitting, is around 33(1) MeV smaller than observed experimentally, as determined from the values in Table~\ref{tab:masses} and Ref.~\cite{ParticleDataGroup:2022pth}. The deliberate removal of $c\bar{c}$ annihilation likely plays at most a modest role and may contribute to small discrepancies such as the $\chi_{cJ}(1P)$ mass difference with respect to experiment (and of course to these states being stable in this calculation).

Bearing these caveats in mind, we first summarise the extracted amplitudes and then discuss interpretations, comparing the results to prior lattice QCD calculations, to phenomenological models, and to experimental candidate states.
\subsection{State content of amplitudes by $J^{PC}$}

We now summarize and discuss the spectroscopic content of each $J^{PC}$ considered in this work. $J^{PC}=\{1,3\}^{-+}$ amplitudes were also computed and found to be very small in the energy region considered and are not discussed further. There are also indications of a $J^{PC}=4^{++}$ state based on Ref.~\cite{Liu:2012ze} and Fig.~\ref{fig:spec_qqbar}, however this would lie at a slightly higher energy than has been considered in this work.

\subsubsection{$J^{PC}=0^{++}$}

Lying well below $D\bar{D}$ threshold, the $\ccz(1P)$ state is clearly present and, owing to our deliberate removal of $c\bar{c}$ annihilation, it is stable. Its presence plays no significant role in the determination of scattering amplitudes at higher energies.

Below the $\DsDs$ and $\psi\omega$ thresholds, \emph{no other poles are found} in $0^{++}$, either as bound states or as  resonances in $\DD$ or $\eta_c\eta$. Distant virtual poles occur well below threshold in some parameterizations, but they have negligible impact on the physical scattering region, and are likely to be artefacts of extrapolating far outside the region of constraint.
The small negative energy shifts in $[000]A_1^{+}$ relative to non-interacting $D_{[000]}\bar{D}_{[000]}$ energies are explained in amplitude terms by very mild attraction at threshold, at a level far below that needed for a bound-state to be present.

Around $\DsDs$ threshold, a similar but slightly larger negative energy shift is observed, but again description in terms of (coupled-channel) amplitudes indicates insufficient strength to require a nearby pole.\footnote{In certain extreme cases, where a very limited set of energy levels were used, we were able to produce a virtual bound-state pole. Further details can be found in Appendices~\ref{app:amps_Swave_below_psiom} and \ref{app:amps_Swave_below_DstDst}.}

In summary, our findings suggest no strong features close to $\DD$ or $\DsDs$ thresholds, with only modest attraction appearing there.

In the energy region above the $\psi\omega$ threshold near 3900 MeV, more significant departures from the non-interacting energy spectrum are present, which amplitude analysis shows are due to the presence of \emph{a single narrow scalar resonance}. Large couplings to the open $\DD$ and $\DsDs$ channels are found, along with a large coupling to the kinematically closed $\DstDst$ channel. Only upper limits were found for the coupling to the $\psi\omega$ channel, while no evidence was found for coupling to the $\etce$ and $\etcep$ channels.

\subsubsection{$J^{PC}=2^{++}$}

In $2^{++}$ the $\cct(1P)$ state is well below $D\bar{D}$ threshold and plays no significant role in the scattering amplitudes at higher energies.

 The $D$-wave nature of $\DD\SLJc{1}{D}{2}$ and $\DsDs\SLJc{1}{D}{2}$ suppresses any near-threshold interaction, with the first significant feature being a peak in the diagonal $\DD\SLJc{1}{D}{2}$ amplitude followed by a peak roughly 50 MeV higher in energy in the diagonal $\DDst\SLJc{3}{D}{2}$ amplitude. The off-diagonal $\DD\SLJc{1}{D}{2}\to\DDst\SLJc{3}{D}{2}$ amplitudes peak roughly in the middle, and these observations likely reflect the different peak-shaping effects of the \mbox{$D$-wave} barrier factors for the displaced thresholds.

As the $S$-wave $\DstDst$ channel opens, sharp features are observed in all open-charm amplitudes, with the diagonal $\DstDst\SLJc{5}{S}{2}$ amplitude turning on rapidly as was seen for the corresponding wave in the $0^{++}$ case. Only a very weak coupling to the $\DsDs\SLJc{1}{D}{2}$ channel is observed, but this may reflect, at least in part, the $D$-wave barrier suppression, $(k_{D_s \bar{D}_s} / k_{D\bar{D}})^2 \sim 0.3$ in the peak region. 

These features are found to be due to \emph{a single narrow resonance} lying between $D_s \bar{D}_s$ threshold and $D^* \bar{D}^*$ threshold.

\subsubsection{$J^{PC}=3^{++}$}

Our results suggest the existence of an as-yet-unobserved narrow $J^{PC}=3^{++}$ resonance coupled dominantly to $\DDst\SLJc{3}{D}{3}$, with a possible coupling to $\DstDst\SLJc{5}{D}{3}$ and only small couplings to closed-charm final states.

\subsubsection{$J^{PC}=2^{-+}$}

We find a $J^{PC}=2^{-+}$ bound state $\eta_{c2}$ around 3830 MeV. In the computed amplitudes its presence is not obviously indicated by any strong scattering behavior above threshold, but it is clearly present as a nearly volume-independent energy level well below threshold.

At the physical light quark mass, it is likely that this state remains below the relevant $\DDst$ open-charm threshold, and will only generate a non-zero width through $c\bar{c}$ annihilation. On these grounds we'd expect it to be rather narrow, and it might be observable in radiative transitions.

\subsection{Comparisons with Prelovsek et al, Ref.~\cite{Prelovsek:2020eiw}}

The most complete previous attempt to study the charmonium scalar and tensor sectors in lattice QCD appears in Ref.~\cite{Prelovsek:2020eiw} where $\DD$ and $\DsDs$ channels are investigated using light and strange quark masses somewhat lighter than those used in this study (the pion mass is 280 MeV). The lightest channel that can couple, $\eta_c\eta$, is assumed decoupled by fiat and is ignored completely, while $J/\psi \, \omega$ is investigated but ultimately not included in determinations of scattering amplitudes.

In $0^{++}$, Ref.~\cite{Prelovsek:2020eiw} claims that \emph{three} states are required to describe their computed finite-volume spectrum: a stable bound-state lying 4 MeV below $D\bar{D}$ threshold, an extremely narrow resonance lying less than 1 MeV below $D_s\bar{D}_s$ threshold, and a resonance with a width of around 60 MeV lying some way above $D_s\bar{D}_s$ threshold, but well below $D^* \bar{D}^*$ threshold.

Only limited consideration of the $2^{++}$ sector is made. A single resonance is claimed, lying some way above $D_s\bar{D}_s$ threshold, and only slightly below $D^* \bar{D}^*$ threshold, a channel which is not included in the analysis.

The authors compute finite volume spectra in the $[000]\,A_1^+$, $[100]\,A_1$, $[110]\,A_1$ and $[100]\,B_1$ irreps on two volumes using operator bases that feature single-hadron-like $c\bar{c}$ operators and $D\bar{D}$, $D_s \bar{D}_s$ meson-meson operators with relevant momenta.\footnote{Unlike in the meson-meson operators used in this paper, Ref.~\cite{Prelovsek:2020eiw} does not make use of optimized single-hadron operator constructions, which may lead to slower relaxation of correlation functions to the relevant energy eigenstates with increasing Euclidean time.} The lowest energy $D^* \bar{D}^*$ operator is included in the rest-frame only. $J/\psi \, \omega$ operators are included, but the energy levels found to have overlap with them are discarded. No $\eta_c\eta$ operators are included, despite this being nominally the lowest threshold channel in the problem.\footnote{In our calculation we have observed complete decoupling of the $\eta_c \eta$ from the rest of the scattering problem, and have found that the spectrum outside those levels with overlap onto $\eta_c \eta$ operators remains unchanged if the $\eta_c \eta$ operators are excluded. As such, it may be the case that Ref.~\cite{Prelovsek:2020eiw}'s exclusion of $\eta_c \eta$ operators has not introduced a significant error.}

Ref.~\cite{Prelovsek:2020eiw} opts to adjust energy shifts to account for the difference between computed single hadron energies and those predicted by the relativistic dispersion relation, and these shifts can be of order 10 MeV, reflecting significant discretisation effects warranting further investigation~\cite{Hudspith:2021iqu}. The authors choose not to associate any systematic error with this process. In contrast, in the current paper, we propagate conservative errors in $m$ and $\xi$ coming from the slightly different dispersion relations for different species of single hadron into the $E_\mathsf{cm}$ values which go into the L\"uscher analysis, and we also implement an additional systematic error onto every energy level to reflect the modest observed departures from relativistic dispersion (see Appendix~\ref{app:disp_extras}). Hence, to a certain extent we are placing part of the discretization uncertainty into the amplitude errors, and offering a more conservative estimate of the precision of determination of the scattering process.

The different light and strange quark masses and volumes make a direct comparison of spectra presented in Ref.~\cite{Prelovsek:2020eiw} to those presented in the current paper impossible, but certain key features can be considered. Focussing on the $[000]\, A_1^+$ spectrum, a difference is immediately apparent, with the energy levels nearest to $D\bar{D}$ threshold in Ref.~\cite{Prelovsek:2020eiw} being found \emph{significantly below} threshold, suggesting strong attraction, while in this paper, the corresponding levels lie very close to the threshold.

\begin{figure}
\includegraphics[width=0.99\columnwidth]{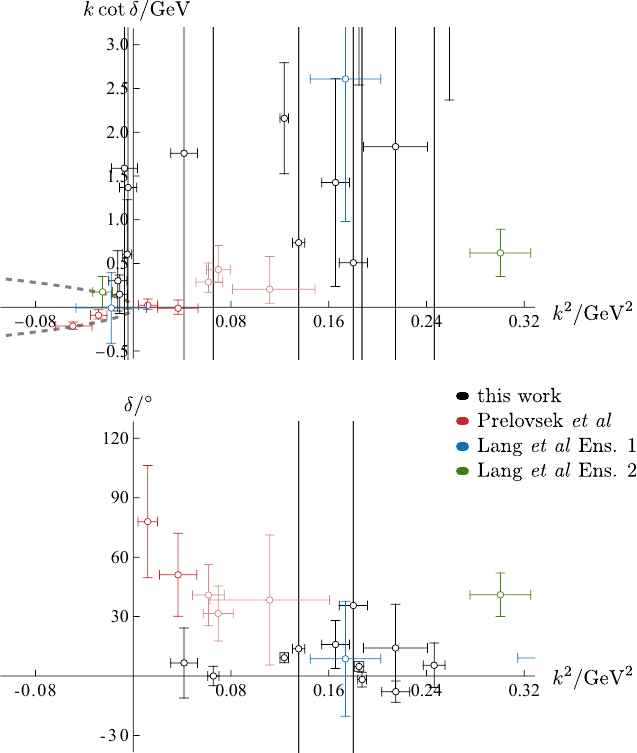}
\caption{Comparison of the amplitudes extracted in the region near $\DD$-threshold region from energy levels in this work (with $m_\pi \approx 391$ MeV), and other lattice calculations, Lang et al~\cite{Lang:2015sba} ($m_\pi \approx 156$ MeV and 266 MeV) and Prelovsek et al~\cite{Prelovsek:2020eiw} ($m_\pi \approx 280$ MeV). Presented as $k \cot \delta$ which has an effective range expansion (upper panel) and $S$-wave elastic phase-shift $\delta$ (lower panel).}
\label{fig:lattice_compare}
\end{figure}

The large downward shifts of these levels, when analysed using the L\"uscher approach, lead to the claim of a bound state in this scattering system. Figure~\ref{fig:lattice_compare} shows the $D\bar{D}$ elastic scattering phase-shift corresponding to low-lying energy levels from Ref.~\cite{Prelovsek:2020eiw}, the current paper, and an earlier calculation at two differing light quark masses~\cite{Lang:2015sba}. The levels below $D\bar{D}$ threshold in Ref.~\cite{Prelovsek:2020eiw} generate the two red points at negative values of $k \cot \delta$ requiring a fit curve that crosses $-\sqrt{-k^2}$ below threshold and hence a bound-state. In contrast, the black points from the current calculation indicate weak interaction near threshold and no bound-state.

Ref.~\cite{Prelovsek:2020eiw} considers different energy regions separately and determines amplitudes in the coupled-channel region using piecewise-in-energy forms rather than the continuous forms with good analytic properties used in the current paper.\footnote{In Appendix E. of Ref.~\cite{Prelovsek:2020eiw}, a continuous function is shown. However, this is formed by joining the piecewise analyses together using smoothed-step or sigmoid functions. These functions contain essential singularities which make an analytic continuation into the complex energy plane questionable.}
The pole lying near the $\DD$ threshold is not present in the amplitude from which the higher two poles are extracted. A coupled-channel analysis is performed above $\DsDs$ threshold, but only a single parameterization is considered that restricts the possible pole content -- it can support two poles decoupled from each other, but cannot straightforwardly describe a single resonance coupled to both the $\DD$ and $\DsDs$ channels.

The current paper presents a more complete study of the scattering system, considering all possible channels, constrained by an order of magnitude more energy levels, and using a variety of analytically well-behaved amplitudes. We come to a completely different conclusion about the number of poles present, and while the pion mass is different, it would be very surprising if \emph{two} additional poles move into the studied energy region under a modest change in the light quark mass.


\subsection{Interpretation and comparisons to other theoretical work}

Our finding of a single relatively narrow resonance in each of $J^{PC}=0^{++}$ and $2^{++}$ can be compared to previous model-based predictions of the state content of the energy region around the $D\bar{D}$ and $D_s \bar{D}_s$ thresholds.

Before the observation of the XYZ candidates, a leading picture of the observed states above 3 GeV was in terms of charm-anticharm bound states formed of heavy quarks moving in a static potential. 
Our results appear to agree with the state counting of this picture, with the single $\chi_{c0}$ and $\chi_{c2}$ resonances corresponding to the $2P$ radial excitations. The large overlap of energy-levels near to the masses of these states with $c\bar{c}$-like operators also supports a dominant role for charmonium-like wavefunction components in these states.
The fact that our scalar state is slightly heavier than the tensor state\footnote{albeit the effect is of limited statistical significance.} is in opposition to the prediction of short-distance spin-orbit effects in this picture. This might indicate that the physics of coupling to the open-charm decay channels is stronger than the relativistic corrections to the interquark potential.

Potential models also predict a single nearby $2^{-+}$ state and a single nearby $3^{++}$ state, as we found, and our determined masses lie within the spread of predictions made by various model implementations.

In order to include some of the physics of hadron strong decay to pairs of lighter hadrons, potential models are sometimes augmented with an application of first-order perturbation theory in which an operator that produces a quark-antiquark pair is introduced. The form of this operator is an assumption of the approach, and a choice, known as the ``$^3\!P_0$ model'', which successfully describes some experimental meson decays, has the pair produced with vacuum quantum numbers. 
The relative strengths of open-charm decays predicted by the $^3\!P_0$ model~\cite{Barnes:1996ff, Barnes:2005pb} are not consistently reflected in our extracted pole couplings. The large coupling to $\DstDst$ of the tensor resonance relative to the scalar resonance \emph{is} a feature of the $^3\!P_0$ model, its spin-recoupling factors giving the amplitudes a ratio of 2. However, these factors do not work universally, as the ratio of couplings for $\chi_{c0} \to \DstDst$ and $\chi_{c0} \to \DD$ is predicted to take value $1/\sqrt{3}$, which is in poor agreement with our extracted couplings. 
The smallness of the tensor resonance coupling to $D_s \bar{D}_s$ \emph{may} have an explanation in the $D$-wave threshold factor, but a very similar reduction would be expected for $\DDst$ decays which is not seen, and the $^3\!P_0$ model does not provide any compensating factor here. 

\vspace{3mm}

The observation of XYZ candidate states spurred theoretical consideration of possible state constructions going beyond just $c\bar{c}$. In particular, the $X(3872)$ at $\DDst$ threshold and the $Z_c(3900)$ nearby have been interpreted as providing evidence for strong long-distance meson-meson interactions in $S$-wave, potentially strong enough to induce binding of molecular-like meson-meson configurations.

Heavy quark spin symmetry applied to the charm quarks suggests similar strong effects in the $\DstDst$ $S$-wave, and potentially $0^{++}, 1^{+-}, 2^{++}$ partners of the $1^{++}$ $X(3872)$~\cite{Albaladejo:2015dsa,Baru:2016iwj}. These states may be bound relative to $\DstDst$, but because they lie above $D\bar{D}$ and $\DDst$ thresholds, they may manifest as resonances.\footnote{The longest-range process of one-pion exchange is not present in elastic $D\bar{D}$ scattering, and hence a bound-state in $D\bar{D}$ must be generated by some other process.}
It is suggested that these molecular states appear \emph{in addition} to the $c\bar{c}$ states discussed above (or for the physical eigenstates to be admixtures). The scalar and tensor resonances found in the current calculation do have numerically significant couplings to the kinematically closed $\DstDst$ channel, which may imply they have significant $\DstDst$ components. However, the state counting suggests that $\DstDst$ $S$-wave interactions are not strong enough to generate additional states (at a pion mass of 391 MeV).

\vspace{3mm}

An approach to explaining at least some of the XYZs that does not directly connect them to meson-meson thresholds is the suggestion that they contain significant \emph{compact tetraquark} components. While the dynamics assumed in models to get these states to bind varies~\cite{Maiani:2004vq,Maiani:2014aja}, inevitably such pictures lead to many states beyond those expected in a $c\bar{c}$ only picture. Tetraquark states are often proposed to lie within a few tens of MeV of meson-meson thresholds with the same quark content and thus unambiguously demonstrating such components is challenging~\cite{Esposito:2021vhu}.
The results presented in this paper do not seem to support additional states of tetraquark origin, but a natural criticism would be that the calculation did not include operators resembling compact tetraquark configurations.\footnote{Our meson-meson operators have a spatial structure that is not compact, rather each meson samples the entire volume of the lattice.} An earlier calculation~\cite{Cheung:2017tnt} performed on the smallest volume lattice used here did include a basis of compact tetraquark operators as well as meson-meson operators. This calculation found no difference in the extracted finite volume spectrum when the tetraquark operators were removed, suggesting that tetraquark components may not be important.

\subsection{Experimental comparisons}

The experimental status of the channels studied in this paper is at present unclear. Peaks are seen in several  processes but often $J^{PC}$ quantum numbers are not known. Nor is it known how peaks in different final states relate to each other.

It is not possible to directly compare the present work to experiment due to the larger light quark mass, the known discretization effects illustrated by the incorrect $J/\psi-\eta_c$ hyperfine splitting, leading to expected differences of a few tens of MeV, and the other systematic uncertainties discussed above. Nevertheless, we can present some discussion assuming plausible extrapolations to the physical light and strange quark masses.

It has been observed in several studies that typically resonance properties vary smoothly with changes in  quark mass~\cite{Nebreda:2010wv,Wilson:2015dqa,Bolton:2015psa,Wilson:2019wfr,Molina:2020qpw,Cheung:2020mql,Gayer:2021xzv,Rodas:2023twk}.\footnote{Although there are notable exceptions, for example Ref.~\cite{Rodas:2023gma}} It has proven to be reasonable in many cases to perform extrapolations based upon the idea that the reduced couplings (pole couplings with the angular momentum barrier divided out) are constant with changing quark mass. Predictions have been made using this approach for $f_2$ resonances~\cite{Briceno:2017qmb, Woss:2020ayi}, the $b_1$ resonance~\cite{Woss:2019hse}, $\rho_J, \omega_J$ resonances~\cite{Johnson:2020ilc}, and a hybrid $\pi_1$~\cite{Woss:2020ayi}. Typically these extrapolations assume that we know the physical mass of the resonance from experiment.

For light quark resonances decaying to final states featuring a pion, there can be a large increase in phase-space with reduction of light quark mass, and a corresponding rapid growth in the decay width of resonances. However, in the current case, consulting Table~\ref{tab:stable_MeV}, we see that even though the light-quark masses are high, because the charm-quark mass is much larger than the light-quark mass, the differences with respect to experiment of the stable hadron masses remain relatively small, and hence we do not expect particular large changes in the resonance properties.

For the case of the single extracted scalar resonance, we might propose two possible extrapolations: (a) if the resonance mass stays where it is (or decreases slightly), there would be only a modest change in the $D\bar{D}$ and $D_s \bar{D}_s$ phase-spaces, and the state would remain an isolated relatively narrow resonance with decays to $D\bar{D}$, to $D_s \bar{D}_s$ (if this is still an open channel) and possibly to $J/\psi \, \omega$. (b) If the resonance mass moves up slightly, getting close to or even above the $D^* \bar{D}^*$ threshold, the large coupling to that channel in \mbox{$S$-wave} \emph{could} generate a large total width for the state. In either case there will be just a single scalar resonance.

Compared to this result there would appear to be a surfeit of experimental scalar candidate states, as discussed in the introduction. What is not clear is whether the features observed experimentally in differing production processes and final states could actually be due to just a single resonance appearing in the coupled-channel system. Production processes must share the same pole singularities as the scattering $t$-matrix, but the real-energy axis lineshape can be sculpted by polynomial energy-dependence from the production factor,\footnote{or even more rapid energy dependence if the Born-term has a nearby singularity.} which, if not accounted for, could lead to slightly differing resonance masses and widths in different processes. 
It remains to be seen if sufficiently rigorous coupled-channel analysis could resolve the experimental $\chi_{c0}(3930), \chi_{c0}(3960)$ peaks, the broad $\chi_{c0}(3860)$ enhancement, and possibly the $X(3915)$ as ultimately being due to amplitudes featuring only a single scalar resonance pole.

For the tensor resonance, similar observations can be made. However, the $D$-wave nature of the open decay modes means that the state will rapidly become narrow if its mass decreases, while the large $S$-wave coupling to $D^* \bar{D}^*$ might make it broad should it go up in mass. The results are consistent with there being a single $\chi_{c2}(3930)$ resonance coupled to $\DD$~\cite{BaBar:2010jfn,LHCb:2020pxc,Belle:2021nuv}, and the current experimental data is not inconsistent with this having at most a small coupling to $D_s \bar{D}_s$~\cite{LHCb:2022aki}.

\section{Summary}                           

\label{sec:summary}

\begin{figure*}
\includegraphics[width=0.97\textwidth]{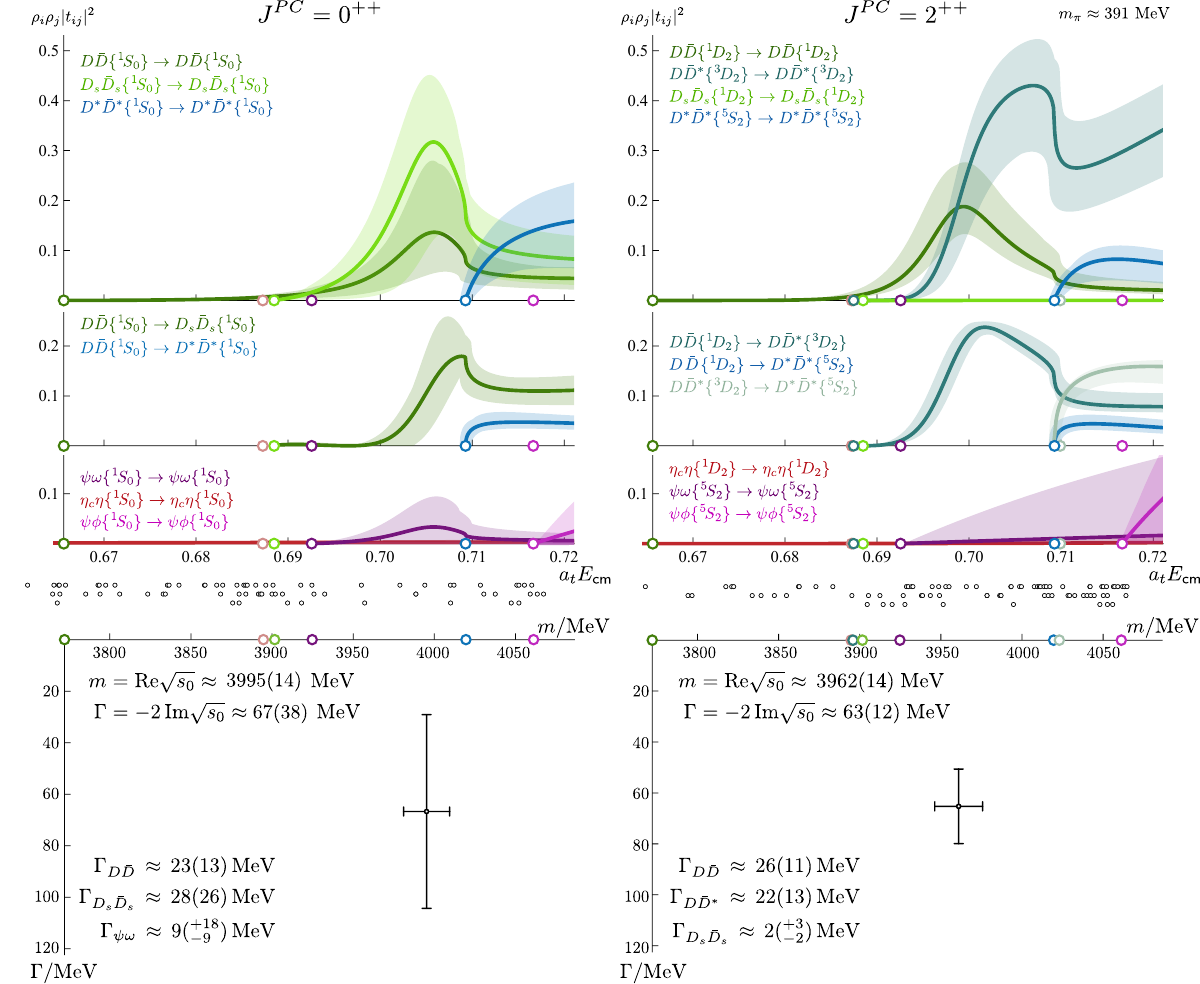}
\caption{
Results from $J^{PC}=0^{++}$ (left column) and $J^{PC}=2^{++}$ (right column).
Top panels: scattering $t$-matrix elements plotted as $\rho_i\rho_j |t_{ij}|^2$. Middle: dots show the energy levels used to constrain scattering amplitudes, with further energy levels at lower energies that mostly constrain the $\eta_c\eta$ amplitudes. Bottom panels: resonance pole positions on the ``proximal'' sheet, the closest unphysical sheet to the real energies at which $s$-channel scattering occurs. The most significant partial widths are also indicated, as determined from pole residues using Eq.~\eqref{eq:partial}.
}
\label{fig:big_summary}
\end{figure*}

We have presented an investigation of the $\ccz$ and $\cct$ channels above $\DD$ threshold where resonant effects are observed in experiment. Working in the approximation where charm-quark annihilation is forbidden, for the first time we have been able to consider all of the necessary channels up to $\psi\phi$ threshold. A summary of our key findings is presented in Figure~\ref{fig:big_summary}.

Working at $m_\pi\approx 391$ MeV, we find a quite simple picture with a single resonance in both $J^{PC}=0^{++}$ and $2^{++}$ strongly coupled to open-charm  decay modes. Both resonances are found just below $\DstDst$ threshold around 4000 MeV with relatively narrow widths of around 60 MeV, and both have a significant coupling to the kinematically closed $\DstDst$ channel in $S$-wave. A key difference between the resonances is that the $\DsDs$ coupling is very small for the tensor resonance, but for the scalar state it is of approximately equal strength to the coupling to $\DD$. 

As a by-product of this work, in order to determine ``background'' partial waves that appear in our lattice QCD calculation, we have found an $\eta_{c2}$ bound state, and a $\cce$ resonance, both of which have coupling to the $\DDst$ channel. Exotic $J^{PC}=1^{-+}$ and $3^{-+}$ amplitudes were found to be small below 4100 MeV. 

Our results are in disagreement with other theoretical work reporting bound or near-threshold states in $\DD$ in $S$-wave~\cite{Tornqvist:1993ng,Pennington:2007xr,Guo:2012tv,Maiani:2014aja,Wang:2019evy,Wang:2020elp,Deineka:2021aeu,Dong:2021bvy}, including a prior lattice QCD calculation~\cite{Prelovsek:2020eiw}. 

The methods used in this paper may be applied to other sectors featuring scattering of hadrons containing charm quarks. Particularly attractive targets are the near-threshold vector-pseudoscalar enhancements, $X/\cco(3872)$, $T_{cc}(3875)^+$, and ${Z}_c/{T}^b_{\psi1}(3900)^+$, whose interaction dynamics are likely related to the states observed in the current paper. A more complete calculation in a robust lattice QCD framework of these systems will aid in understanding the inner workings of QCD at these energies.



%
\acknowledgments

We thank our colleagues within the Hadron Spectrum Collaboration (\url{www.hadspec.org}), in particular Raul Brice\~{n}o, Andrew Jackura and Arkaitz Rodas, and also acknowledge useful discussions with Igor Danilkin, Feng-Kun Guo, Christoph Hanhart, Sasa Prelovsek, Steve Sharpe and Adam Szczepaniak.
DJW acknowledges support from a Royal Society University Research Fellowship. DJW \& CET acknowledge support from the U.K. Science and Technology Facilities Council (STFC) [grant number ST/T000694/1].
JJD acknowledges support from the U.S. Department of Energy contract DE-SC0018416 at William \& Mary, and JJD \& RGE from contract DE-AC05-06OR23177, under which Jefferson Science Associates, LLC, manages and operates Jefferson Lab. 

The software codes
{\tt Chroma}~\cite{Edwards:2004sx}, {\tt QUDA}~\cite{Clark:2009wm,Babich:2010mu}, {\tt QUDA-MG}~\cite{Clark:SC2016}, {\tt QPhiX}~\cite{ISC13Phi}, {\tt MG\_PROTO}~\cite{MGProtoDownload}, {\tt QOPQDP}~\cite{Osborn:2010mb,Babich:2010qb}, and {\tt Redstar}~\cite{Chen:2023zyy} were used. 
Some software codes used in this project were developed with support from the U.S.\ Department of Energy, Office of Science, Office of Advanced Scientific Computing Research and Office of Nuclear Physics, Scientific Discovery through Advanced Computing (SciDAC) program; also acknowledged is support from the Exascale Computing Project (17-SC-20-SC), a collaborative effort of the U.S.\ Department of Energy Office of Science and the National Nuclear Security Administration.

This work used the Cambridge Service for Data Driven Discovery (CSD3), part of which is operated by the University of Cambridge Research Computing Service (www.csd3.cam.ac.uk) on behalf of the STFC DiRAC HPC Facility (www.dirac.ac.uk). The DiRAC component of CSD3 was funded by BEIS capital funding via STFC capital grants ST/P002307/1 and ST/R002452/1 and STFC operations grant ST/R00689X/1. Other components were provided by Dell EMC and Intel using Tier-2 funding from the Engineering and Physical Sciences Research Council (capital grant EP/P020259/1). This work also used the earlier DiRAC Data Analytic system at the University of Cambridge. This equipment was funded by BIS National E-infrastructure capital grant (ST/K001590/1), STFC capital grants ST/H008861/1 and ST/H00887X/1, and STFC DiRAC Operations grant ST/K00333X/1. DiRAC is part of the National E-Infrastructure.
This work also used clusters at Jefferson Laboratory under the USQCD Initiative and the LQCD ARRA project.

Propagators and gauge configurations used in this project were generated using DiRAC facilities, at Jefferson Lab, and on the Wilkes GPU cluster at the University of Cambridge High Performance Computing Service, provided by Dell Inc., NVIDIA and Mellanox, and part funded by STFC with industrial sponsorship from Rolls Royce and Mitsubishi Heavy Industries. Also used was an award of computer time provided by the U.S.\ Department of Energy INCITE program and supported in part under an ALCC award, and resources at: the Oak Ridge Leadership Computing Facility, which is a DOE Office of Science User Facility supported under Contract DE-AC05-00OR22725; the National Energy Research Scientific Computing Center (NERSC), a U.S.\ Department of Energy Office of Science User Facility located at Lawrence Berkeley National Laboratory, operated under Contract No. DE-AC02-05CH11231; the Texas Advanced Computing Center (TACC) at The University of Texas at Austin; the Extreme Science and Engineering Discovery Environment (XSEDE), which is supported by National Science Foundation Grant No. ACI-1548562; and part of the Blue Waters sustained-petascale computing project, which is supported by the National Science Foundation (awards OCI-0725070 and ACI-1238993) and the state of Illinois. Blue Waters is a joint effort of the University of Illinois at Urbana-Champaign and its National Center for Supercomputing Applications.


\clearpage
\appendix

\label{sec:appendices}

\onecolumngrid

\section{Dispersion relations and additional systematic uncertainty}
\label{app:disp_extras}

\begin{figure*}[hb]
\includegraphics[width=0.99\textwidth]{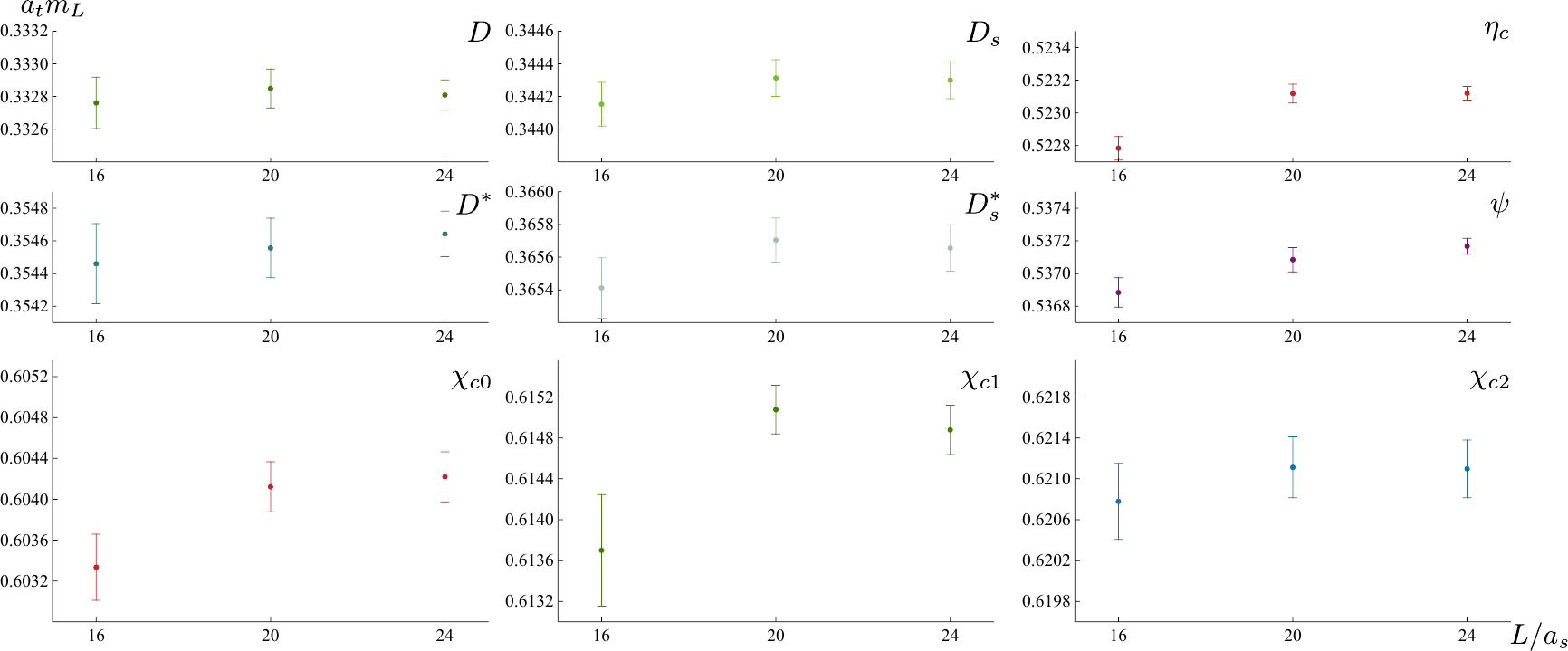}
\caption{Masses of stable hadrons obtained from dispersion relation fits to each lattice volume individually, $a_t m_L$, using Eq.~\ref{eq:disp}. Fits include points with momentum $|\vec{n}|^{\,2} \le {3,4,6}$ for $L/a_s={16,20,24}$ respectively.}
\label{fig:disp_L20}
\end{figure*}

In this appendix we give further details of the dispersion relations, stable hadron masses and the additional systematic uncertainty included in the spectrum. Figure~\ref{fig:disp_L20} shows the result of determining stable hadron masses from dispersion relation fits to each lattice volume individually. 

\begin{figure*}[hb]
\includegraphics[width=0.99\textwidth]{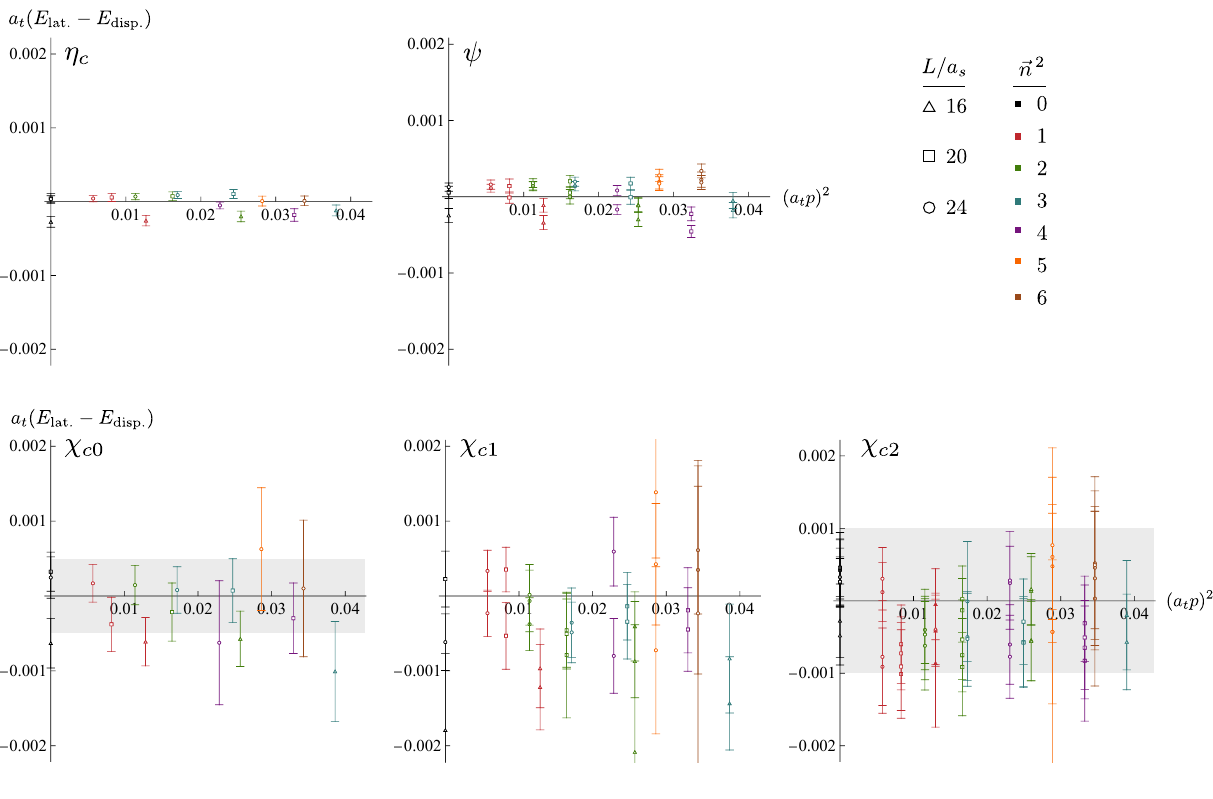}
\caption{Residuals upon fitting stable hadron energies to the dispersion relation given in Eq.~\ref{eq:disp} with independent values of $m$, $\xi$ for each hadron. Grey bands indicate our selection of modest additional systematic uncertainties added to energy levels for $J=0,1$ (left) and $J=2,3$ (right).}
\label{fig:disp_residuals}
\end{figure*}

Figure~\ref{fig:disp_residuals} shows the deviations of the computed energies from the dispersion relation fitted to all three volumes simultaneously. The size of these residuals, particularly for the $\chi_{cJ}$ states, motivates our addition of a systematic error as described in the main text.
Without this additional systematic, the $\chi^2/N_{\mathrm{dof}}$ values for descriptions of computed finite-volume spectra using scattering amplitudes in many cases are large. By adding this systematic error, which may reflect discretization effects or some other unaccounted-for systematic, we make the uncertainty on the amplitude descriptions more accurately reflect the uncertainty in the calculation. Additional details in the context of $DK$ scattering are given in Ref.~\cite{Cheung:2020mql}.

In Table~\ref{tab:disp_details} we provide details of the dispersion relation fits for stable charmed and charmonium mesons using Eq.~\ref{eq:disp} that are shown in Fig.~\ref{fig_disp} and Table~\ref{tab:masses}. 

\begin{table}[hb]
  \begin{tabular}{c|cl|cl|ccc}
  hadron & & \multicolumn{1}{c|}{$a_t m$} & &\multicolumn{1}{c|}{$\xi$} & $\chi^2/N_{\mathrm{dof}}$ \\[0.2ex]
  \hline
  $D$            & & 0.33281(9)  & & 3.466(4) & $\frac{7.79}{7 - 2} =  1.56$ \\
  $D_s$          & & 0.34424(11) & & 3.457(4) & $\frac{9.76}{7 - 2} =  1.95$ \\
  $D^\ast$       & & 0.35464(14) & & 3.477(7) & $\frac{28.6}{16 - 2} =  2.04$\\
  $D_s^\ast$     & & 0.36566(14) & & 3.483(7) & $\frac{28.5}{16 - 2} =  2.04$\\[0.2ex]
  $\eta_c$       & & 0.52312(4)  & & 3.491(2) & $\frac{50.5}{7 - 2}  =  10.1^\dagger$\\
  $\psi$        &  & 0.53715(5) & & 3.491(2) & $\frac{128}{16 - 2}  =  9.17^\ddagger$\\
  $h_c$           & & 0.61662(26) & & 3.450(22) & $\frac{31.4}{16 - 2}  =  2.24$ \\
  $\chi_{c0}$     & & 0.60422(25) & & 3.478(25) & $\frac{2.69}{7 - 2}  =  0.54$ \\
  $\chi_{c1}$     & & 0.61488(46) & & 3.462(26) & $\frac{20.9}{16 - 2}  =  1.49$ \\
  $\chi_{c2}$     & & 0.62110(28) & & 3.454(25) & $\frac{23.8}{27 - 2} =  0.95$ \\
  $\eta_c^\prime$ & & 0.64160(55) & & 3.647(115) & $\frac{8.58}{7 - 2}  =  1.72$\\
  $\psi^\prime$   & & 0.64566(110) & & 3.484(141) & $\frac{7.99}{10 - 2}  =  1.00^\sharp$
  \end{tabular}
  \caption{Results of dispersion relation fits on the $L/a_s=24$ lattice with  $|\vec{n}|^2\le 6$.
    Notes: $^\dagger$the largest individual deviation here is 1.95$\sigma$, and most are below 1$\sigma$, suggesting that correlations are playing a significant role. $^\ddagger$a single level, in $[002]A_1$,  contributes most of the tension here, having a $4\sigma$ deviation -- removing this level results in $\chi^2/N_{\mathrm{dof}} = \frac{27.7}{15 - 2} = 2.13$ and compatible values of $a_tm=0.53720(5)$ and $\xi = 3.489(2)$. $^\sharp$in this case only momenta, $|\vec{n}|^2\le 4$ were used.}
  \label{tab:disp_details}
\end{table}

\clearpage


\section{Including the $\chi_{c0}(1P)$ bound-state in amplitudes}
\label{app:bs_pole}

In this appendix we consider the two-channel rest-frame analysis of Section~\ref{sec:Swave_rest_DDthreshold}, but additionally include those energy levels lying far below $\etce$ threshold which we have identified as being due to the stable $\ccz(1P)$ state. The purpose is to show that the amplitude behavior above threshold is unchanged upon this inclusion.

\begin{figure*}[h]
\includegraphics[width=0.9\textwidth]{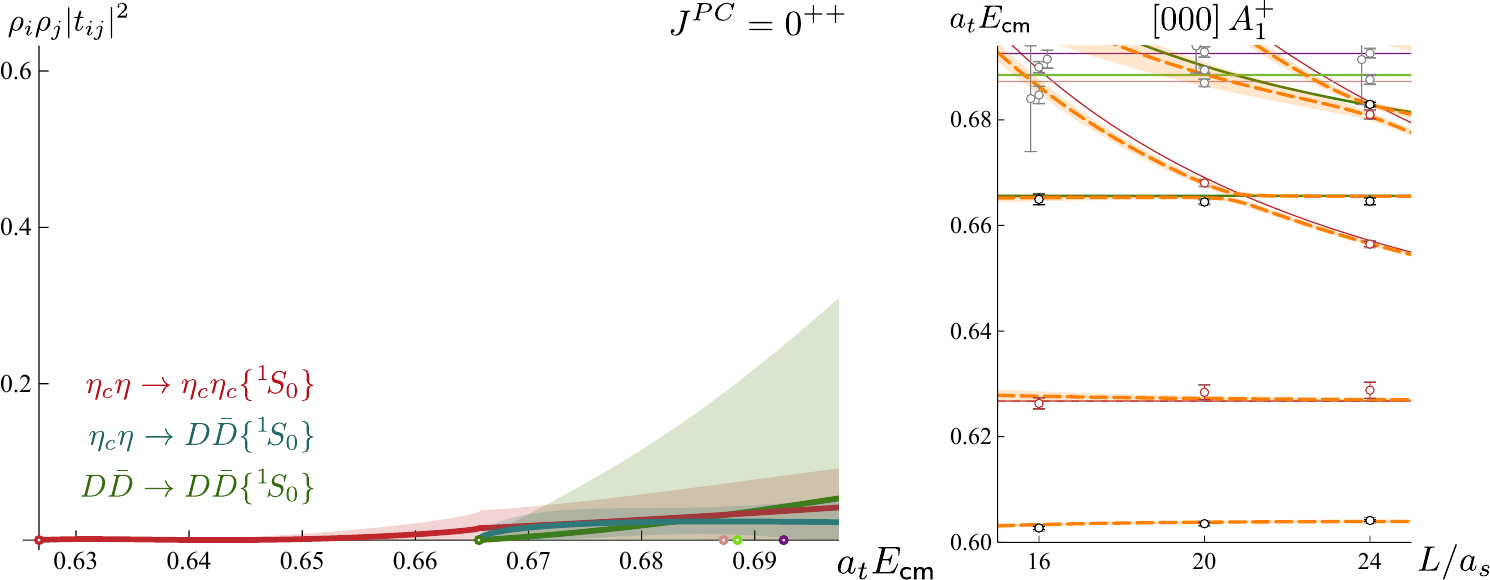}
\caption{As Figure~\ref{fig:amp_spec_Swave_two_channel}, but including levels well below $\eta_c\eta$ threshold due to the stable $\chi_{c0}(1P)$ state. Amplitude as given by Eq.~\ref{eq:fit_0pP_with_pole}.}
\label{fig:amp_spec_Swave_two_channel_with_pole}
\end{figure*}

To describe the scattering system including the deep bound-state, a $K$-matrix with a pole term and a matrix of constants is used, with a Chew-Mandelstam phase-space subtracted at the $K$-matrix pole location. Describing 13 energy levels results in amplitude parameters,

\begin{center}
\begin{tabular}{rll}
$a_t m=$ & $(0.60402 \pm 0.00037 \pm 0.00004)$ & \multirow{6}{*}{ $\begin{bmatrix*}[r]   1.00 &   0.52 &   0.02 &   0.53 &   0.08 &  -0.09\\
&  1.00 &   0.45 &   0.91 &   0.54 &   0.37\\
&&  1.00 &   0.19 &   0.99 &   0.98\\
&&&  1.00 &   0.28 &   0.10\\
&&&&  1.00 &   0.96\\
&&&&&  1.00\end{bmatrix*}$ } \\ 
$a_t g_{\eta_c\eta}=$ & $(0.23 \pm 0.09 \pm 0.02)$ & \\
$a_t g_{\DD}=$        & $(0.23 \pm 0.63 \pm 0.15)$ & \\
$\gamma_{\eta_c\eta\to\eta_c\eta}=$ & $(0.97 \pm 0.53 \pm 0.08)$ \\
$\gamma_{\eta_c\eta\to\DD}=$  & $(0.05 \pm 1.54 \pm 0.31)$ & \\
$\gamma_{\DD\to\DD}=$  & $(1.11 \pm 3.24 \pm 0.07)$ & \\[1.3ex]
&\multicolumn{2}{l}{ $\chi^2/ N_\mathrm{dof} = \frac{8.19}{13-6} = 1.17$\,,}
\end{tabular}
\end{center}
\vspace{-8mm}
\begin{equation}\label{eq:fit_0pP_with_pole}\end{equation}
and the resulting amplitude is presented in Fig.~\ref{fig:amp_spec_Swave_two_channel_with_pole}, which is comparable to Fig.~\ref{fig:amp_spec_Swave_two_channel}.

We performed a limited exploration of the space of possible parameterization variations, finding amplitudes with $g_{\DD}=0$ that described the data as well as the above description. We were not able to find amplitudes with a pole term but with $g_{\eta_c\eta}$ set to zero that described the data well. Including moving-frame energy levels do not change these conclusions. 

\clearpage
\pagebreak

\section{$J^{PC}=0^{++}$ amplitude determinations with 3 and 5 channels}

In this appendix we investigate scalar amplitudes determined using highest energies that lie between the two channel region in sections~\ref{sec:Swave_rest_DDthreshold} \& \ref{sec:Swave_threshold_mov} and the 7 channel region in sections~\ref{sec:Swave_scattering_coupled_rest} \& \ref{sec:Swave_scattering_final}. The aim is to more closely inspect the $\DsDs$ near-threshold region, and to show that inclusion of the $\DstDst$ channel is not essential to extract the resonance pole (although the extra levels and wider energy coverage are helpful).

\subsection{$J^{PC}=0^{++}$ below $\psi\omega$ threshold}
\label{app:amps_Swave_below_psiom}

We determine scattering amplitudes in the region where $\etce$, $\DD$ and $\DsDs$ are active, below the $\psiom$ threshold at $a_t E_\cm=0.690$. In this case we opt to neglect the $\etcep$ channel which is assumed to be decoupled, and exclude those levels dominated by overlap with operators resembling $\etcep$.
One motivation for considering only these levels is that the description in this energy region appears to have some tension in the larger energy region $S$-wave amplitude determinations, for example see the left panel of Fig.~\ref{fig:rest_irreps_fvs}.

We make use of rest-frame irreps ($L/a_s=16,20,24$) and moving-frame irreps ($L/a_s=20,24$) with energies up to $a_t E_\cm=0.690$. This includes 5 levels dominated by overlap with $\DsDs$ operators very close to $\DsDs$ threshold from $[000]A_1^{+}$ and $[002]A_1$. The $[001]A_1$ and $[111]A_1$ irreps are included but they do not have any levels dominated by $\DsDs$ operators in this energy region.

\begin{figure*}
\includegraphics[width=1.0\textwidth]{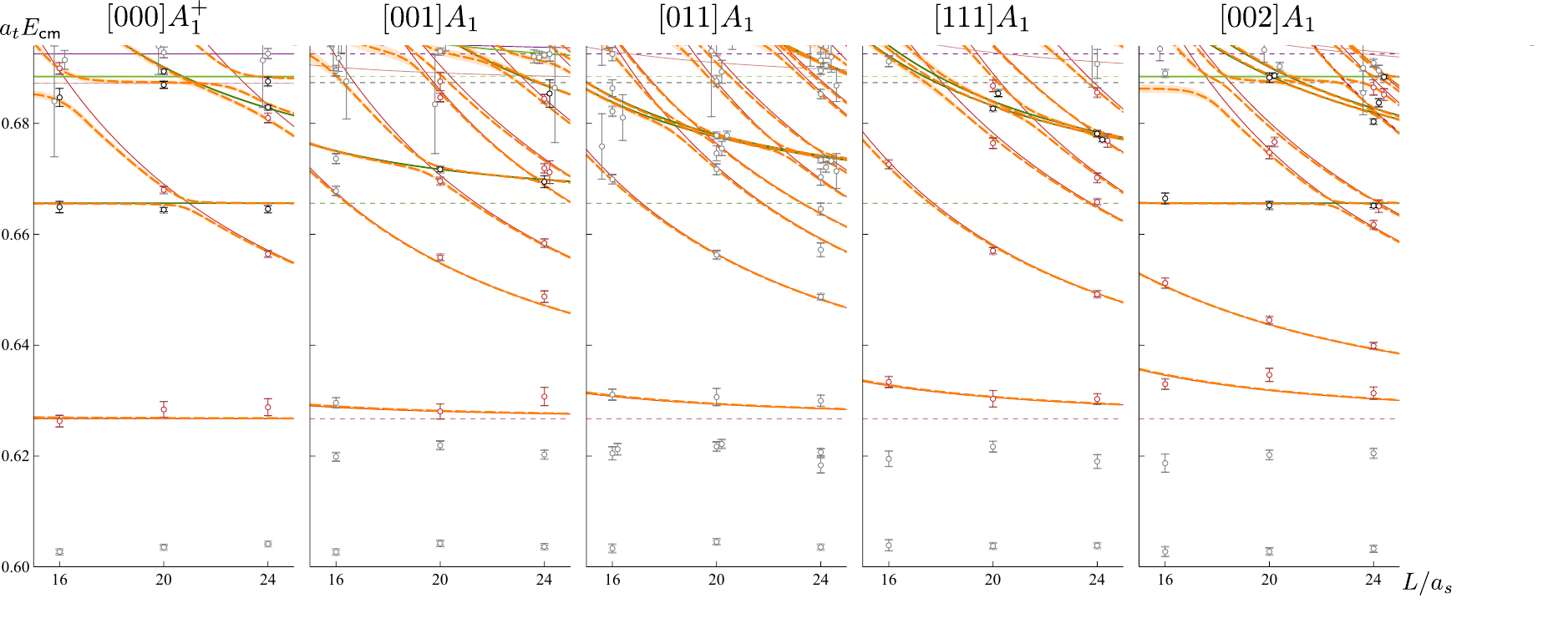}
\caption{As Figure~\ref{fig:amp_spec_Swave_two_channel}, except solutions from the three-channel amplitudes in Eq.~\ref{eq:fit_0pP_3chan} determined from $[000]\,A_1^{+}$ and moving frames $A_1$ irreps are shown as the orange curves.}
\label{fig:specs_A1s_fitted_upto_0p690}
\end{figure*}

\begin{figure}
\includegraphics[width=0.5\textwidth]{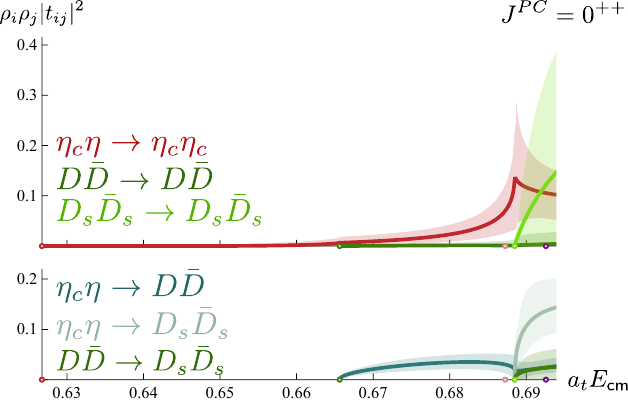}
\caption{Three-channel $J^{PC}=0^{++}$ scattering amplitudes in Eq.~\ref{eq:fit_0pP_3chan} determined from $[000]A_1^{+}$ and moving frame $A_1$ irreps.}
\label{fig:amps_3chan_upto_0p690}
\end{figure}

An example amplitude determined from these energies is given by a $K$-matrix of constants $K_{ij}=\gamma_{ij}$. In this case, we subtract the Chew-Mandelstam function at the threshold of each channel. The resulting parameter values are,

\begin{center}
\begin{tabular}{rll}
$\gamma_{\etce\to\etce} =$     & $(1.39 \pm 0.58 \pm 0.40)$ & \multirow{6}{*}{ $\begin{bmatrix*}[r]   1.00 &   0.23 &   0.70 &   0.10 &   0.92 &   0.07\\
&  1.00 &   0.32 &  -0.28 &   0.26 &  -0.55\\
&&  1.00 &   0.18 &   0.91 &   0.13\\
&&&  1.00 &   0.14 &   0.84\\
&&&&  1.00 &   0.14\\
&&&&&  1.00\end{bmatrix*}$ } \\ 
$\gamma_{\DD\to\DD}=$          & $(0.13 \pm 0.35 \pm 0.22)$ & \\
$\gamma_{\DsDs\to\DsDs}=$      & $(4.87 \pm 3.84 \pm 3.28)$ & \\
$\gamma_{\etce\to\DD}=$        & $(0.57 \pm 0.32 \pm 0.08)$ & \\
$\gamma_{\etce\to\DsDs}=$      & $(3.18 \pm 1.35 \pm 0.96)$ & \\
$\gamma_{\DD\to\DsDs}=$        & $(-0.87 \pm 1.07 \pm 0.45)$ & \\[1.3ex]
&\multicolumn{2}{l}{ $\chi^2/ N_\mathrm{dof} = \frac{53.0}{60 - 6 - 8} =  1.15$\,,}
\end{tabular}
\vspace{-0.5cm}
\begin{equation}\label{eq:fit_0pP_3chan}\end{equation}
\end{center}
where the central value of the $\gamma_{\DsDs\to\DsDs}$ parameter is numerically larger than in the other amplitudes in this work, although with a large uncertainty. The description of the finite-volume energy levels is improved in the region of $\DsDs$ threshold, as can be seen in Fig.~\ref{fig:specs_A1s_fitted_upto_0p690}. The amplitudes are plotted in Fig.~\ref{fig:amps_3chan_upto_0p690}.

With this selection of levels, stronger features can be seen at $\DsDs$ threshold than when considering a wider energy range. Amplitudes determined from this limited selection of energies feature poles on unphysical sheets in the region of $a_t\sqrt{s}\approx 0.68$, which may be at complex locations in coupled-channel cases, or on the real axis below $\DsDs$ threshold in decoupled cases. In the case presented in Eq.~\ref{eq:fit_0pP_3chan}, there is a pole on a ``hidden'' sheet, although within uncertainties the imaginary part is consistent with zero, as summarized in Table~\ref{tab:3chan_poles}.

\begin{table}[!h]
\begin{tabular}{c|c}
$\mathrm{sign}\:\mathrm{Im}$ & \\
$(k_{\etce},k_{\DD},k_{\DsDs})$ & $a_t\sqrt{s_0}$\\
\hline
$(+,-,-)$ & $(0.675 \pm 0.022) \pm \frac{i}{2}(0.027 \pm 0.028)$\\
$(+,+,-)$ & $(0.681 \pm 0.012) \pm \frac{i}{2}(0.029 \pm 0.045)$
\end{tabular}
\caption{Poles in three channel $\etce-\DD-\DsDs$ amplitudes, Eq.~\ref{eq:fit_0pP_3chan}.}
\label{tab:3chan_poles}
\end{table}

The large uncertainties on the parameters and the plotted amplitudes indicate that the constraint on $\DsDs$ is not particularly strong. We have highlighted this because it is the most extreme amplitude behavior near $\DsDs$ threshold that the finite-volume spectra can tolerate. However, no poles have been found \emph{close to} $\DsDs$ threshold in this work, in contrast to Ref.~\cite{Prelovsek:2020eiw} which finds one within a few MeV. The energy shifts seen in the current calculation are smaller and the corresponding interactions are weak. Adding further levels at higher energies pushes the solution towards a smaller $\DsDs$ amplitude just above threshold, the pole then moves away, and the coupling weakens. The poles in Table~\ref{tab:3chan_poles} thus appear to be a property of describing energy levels in too small a region, and are not a good overall reflection of the findings in this work.

\vspace{3mm}

\subsection{Determining $J^{PC}=0^{++}$ up to $\DstDst$ threshold only}
\label{app:amps_Swave_below_DstDst}

The $0^{++}$ resonance identified in sections~\ref{sec:Swave_scattering_coupled_rest} \& \ref{sec:Swave_scattering_final} lies below $\DstDst$ threshold and so, in principle, we should be able to extract it without considering the $\DstDst$ and $\psi\phi$ channels, although the strong overlap with $\DstDst$ operators of several states below threshold (as seen in Fig~\ref{fig:histos_A1pP_L24}) suggest there is merit in including the kinematically closed $\DstDst$ channel.

In this section we consider a coupled $\etce-\DD-\etcep-\DsDs-\psiom$ scattering system below $\DstDst$ threshold, taking the opportunity to test some of the properties of the amplitudes at lower energies. As with the main analysis, we begin by using only at-rest energies before making use of both rest-frame and moving-frame energies together.

\subsubsection{Rest-frame energies only}
\label{app:DstDst_rest}

\begin{figure*}
\includegraphics[width=0.66\textwidth]{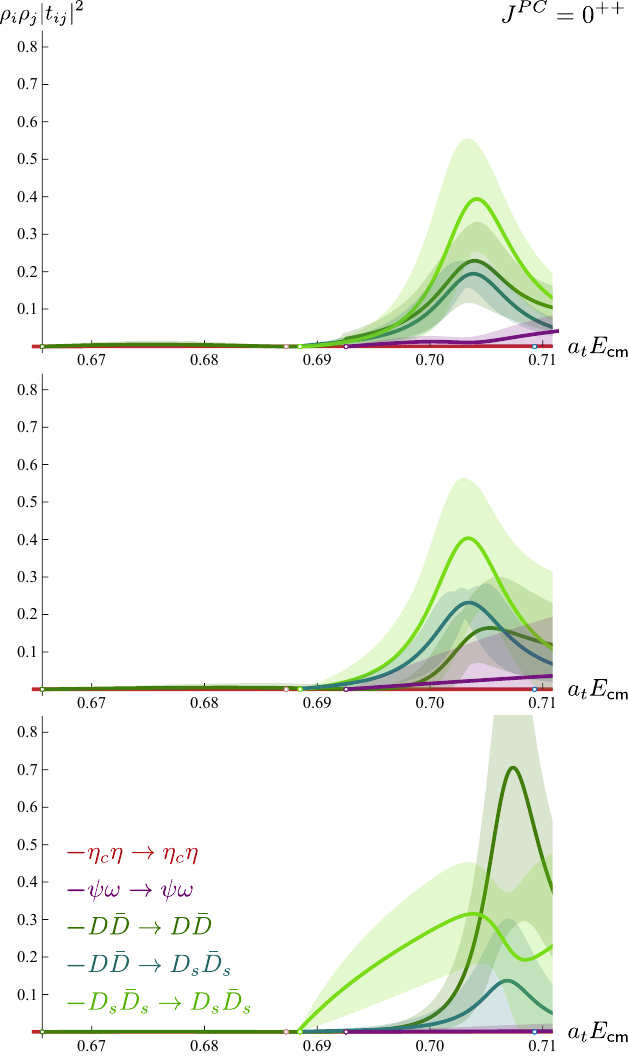}
\caption{A selection of scattering amplitudes determined below $a_t E_\cm=0.709$ (just below $\DstDst$ threshold) in $J^{PC}=0^{++}$. Peaks are seen across all parameterizations in $\DD$ that correspond to a resonance pole consistent with that given in the main text, strongly coupled to $\DD$ and $\DsDs$. 
The top panel shows amplitudes obtained from 30 rest frame energies. 
The middle panel shows amplitudes obtained from 75 energies including moving frames.
The bottom panel highlights a slightly different amplitude found with hadron masses set to 1$\sigma$ below their best-fit values.}
\label{fig:amps_compared_upto_0p709}
\end{figure*}

Working below $a_tE_\cm=0.709$, there are 30 levels over the three volumes (including three $\eta_c\etap$ levels) from $[000]A_1^+$. Following the methods outlined in the main text, we make use of a $K$-matrix with a pole and constants, $K_{ij}= \frac{g_ig_j}{m^2-s}+\gamma_{ij}$, and a Chew-Mandelstam phase space subtracted at the $K$-matrix pole. One representative result is,
\begin{center}
\begin{tabular}{rll}
$a_t m=$ & $(0.70398 \pm 0.00128 \pm 0.00035)$ & \multirow{6}{*}{ $\begin{bmatrix*}[r]   1.00 &  -0.10 &  -0.57 &  -0.27 &  -0.16 &  -0.10\\
&  1.00 &   0.14 &  -0.07 &   0.28 &   0.01\\
&&  1.00 &   0.09 &   0.01 &   0.00\\
&&&  1.00 &  -0.14 &   0.17\\
&&&&  1.00 &  -0.03\\
&&&&&  1.00\end{bmatrix*}$ } \\ 
$a_t g_{\DD}=$             & $(0.0867 \pm 0.0155 \pm 0.00143)$ & \\
$a_t g_{\DsDs}=$           & $(0.1281 \pm 0.0222 \pm 0.00393)$ & \\
$\gamma_{\eta_c\eta\to\eta_c\eta} =$              & $(0.07 \pm 0.098 \pm 0.048)$ & \\
$\gamma_{\DD\to\psiom} = $                        & $(2.04 \pm 0.620 \pm 0.208)$ & \\
$\gamma_{\eta_c\etap\to\eta_c\etap}  =$  & $(3.17 \pm 1.49 \pm 0.66)$ & \\[1.3ex]
&\multicolumn{2}{l}{ $\chi^2/ N_\mathrm{dof} = \frac{41.9}{30-6} = 1.75$\,,}
\end{tabular}
\vspace{-0.4cm}\begin{equation}\label{eq:amp_below_DstDst_rest}\end{equation}
\end{center}
where as previously all parameters not listed are set equal to zero. This amplitude is plotted in the top panel of Fig.~\ref{fig:amps_compared_upto_0p709}.

The limited set of energy levels provides relatively little constraint for the $\psi\omega$ channel, with only three levels dominated by $\psi\omega$ operators being present very close to $\psi\omega$ threshold. However, in this case, the $\psi\omega$ amplitude has an interesting shape, it produces a dip around the pole position and has a relatively sharp rise from threshold, resulting in a shallow peak slightly \emph{below} the resonance mass. The uncertainties are large and this feature does not survive the addition of more energies, but it does show how small strength features in very weakly coupled channels do not always resemble the dominant resonance.

Adding more free parameters can produce a lower $\chi^2/N_\mathrm{dof}$, in particular when allowing non-zero $\gamma_{\DD\to\DD}$ and $\gamma_{\DsDs\to\DsDs}$. However these also result large uncertainties on the determined amplitudes. Freedom in the $\gamma_{\DD\to\psiom}$ parameter can be interchanged with $\gamma_{\psiom\to\psiom}$ or $g_\psiom$ with little effect on the amplitudes within uncertainties, and only small changes in the $\chi^2/N_\mathrm{dof}$.

\subsubsection{Including moving-frame energies}
\label{app:DstDst_mov}

Using rest-frame energies up to $a_t E_\cm=0.709$ and the same selection of moving-frame energies up to $a_t E_\cm=0.690$ that were previously used in section~\ref{app:amps_Swave_below_psiom}, results in 75 levels to constrain the $S$-wave amplitudes. We fix the $J^{PC}=2^{++}$ result to the reference amplitude given in Eq.~\ref{eq:fit_2pP_rest+flight}, and for simplicity we fix all other contributing partial-waves to zero. We also fix the strength of the decoupled $\etcep$ channel via $\gamma_{\etcep\to\etcep}=3$, consistent with the determination in Eqs.~\ref{eq:fit_0pP_rest+flight} and \ref{eq:amp_below_DstDst_rest}. An example amplitude determined from these levels is,
\begin{center}
\begin{tabular}{rll}
$a_t m=$ & $(0.7037 \pm 0.0013 \pm 0.0007)$ & \multirow{7}{*}{ $\begin{bmatrix*}[r]   1.00 &   0.31 &  -0.46 &  -0.22 &   0.08 &   0.47 &  -0.06\\
&  1.00 &  -0.06 &  -0.05 &  -0.32 &   0.42 &  -0.01\\
&&  1.00 &   0.21 &  -0.11 &   0.06 &   0.00\\
&&&  1.00 &   0.14 &   0.10 &   0.09\\
&&&&  1.00 &   0.04 &   0.05\\
&&&&&  1.00 &   0.03\\
&&&&&&  1.00\end{bmatrix*}$ } \\ 
$a_t g_{\DD}=$              & $(0.081 \pm 0.016 \pm 0.001)$ & \\
$a_t g_{\DsDs}=$           & $(0.133 \pm 0.022 \pm 0.009)$ & \\
$\gamma_{\eta_c\eta\to\eta_c\eta} =$               & $(0.02 \pm 0.08 \pm 0.05)$ & \\
$\gamma_{\DD\to\DD} = $                            & $(-0.53 \pm 0.21 \pm 0.11)$ & \\
$\gamma_{\DsDs\to\DsDs} = $                        & $(0.16 \pm 1.02\pm 0.34)$ & \\
$\gamma_{\psiom\to\psiom} = $                      & $(1.02 \pm 2.01 \pm 0.26)$ & \\
$\gamma_{\eta_c\etap\to\eta_c\etap}  =$  & $3$ (fixed) & \\[1.3ex]
&\multicolumn{2}{l}{ $\chi^2/ N_\mathrm{dof} = \frac{89.9}{75 - 7 - 11} =  1.58$\,.}
\end{tabular}
\end{center}
In this case, the second uncertainties are obtained by varying the scattering hadron masses to their upper values in Table~\ref{tab:masses} ($m_i\to m_i+\delta m_i$), but only half a sigma in the negative direction ($m_i\to m_i-\delta m_i/2$). When varying the masses to their lower values ($m_i\to m_i-\delta m_i$), we find qualitatively slightly different solutions which are discussed below. This amplitude is plotted in the central panel of Fig.~\ref{fig:amps_compared_upto_0p709}.

In this reduced energy region below $\DstDst$ threshold, when performing amplitude determinations at the extreme lower end of the ranges of the hadron masses ($m_i\to m_i-\delta m_i$ in Table~\ref{tab:masses}), we observe a second class of solution for the $\DsDs$ amplitude, shown in the bottom panel of Fig.~\ref{fig:amps_compared_upto_0p709}. We see that this solution has a significantly stronger turn-on of the $D_s\bar{D}_s \to D_s\bar{D}_s$ amplitude at threshold, and an atypical behavior at higher energies where a strong peak appears in $D\bar{D} \to D\bar{D}$.

The parameters corresponding to this solution using the rather extreme mass values $m_i-\delta m_i$ are,
\begin{center}
\begin{tabular}{rll}
$a_t m=$  & $(0.7069 \pm 0.0010)$ & \multirow{7}{*}{ $\begin{bmatrix*}[r]   1.00 &  -0.17 &  -0.62 &  -0.32 &   0.00 &  -0.04 &  -0.12\\
&  1.00 &  -0.14 &  -0.08 &  -0.71 &   0.08 &   0.00\\
&&  1.00 &   0.14 &   0.26 &  -0.18 &   0.07\\
&&&  1.00 &   0.21 &   0.27 &   0.11\\
&&&&  1.00 &   0.13 &   0.11\\
&&&&&  1.00 &   0.09\\
&&&&&&  1.00\end{bmatrix*}$ } \\ 
$a_t g_{\DD}=$         & $(0.102 \pm 0.019)$ & \\
$a_t g_{\DsDs}=$       & $(0.065 \pm 0.057)$ & \\
$\gamma_{\eta_c\eta\to\eta_c\eta} =$ & $(-0.04 \pm 0.10)$ & \\
$\gamma_{\DD\to\DD} =$ & $(-0.39 \pm 0.23)$ & \\
$\gamma_{\DsDs\to\DsDs} =$ & $(\;\;2.87 \pm 1.02)$ & \\
$\gamma_{\psiom\to\psiom} =$ & $(-0.22 \pm 1.19)$ & \\
$\gamma_{\etcep\to\etcep} = $ & 3 (fixed) & \\[1.3ex]
&\multicolumn{2}{l}{ $\chi^2/ N_\mathrm{dof} = \frac{91.9}{75 - 7 - 11} =  1.61$\,,}
\end{tabular}
\end{center}
and this amplitude has a virtual bound-state pole strongly coupled to the $\DsDs$ channel at ${a_t\sqrt{s_0}=0.6656 \pm 0.0157}$ with $a_t|c_{\DsDs}|=0.413 \pm 0.056$. We highlight this result in part because it is resembles somewhat the solution found in Ref.~\cite{Prelovsek:2020eiw}, but only qualitatively, as the pole is roughly 125 MeV below $\DsDs$ threshold while Ref.~\cite{Prelovsek:2020eiw} reports a pole within a few MeV of the $\DsDs$ threshold. 

We consider this solution to be disfavored as it only appears when an extreme choice is made for all scattering  hadron masses, but even here a scalar resonance appears, in reasonable agreement with our other determinations, with a pole position $a_t\sqrt{s_0}=0.7068(9)\pm \frac{i}{2}0.0059(26)$ on the proximal sheet. Large couplings to $\DD$ and $\DsDs$ and small couplings to the $\etce$ and $\psiom$ channels are found.

It should also be noted that while the central values look quite different, the uncertainty bands are largely consistent across sections~\ref{app:DstDst_rest}, \ref{app:DstDst_mov}, and also the main results given above, as can be seen in Fig.~\ref{fig:amps_compared_upto_0p709}.

\clearpage
\pagebreak

\section{A $J^{PC}=2^{++}$ toy-model study}
\label{app:toy_D}

The purpose of this appendix is to show that the couplings to the kinematically closed $\DstDst$ channel can be determined reliably from the volume-dependence of energy levels. We illustrate the sensitivity using a simplified two-channel system with a resonance coupling to an open $D$-wave channel (which we call $\DD$) and a closed $S$-wave channel ($\DstDst$). For an approximately fixed resonance mass and width, we show that the spectra are sensitive to the value of the coupling to $\DstDst$. This toy-model also contains a further example of the $K$-matrix pole ``coupling-ratio phenomenon'' described in Section~\ref{sec:2pP_rest}.

We utilize a two-channel version of the Flatt\'e amplitude Eq.~\ref{eq:flatte}, where the lower channel is $\DD\SLJc{1}{D}{2}$ (with $D$-wave suppression close to threshold) and the higher channel is $\DstDst\SLJc{5}{S}{2}$ (an $S$-wave channel that can open rapidly). The pole parameter is set to $a_tm = 0.7$, however we subtract the real correction from $-ig_{\DstDst}^2\rho_{\DstDst}(m^2)$, as described after Eq.~\ref{eq:flatte}, so that the pole parameter $m$ retains its meaning.
We initially fix $a_t g_{\DstDst}=1.6$ which is a representative value giving $\DstDst\SLJc{5}{S}{2}$ amplitudes similar to those found throughout this work. The $g_{\DD}$ coupling is then chosen so that a $t$-matrix pole width $a_t\Gamma=-2 \, \mathrm{Im}\, a_t\sqrt{s_0} = 0.0116$ is obtained (corresponding to $\approx$ 66 MeV). We then reduce $g_{\DstDst}$ and adjust $g_{\DD}$ in order to maintain an approximately constant $t$-matrix pole position $a_t\sqrt{s_0}=0.697\pm \frac{i}{2}0.0116$. 

\begin{figure*}
\includegraphics[width=1.0\textwidth]{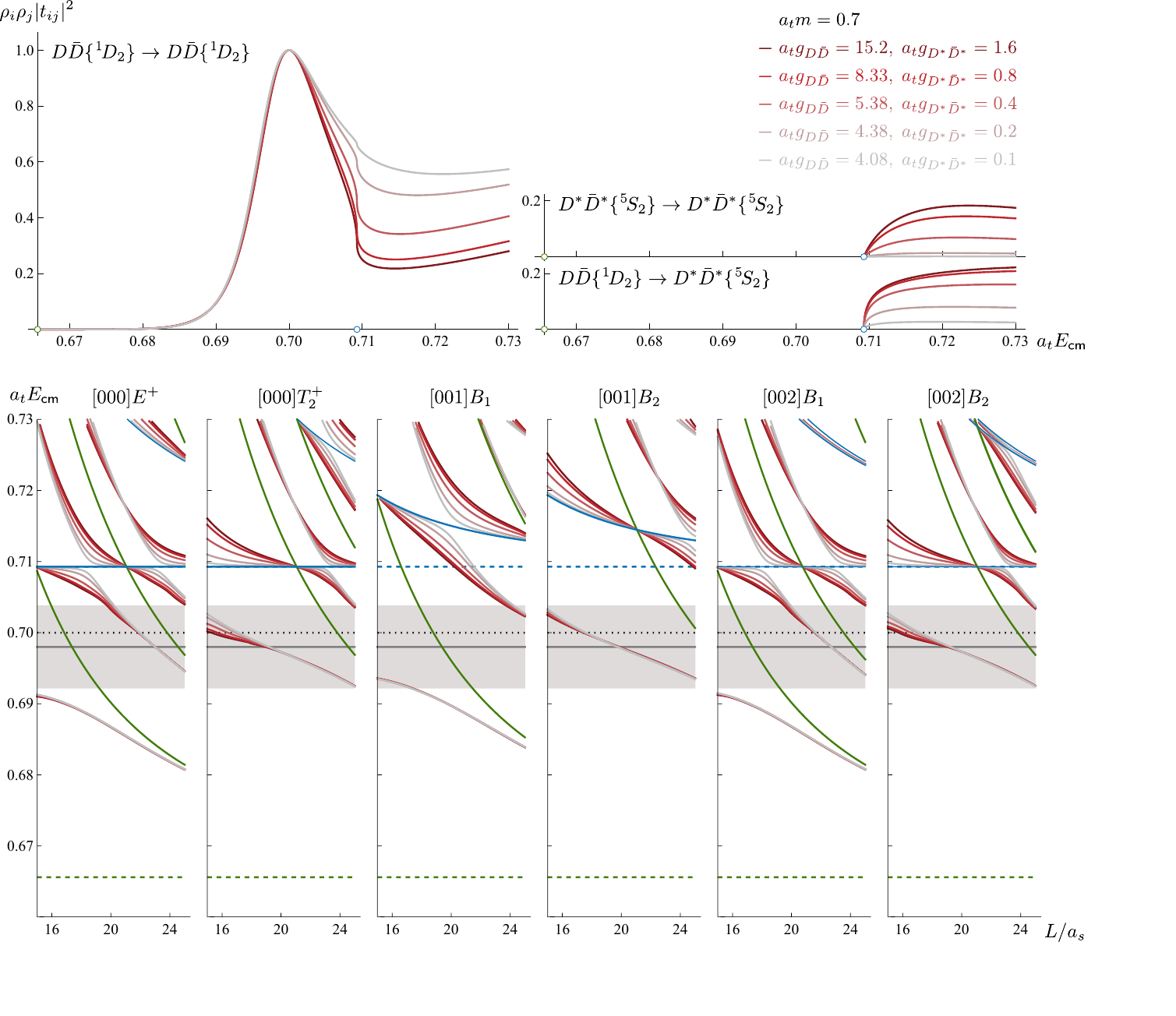}
\caption{Toy model amplitudes consisting of $\DD\SLJc{1}{D}{2}$ and $\DstDst\{\SLJ{5}{S}{2}\}$ as described in the text. 
Upper panel: Scattering amplitudes plotted as $\rho_i\rho_j|t_{ij}|^2$ for each of the parameters given in the top right. Circles on the horizontal axis indicate threshold energies.
Lower panel: The solid green and blue curves show non-interacting energies corresponding to $\DD$ and $\DstDst$ respectively. Degeneracies are not indicated since only a single level is expected for each hadron-hadron pair when only a single combination is present in the L\"uscher determinant condition Eq.~\ref{eq:det}. The dashed horizontal lines indicate kinematic thresholds. The grey band and horizontal solid grey line indicate the mass and width of the resonance pole. The dotted horizontal line indicates both the position of the mass parameter $a_tm$, and the centre of the peak seen in the $\DD\SLJc{1}{D}{2}$ amplitudes. The red and grey spectrum of curves show the finite volume spectra obtained from the L\"uscher determinant condition Eq.~\ref{eq:det} corresponding to the same-colored amplitudes from the upper panel.
}
\label{fig:toy}
\end{figure*}

In Fig.~\ref{fig:toy} we show the amplitudes and the finite volume spectra resulting from this procedure. Below $a_tE_\cm=0.7$, on the lower half of the resonance hump, we see almost no variation as these parameters are changed. Similarly, the finite volume spectra in this energy region show little sensitivity.

On the other hand \emph{above} $a_tE_\cm=0.7$, significant differences are observed. An avoided level crossing occurs in every irrep around the position of the lowest $\DstDst$ non-interacting energy with departures proportional to the size of $g_{\DstDst}$. These deviations are significantly larger than typical uncertainties in the computed spectrum and so it is plausible that the coupling to this channel can be well-determined.

We also observe the beginning of the onset of the coupling-ratio phenomenon in this toy model given that there is only a relatively small difference between the amplitudes with $a_tg_{\DstDst}=0.8$ and $a_tg_{\DstDst}=1.6$, which correspond to very similar coupling ratios, $g_{\DstDst}/ g_{\DD} = 1.05, 0.096$ respectively. 

\clearpage
\pagebreak


\section{Summary of amplitude parameterizations}

In this appendix we summarise the parameterizations of the $J^{PC}=3^{++}$, $2^{++}$ and $0^{++}$ amplitudes.

\subsection{Summary of $J^{PC}=3^{++}$ parameterizations}
\label{app:Fwave_par_vars}

In Table~\ref{tab:J3pP_par_var} we summarize 10 $J^{PC}=3^{++}$ parameterization variations, describing 16 energy levels in the $[000]A_2^+$ irrep as introduced in section~\ref{sec:3pP_scattering}. All eigenvalues of the data correlation matrix were above the cutoff of $\Lambda=0.02$. The amplitudes are shown in Fig.~\ref{fig:amp_F_par_var}.

\begin{table*}[h]
\begin{tabular}{l|r|l}
amplitude parameters & \multicolumn{1}{c|}{$\chi^2/N_{\mathrm{dof}}$} & other details\\
\hline
\multicolumn{3}{l}{$m$, $g_{\DDst\SLJc{3}{D}{3}}$, $\gamma_{\psi\omega\to\psi\omega\SLJc{3}{D}{3}}$, $\gamma_{\psi\omega\to\psi\omega\SLJc{5}{D}{3}}$, $\gamma_{\psi\phi\to\psi\phi\SLJc{5}{D}{3}}$}\\
$\quad \& \, g_{\DsDsst\SLJc{3}{D}{3}}, g_{\DstDst\SLJc{5}{D}{3}}, \gamma_{\DDst\to\DDst\SLJc{3}{D}{3}}, \gamma_{\DsDsst\to\DsDsst\SLJc{3}{D}{3}}, \gamma_{\DstDst\to\DstDst\SLJc{5}{D}{3}}$   &  $\frac{8.34}{16 - 10} =  1.39$ & reference amp\\
$\quad \&\, \gamma_{\DsDsst\to\DsDsst\SLJc{3}{D}{3}}, \gamma_{\DstDst\to\DstDst\SLJc{5}{D}{3}}$  & $\frac{8.47}{16 - 7} =  0.94$ \\
$\quad \&\, \gamma_{\DsDsst\to\DsDsst\SLJc{3}{D}{3}}, \gamma_{\DstDst\to\DstDst\SLJc{5}{D}{3}}$  & $\frac{8.45}{16 - 7} =  0.94$ & simple phase space\\
$\quad \&\, \gamma_{\DDst\to\DDst\SLJc{3}{D}{3}}, \gamma_{\DsDsst\to\DsDsst\SLJc{3}{D}{3}}, \gamma_{\DstDst\to\DstDst\SLJc{5}{D}{3}}$   &  $\frac{8.34}{16 - 8} =  1.04$\\
$\quad \&\, g_{\DstDst\SLJc{5}{D}{3}}, \gamma_{\DDst\to\DDst\SLJc{3}{D}{3}}, \gamma_{\DsDsst\to\DsDsst\SLJc{3}{D}{3}}, \gamma_{\DstDst\to\DstDst\SLJc{5}{D}{3}}$   &  $\frac{8.50}{16 - 8} =  1.06$ \\
$\quad \&\, g_{\DstDst\SLJc{5}{D}{3}}, \gamma_{\DDst\to\DDst\SLJc{3}{D}{3}}, \gamma_{\DsDsst\to\DsDsst\SLJc{3}{D}{3}}, \gamma_{\DstDst\to\DstDst\SLJc{5}{D}{3}}$   &  $\frac{8.48}{16 - 8} =  1.06$ & simple phase space\\
$\quad \&\, g_{\DstDst\SLJc{5}{D}{3}}, \gamma_{\DDst\SLJc{3}{D}{3}\to\DstDst\SLJc{5}{D}{3}}, \gamma_{\DsDsst\to\DsDsst\SLJc{3}{D}{3}}$   &  $\frac{4.65}{16 - 8} =  0.58$ \\
$\quad \&\, g_{\DstDst\SLJc{5}{D}{3}}, \gamma_{\DDst\SLJc{3}{D}{3}\to\DstDst\SLJc{5}{D}{3}}, \gamma_{\DsDsst\to\DsDsst\SLJc{3}{D}{3}}, \gamma_{\DstDst\to\DstDst\SLJc{5}{D}{3}}$   &  $\frac{ 4.33}{16 - 9} =  0.62$ \\
$\quad \&\, g_{\DsDsst\SLJc{3}{D}{3}}, \gamma_{\DsDsst\to\DsDsst\SLJc{3}{D}{3}}, \gamma_{\DstDst\to\DstDst\SLJc{5}{D}{3}}$   &  $\frac{8.47}{16 - 8} =  1.06$\\
$\quad \&\, g_{\DsDsst\SLJc{3}{D}{3}}, g_{\DstDst\SLJc{5}{D}{3}}, \gamma_{\DsDsst\to\DsDsst\SLJc{3}{D}{3}}, \gamma_{\DstDst\to\DstDst\SLJc{5}{D}{3}}$   &  $\frac{8.47}{16 - 9} =  1.21$

\end{tabular}
\caption{$J^{PC}=3^{++}$ parameterization variations. Parameters not listed are fixed to zero. A $K$-matrix constant parameter $\gamma$ is diagonal in $\SLJ{2S+1}{\ell}{J}$ unless otherwise stated. A Chew-Mandelstam phase space with a pole subtraction point is used unless otherwise stated.}
\label{tab:J3pP_par_var}
\end{table*}

\clearpage
\pagebreak


\subsection{Summary of $J^{PC}=2^{++}$ parameterizations}
\label{app:Dwave_par_vars}

In Table~\ref{tab:J2pP_par_var} we summarize 24 $J^{PC}=2^{++}$ parameterization variations obtained including energies from moving frame irreps. Many more parameterization forms were attempted but only those with good $\chi^2$ minima are retained. Four additional parmeterizations were obtained with the same free parameters as the reference amplitude, two with the scattering hadron masses set to their central values $\pm 1\sigma$, as given in Table~\ref{tab:masses}, and two with the anisotropy set to its upper and lower values $3.444\pm0.006$ as determined from the pion. By default we apply a cutoff on data correlation eigenvalues of $\Lambda=0.02$ as mentioned in the text, with further details given in Appendix~\ref{app:svd}.

\begin{table*}[h]
\begin{tabular}{l|r|l}
amplitude parameters & \multicolumn{1}{c|}{$\chi^2/N_{\mathrm{dof}}$} & other details\\
\hline
\multicolumn{3}{l}{$m$, $g_{\DD}$, $g_{\DDst}$, $g_{\DsDs}$, $g_{\DstDst}$, $\gamma_{\eta_c\eta\to\eta_c\eta}$, $\gamma_{\psiom\to\psiom}$, $\gamma_{\psi\phi\to\psi\phi}$}\\
                                                                                                              & $\frac{65.3}{86 - 7 - 23} =  1.17$\\
$\quad \& \quad g_{\DD}$ (freed)                                                                                       & $\frac{65.1}{86 - 7 - 23} =  1.16$ & $g_{\DDst}=-40$ (fixed)\\
$\quad \& \quad \gamma_{\DD\to\DsDs}$                                                                          & $\frac{62.8}{86 - 8 - 23} =  1.14$ & reference amp Eq.~\ref{eq:fit_2pP_rest+flight}\\
$\quad \& \quad \gamma_{\DD\to\DD},\gamma_{\DDst\to\DDst}, \gamma_{\DsDs\to\DsDs}, \gamma_{\DstDst\to\DstDst}$ & $\frac{64.0}{86 - 11 - 23} =  1.23$\\
$\quad \& \quad \gamma_{\DD\to\DsDs},\gamma_{\DD\to\DDst},\gamma_{\DDst\to\DsDs}, \gamma_{\DsDs\to\DsDs}$                           & $\frac{56.9}{86 - 11 - 23} =  1.09$\\      
$\quad \& \quad \gamma_{\DD\to\DsDs},\gamma_{\DDst\to\DDst}, \gamma_{\DstDst\to\DstDst}$                       & $\frac{65.3}{86 - 10 - 23} =  1.23$\\
$\quad \& \quad \gamma_{\DDst\to\DDst}, \gamma_{\DstDst\to\DstDst}$                                            & $\frac{65.3}{86 -  9 - 23} =  1.21$\\
$\quad \& \quad \gamma_{\DDst\to\DsDs}$                                                                        & $\frac{60.0}{86 -  8 - 23} =  1.09$\\
$\quad \& \quad \gamma_{\DstDst\to\DstDst}$                                                                    & $\frac{65.3}{86 -  8 - 23} =  1.19$\\
$\quad \& \quad \gamma_{\DDst\to\psiom}$                                                                       & $\frac{65.2}{86 -  7 - 23} =  1.16$ & $\gamma_{\psiom\to\psiom}=0$ (fixed)\\ 
\hline
 & $\frac{99.1}{86 - 7 - 12} =  1.48$ & $\Lambda=0.01$\\
 & $\frac{74.6}{86 - 7 - 19} =  1.24$ & $\Lambda=0.016$\\
 & $\frac{54.4}{86 - 7 - 28} =  1.07$ & $\Lambda=0.024$\\
 & $\frac{45.3}{86 - 7 - 35} =  1.03$ & $\Lambda=0.032$\\
 & $\frac{36.8}{86 - 7 - 40} =  0.94$ & $\Lambda=0.040$\\
%
%
$\quad \& \quad \gamma_{\DD\to\DsDs}$                                                                          & $\frac{63.3}{86 - 8}     =  0.81$ & uncorrelated\\
\hline
& & simple phase space \\
$\quad \& \quad \gamma_{\psiom\SLJc{3}{D}{2}\to\psiom\SLJc{3}{D}{2}}$                                          & $\frac{66.9}{86 - 8 - 23} =  1.22$ & \\
$\quad \& \quad \gamma_{\DD\to\DsDs} + \gamma_{\DsDs\to\DstDst} $                                              & $\frac{70.9}{86 - 9 - 23} =  1.31$ & \\
$\quad \& \quad \gamma_{\DD\to\DsDs} + \gamma_{\DsDs\to\DstDst} $                                              & $\frac{68.1}{86 - 9 - 23} =  1.26$ & $g_\DD$ = 20\\
$\quad \& \quad \gamma_{\DD\to\DsDs} + \quad \gamma_{\DDst\to\DDst} + \quad \gamma_{\DstDst\to\DstDst}$        & $\frac{69.4}{86 - 10 - 23} =  1.31$ & \\
\hline
& & simple phase space \\
& $\frac{106.6}{86 - 7 - 12} =  1.59$ & $\Lambda=0.010$\\
& $\frac{78.8}{86 - 7 - 19} =  1.31$ & $\Lambda=0.016$\\
& $\frac{69.5}{86 - 7 - 23} =  1.24$ & $\Lambda=0.020$\\
& $\frac{57.0}{86 - 7 - 28} =  1.12$ & $\Lambda=0.024$\\
\end{tabular}

\caption{$J^{PC}=2^{++}$ parameterization variations. Parameters not listed are fixed to zero. A Chew-Mandelstam phase space with a $K$-matrix pole subtraction point is used unless otherwise stated. An cutoff on data correlation eigenvalues of $\Lambda=0.02$ is used unless otherwise stated. We fix $g_{\DD}=10 \cdot a_t$ unless indicated otherwise.
}
\label{tab:J2pP_par_var}
\end{table*}

\clearpage
\pagebreak

\subsection{Summary of $J^{PC}=0^{++}$ parameterizations}
\label{app:Swave_par_vars}

\subsubsection{Coupled $\etce$ and $\DD$ scattering below $\etcep$ and $\DsDs$ thresholds}
\label{app:Swave_par_vars_two_chan}

This section gives further details on the amplitudes variations used in Section~\ref{sec:Swave_threshold_mov}, where $K$-matrices are determined using energies from $[000]A_1^{+}$, $[001]A_1$, $[111]A_1$ and $[002]A_1$, resulting in 43 levels. In two of the fits a reduced selection of energies is used, removing $[002]A_1$ levels and resulting in 31 levels. 
A data correlation eigenvalue cutoff of $\Lambda = 0.02$ is used, resulting in 5 resets for the 43 level selection, and 3 resets for the 31 level selection. Using two or three constant parameter terms $\gamma_{ij}$ in the $K$-matrix results in 10 parameterisations in total, as summarized in Table~\ref{tab:Swave_threshold_par_var}.   

\begin{table*}[h]
\begin{tabular}{ccc|l}
$\gamma_{\eta_c\eta\to\eta_c\eta}$ & $\gamma_{\eta_c\eta\to \DD}$ & $\gamma_{\DD\to\DD}$ & $\qquad{\chi^2}/{N_\mathrm{dof}}\qquad$ \\
\hline
\hline
\multicolumn{4}{l}{with Chew-Mandelstam phase space}\\
0.37(16) & -0.65(18) &  0.06(49) & $\frac{28.4}{31-3-3} = 1.14$ $(\ast)$\\
0.37(15) & -0.64(16) &  0.15(34) & $\frac{40.5}{43-3-5} = 1.16$\\
--       & -0.45(14) &  0.14(32) & $\frac{47.6}{43-2-5} = 1.32$\\
0.40(12) & --        & -0.39(24) & $\frac{48.6}{43-2-5} = 1.35$\\
0.37(14) & -0.61(15) & --        & $\frac{40.8}{43-2-5} = 1.13$\\
\hline
\multicolumn{4}{l}{with simple phase space}\\
0.36(11) & -0.51(15) & -0.02(40) & $\frac{45.0}{31-3-3} = 1.80$ $(\ast)$\\
0.36(14) & -0.63(15) &  0.14(31) & $\frac{40.4}{43-3-5} = 1.15$\\
--       & -0.46(14) &  0.13(31) & $\frac{47.4}{43-2-5} = 1.32$\\
0.40(12) & --        & -0.39(25) & $\frac{48.5}{43-2-5} = 1.35$\\
0.36(14) & -0.61(15) & --        & $\frac{40.6}{43-2-5} = 1.13$\\
\end{tabular}
\caption{Parameterization variations for two-channel $\etce-\DD$ $S$-wave scattering amplitudes including moving frame energies. The first row in each block, indicated by $(\ast)$, uses only 31 levels, excluding levels from $[002]A_1$.
`--' indicates that a parameter is fixed to zero.
The number of degrees of freedom is taken to be $N_\mathrm{dof} = N_\mathrm{levels}-N_\mathrm{pars.}-N_\mathrm{reset}$ using a data-correlation eigenvalue cutoff of $\Lambda=0.02$ as discussed in Appendix~\ref{app:svd}.
}
\label{tab:Swave_threshold_par_var}
\end{table*}

\pagebreak
\subsubsection{Coupled-channel scattering up to $\psi\phi$ threshold at rest and $a_tE_\cm=0.69$ in moving frames}

In Table~\ref{tab:Swave_mov_par_var} we summarize $J^{PC}=0^{++}$ parameterization variations working up to $\psi\phi$ threshold while including moving-frame information. One example with parameter values and correlations is given in Eq.~\ref{eq:fit_0pP_rest+flight}. These amplitudes are plotted in Fig.~\ref{fig:Swave_amp_par_var}, and are used when determining $t$-matrix poles in Section~\ref{sec:poles:scalar}.

\begin{table*}[h]
\begin{tabular}{l|r|l}
amplitude parameters & \multicolumn{1}{c|}{$\chi^2/N_{\mathrm{dof}}$} & other details\\
\hline
\multicolumn{3}{l}{$m$, $g_{\DD}$, $g_{\DsDs}$, $g_{\DstDst}$, $\gamma_{\eta_c\eta\to\eta_c\eta}$, $\gamma_{\psi\phi\to\psi\phi}$}\\
$\quad\quad \&\,  g_{\psiom},\,\gamma_{\DD\to\DsDs},\,\gamma_{\eta_c\etap\to\eta_c\etap},\,\gamma_{\psiom\SLJc{5}{D}{4}\to\psiom\SLJc{5}{D}{4}}$   &  $\frac{91.0}{90 - 10 - 16} =  1.42$ & reference amp\\
$\quad\quad \&\, \gamma_{\psiom\to\psiom}$                                                                   & $\frac{91.6}{90 - 7 - 16} =  1.37$\\
$\quad\quad \&\,  g_{\psiom},\,\gamma_{\DD\to\DsDs}$                                                         & $\frac{91.5}{90 - 8 - 16} =  1.39$\\
$\quad\quad \&\, g_{\psiom},\,\gamma_{\DD\to\DsDs},\,\gamma_{\psiom\SLJc{5}{D}{4}\to\psiom\SLJc{5}{D}{4}}$   &  $\frac{91.2}{90 - 9 - 16} = 1.40$\\
$\quad\quad \&\, \gamma_{\DD\to\DD},\,\gamma_{\DD\to\DsDs},\, g_{\psiom}$            &  $\frac{87.1}{90 - 9 - 16} =  1.34$\\
$\quad\quad \&\, \gamma_{\DD\to\DD},\,\gamma_{\DD\to\DsDs},\, \gamma_{\psiom\to\psiom},\, \gamma_{\psiom\to\DstDst}$ &  $\frac{94.7}{90 - 10 - 16} =  1.48$\\
$\quad\quad \&\, \gamma_{\DD\to\DD},\,\gamma_{\DD\to\DsDs},\, \gamma_{\psiom\to\psiom}$            &  $\frac{94.7}{90 - 9 - 16} =  1.46$\\
$\quad\quad \&\, \gamma_{\DsDs\to\DsDs},\,\gamma_{\psiom\to\psiom}$                                &  $\frac{93.6}{90 - 9 - 16} =  1.44$\\
$\quad\quad \&\, \gamma_{\DstDst\to\DstDst},\,\gamma_{\psiom\to\psiom}$                            &  $\frac{92.8}{90 - 9 - 16} =  1.43$\\
$\quad\quad \&\, \gamma_{\DD\to\DsDs},\,\gamma_{\psiom\to\psiom},\,\gamma_{\eta_c\etap\to\eta_c\etap}$  &  $\frac{95.0}{90 - 9 - 16} =  1.46$\\
$\quad\quad \&\,  \gamma_{\psiom\to\psiom}$                                                        & $\frac{92.9}{90 - 7 - 16} =  1.39$ & simple phase space\\
$\quad\quad \&\,  \gamma_{\psiom\to\psiom},\, \gamma_{\DD\to\DsDs}$                                & $\frac{90.6}{90 - 8 - 16} =  1.37$ & simple phase space\\
\hline
$\quad\quad \&\,  \gamma_{\DD\to\DsDs},\,g_{\psiom}$                                                        &  $\frac{143.0}{90 - 8 - 6} =  1.88$ & $\Lambda = 0.01$ ($\ast$)\\
$\quad\quad \&\,  \gamma_{\DD\to\DsDs},\,g_{\psiom}$                                                        &  $\frac{59.4}{90 - 8 - 35} =  1.26$ & $\Lambda = 0.04$ ($\ast$)\\
\hline
$\quad\quad \&\, \gamma_{\DD\to\DsDs},\,\gamma_{\psiom\to\psiom}$                                                         &  $\frac{143.4}{90 - 8 - 6} =  1.89$ & $\Lambda = 0.01$ ($\ast$)\\
$\quad\quad \&\, \gamma_{\DD\to\DsDs},\,\gamma_{\psiom\to\psiom}$                                                         &  $\frac{115.3}{90 - 8 - 11} =  1.62$ & $\Lambda = 0.016$\\
$\quad\quad \&\, \gamma_{\DD\to\DsDs},\,\gamma_{\psiom\to\psiom}$                                                         &  $\frac{82.2}{90 - 8 - 20} =  1.33$ & $\Lambda = 0.024$\\
$\quad\quad \&\, \gamma_{\DD\to\DsDs},\,\gamma_{\psiom\to\psiom}$                                                         &  $\frac{69.8}{90 - 8 - 28} =  1.29$ & $\Lambda = 0.032$\\
$\quad\quad \&\, \gamma_{\DD\to\DsDs},\,\gamma_{\psiom\to\psiom}$                                                         &  $\frac{59.4}{90 - 8 - 35} =  1.26$ & $\Lambda = 0.04$ ($\ast$)\\
$\quad\quad \&\, \gamma_{\DD\to\DsDs},\,\gamma_{\psiom\to\psiom}$                                                         &  $\frac{71.0}{90 - 8} =  0.87$ & uncorrelated \\
\end{tabular}

\caption{$J^{PC}=0^{++}$ parameterization variations. Parameters not listed are fixed to zero. A Chew-Mandelstam phase space with a $K$-matrix pole subtraction point is used unless otherwise stated. A data-correlation eigenvalue cutoff of $\Lambda=0.02$ is used unless otherwise stated. If $\gamma_{\psiom\SLJc{5}{D}{4}\to\psiom\SLJc{5}{D}{4}}$ is not listed, it is fixed to 300. If $\gamma_{\eta_c\etap\to\eta_c\etap}$ is not listed, it is fixed to 3. All meson-meson channels are $\SLJ{1}{S}{0}$ unless otherwise stated. Amplitudes marked ($\ast$) are provided for comparison with Table~\ref{tab:J2pP_par_var}, and they are not included on the plots or used in the analysis.
}
\label{tab:Swave_mov_par_var}
\end{table*}


\clearpage
\pagebreak

\section{Data covariance eigenvalue cutoff}
\label{app:svd}

Relatively large data correlations between the energy levels on each lattice volume are found in this work. For small selections of energy levels, such as those obtained using only rest-frame energies, this does not present a problem, but for larger selections of energy levels, such as those using moving-frames, inverting the data covariance for use in a correlated $\chi^2$ produces an object of questionable validity. Given the use of ensembles of typically $\sim 500$ gauge configurations, one should not expect to be able to reliably determine all components of data covariance matrices of increasingly high rank. This issue is most relevant for the amplitude determinations in Sections~\ref{sec:Swave_scattering_final} and \ref{sec:2pP_rest_and_mov}.

\begin{figure*}[h]
\includegraphics[width=1.0\textwidth]{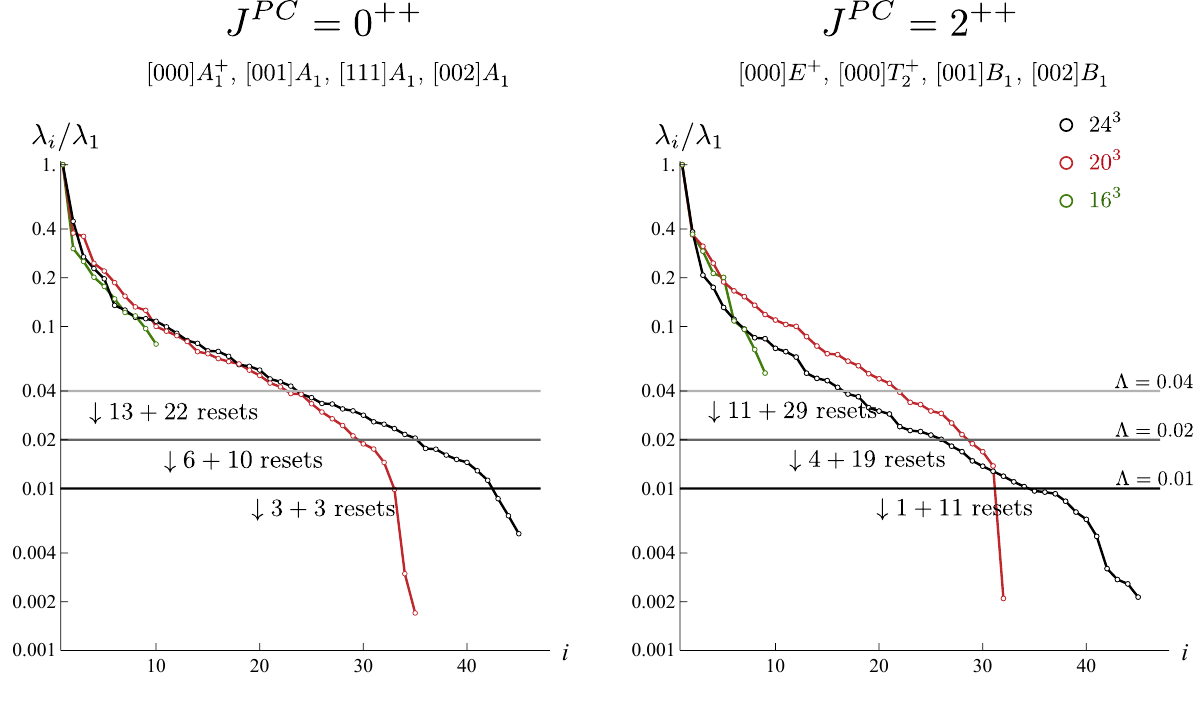}
\caption{The eigenvalues, $\lambda_i$, of the data correlation matrix normalized to the largest eigenvalue $\lambda_1$, ordered in decreasing magnitude. We observe a steep falloff above $i \approx 30$ where the directions in the eigenspace are unlikely to be reliably determined. The number of resets on the plot indicates the number of modes that are discarded on the $20^3$ + $24^3$ volumes, with cuts of $\Lambda=0.04$ (light grey), 0.02 (dark grey) and 0.01 (black). The ``$0^{++}$'' and ``$2^{++}$'' refer to the correlation matrix used in the largest amplitude determinations of these $J^{PC}$ 
from Sections~\ref{sec:Swave_scattering_final} and \ref{sec:2pP_rest_and_mov} respectively.}
\label{fig:eigs}
\end{figure*}

In Fig.~\ref{fig:eigs}, we show the eigenvalues $\lambda_i$ of the data correlation matrices, normalised to the largest eigenvalue, $\lambda_1$ for the two largest sets of spectra relevant to scalar and tensor scattering, including moving-frame energies. 
A steep dropoff in value is observed for the smallest value eigenvalues, and we infer that this is associated with the these modes being poorly determined. We choose to place a cut on the allowed values when performing fits, removing the eigenvectors associated with the cut eigenvalues from the matrix inverse. Our default choice is to retain only those modes with $\Lambda=\lambda/\lambda_1>0.02$. We have explored a range of values of this cut between $\Lambda =0.01$ and 0.04, and have reported the modest sensitivity to this choice in earlier appendices. 

We have also explored other approaches, such as artificially setting the correlations to zero, ``shrinkage'' which interpolates between fully correlated and uncorrelated~\cite{LEDOIT2004365,Rinaldi:2018osy}, and using ``eigenvalue limits'' rather than hard cutoffs~\cite{Dowdall:2019bea}. The outcomes are broadly similar as can be inferred from Figs.~\ref{fig:Dwave_amp_par_var} \&~\ref{fig:Swave_amp_par_var}, which include amplitudes determined using several different values of $\Lambda$ and one where the correlations are set to zero, as shown in Tables~\ref{tab:Swave_mov_par_var} and~\ref{tab:J2pP_par_var} respectively. Provided the smallest eigenmodes or the most extreme correlations are tamed using one of these methods, the results are in good agreement. We consider $\Lambda = 0.02$ to be a conservative choice.

\vspace{5mm}
\section{Additional scattering amplitude poles in $J^{PC}=2^{++}$}
\label{app:extra_poles}

The $J^{PC}=2^{++}$ amplitudes determined in Section~\ref{sec:2pP_rest_and_mov} feature a single narrow resonance pole that is systematically present across many parameterizations, but in addition other poles can be present which vary in location and which do not have obvious interpretations. We explore these in this Appendix.
In particular we investigate the origin of the closest of these additional poles using simplified elastic and two-channel systems that capture the main features of the amplitudes used in this work. We explore the dependence of these poles on the $g_\DDst$ parameter and propose an alternative parameterization where the additional poles do not arise. Ultimately we find that the narrow resonance pole on the proximal $\scriptstyle{(\DD[-],\DDst[-],\DsDs[-],\DstDst[+])}$ sheet is the only nearby pole singularity necessary to describe the finite-volume spectra.

\begin{figure*}[htb]
  \includegraphics[width=0.999\textwidth]{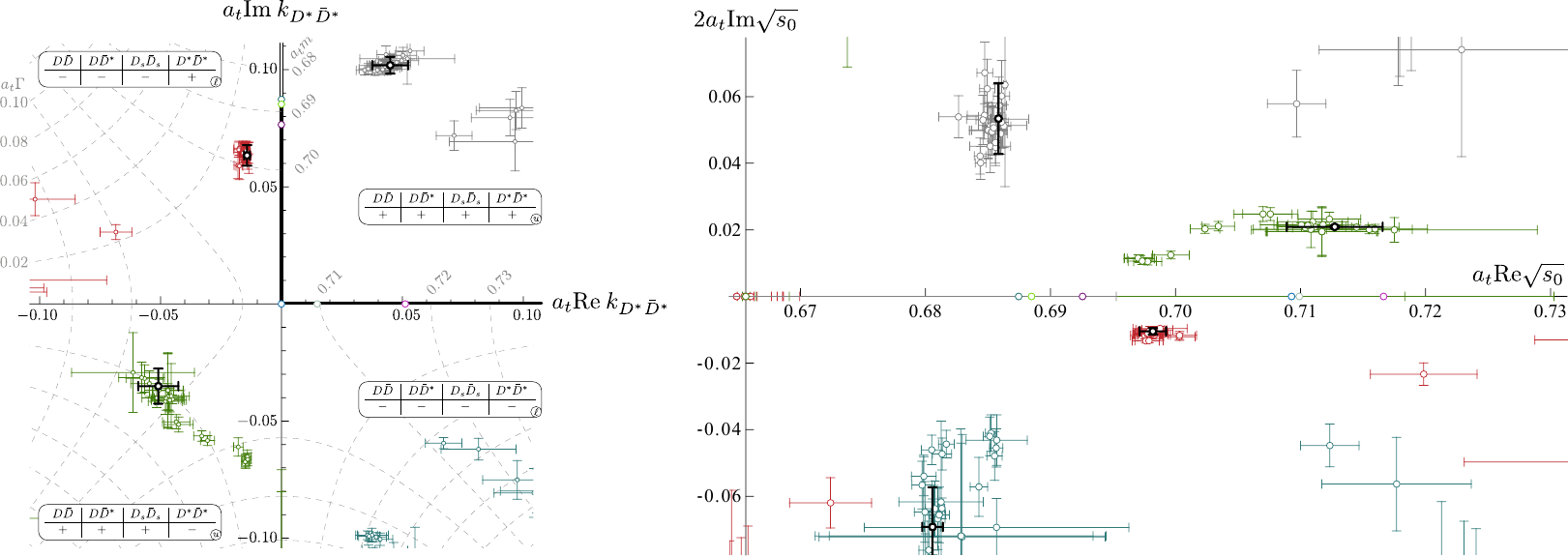}
  \caption{Poles of the $J^{PC}=2^{++}$ amplitudes plotted in the complex $k_{\DstDst}$ and $\sqrt{s}$ planes. The resonance pole on the $(-,-,-,+)$ sheet is shown in red. A second nearby pole observed on the $(+,+,+,-)$ sheet is shown in green. Several other more distant poles are also present as described in the text. In particular, there are virtual bound state poles on several sheets below $\DD$ threshold. For each pole the reference parameterization is highlighted in black.}
 \label{fig:Dwave_extra_pole}
\end{figure*}

Figure~\ref{fig:Dwave_extra_pole} shows the $t$-matrix poles found for a range of parameterizations, where the nearby pole on the proximal sheet (in red) is observed to show very little variation over parameterization. In section~\ref{sec:poles:scalar} we discuss ``mirror'' poles in the context of the scalar amplitudes, and these poles on ``hidden'' sheets are to be expected with a large number of Riemann sheets and several decoupled hadron-hadron channels in each case. Many of these poles can be ignored due to their distance on the Riemann surface from  physical scattering. In Figure~\ref{fig:Dwave_extra_pole} the green and blue points show such ``mirror'' poles, and we observe that they show a greater scatter over parameterization variation than the pole on the proximal sheet.

Figure~\ref{fig:Dwave_extra_pole} also shows, in grey, poles found \emph{on the physical sheet}. Such poles indicate a breakdown of causality, but depending upon how close they are to real scattering energies, the pathology may not be of any practical relevance.
The origin of these poles can be traced back to the presence of the $k_i^{-\ell_i}$ barrier factor in Eq.~\ref{eq:kmat_poles_poly} for $D$-wave channels. Such barriers are necessary to promote the expected behavior of amplitudes at threshold~\footnote{Which matches the behavior of the L\"uscher Zeta functions at threshold.}, but unless some other part of the amplitude suppresses their effect at higher-energies, they can give rise to unwanted energy dependence.

\begin{figure*}
  \includegraphics[width=0.66\textwidth]{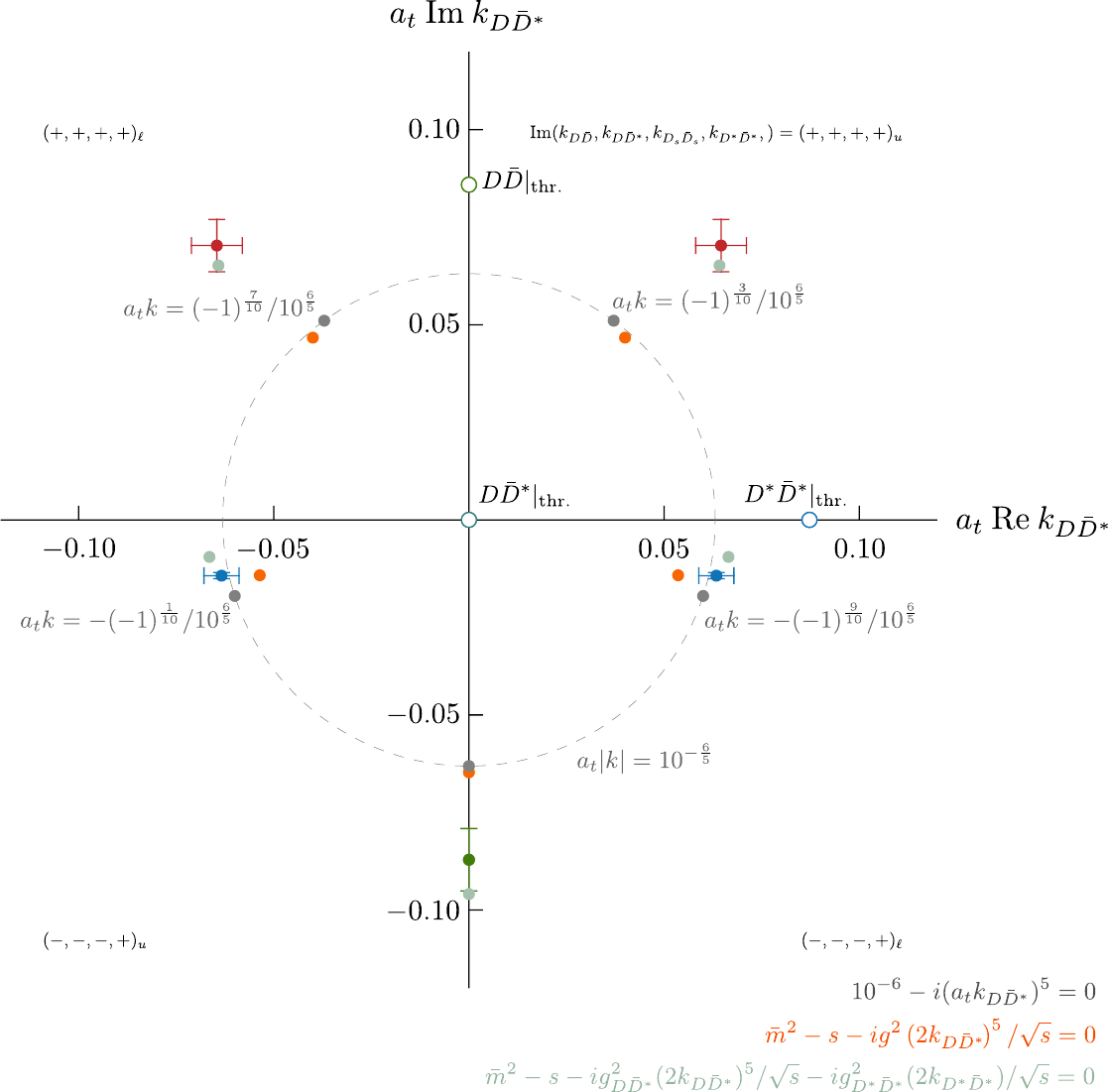}
  \caption{The ``roots of unity''-like phenomena present in the $2^{++}$ amplitudes. The grey circles show the solutions of $10^{-6}-i(a_tk_{\DDst})^5=0$, an arbitrary but simple choice whose zeros approximate the positions of the observed poles, and the large dashed circle shows $|a_tk_{\DDst}|=10^{-6/5}$. The orange circles are the solutions of $m_0^2-s-ig_{D\bar{D}^\ast}^2(2k_{D\bar{D}^\ast})^5/\sqrt{s}=0$. The pale blue-green circles are the solutions of ${m_0^2-s-ig_{D\bar{D}^\ast}^2(2k_{D\bar{D}^\ast})^5/\sqrt{s}-ig_{D^\ast\bar{D}^\ast}^2 (2k_{D^\ast\bar{D}^\ast})/\sqrt{s}=0}$ which closely mimics the observed behavior of the amplitudes determined above. The points with error bars are the relevant poles of the reference parameterization. The open circles on axes are hadron-hadron thresholds.}
  \label{fig:fig_kDDst}
\end{figure*}

The amplitudes presented in Section~\ref{sec:2pP_rest} feature a large contribution to the denominator of the $t$-matrix from the $\DDst$ channel,
\begin{align*}
\sim g^2_{\DDst\SLJc{3}{D}{2}}(2k_{\DDst})^4\rho_{\DDst}\,,
\end{align*}
where values of $g_{\DDst\SLJc{3}{D}{2}}$ are found between $-30\;a_t$ and $-40\;a_t$. The dominance of this term over others in the denominator offers an explanation of the presence of physical sheet poles.
A simple way to see this is by plotting the positions of the poles of the amplitudes in the complex-$k_{\DDst}$ plane, as is done in Fig.~\ref{fig:fig_kDDst}. Using the reference parameterization in Eq.~\ref{eq:fit_2pP_rest+flight}, the position of the complex-conjugate pair (in $s$) of poles due to the resonance are plotted in blue. The physical sheet poles are shown in red, and a virtual bound state pole that also arises is shown in green.

These poles can be compared to those present in a simple toy-amplitude featuring a single elastic $\DDst\SLJc{3}{D}{2}$ amplitude constructed from just a $K$-matrix pole, $K=g^2_{\DDst}/{(m_0^2-s)}$. In this case, the $t$-matrix has a denominator $D=m_0^2-s-ig_{D\bar{D}^\ast}^2(2k_{D\bar{D}^\ast})^5/\sqrt{s}$, and with $g_{\DDst\SLJc{3}{D}{2}}=-40\: a_t$ and $a_tm=0.71$; numerically the final term dominates the behavior. In this case the denominator equals zero at five points shown in orange in Fig.~\ref{fig:fig_kDDst}, which lie very close to the roots of $10^{-6}-i(a_tk_{\DDst})^5$, shown in grey. This ``roots of unity''-like phenomenon is unavoidable with a large coupling and the $D$-wave threshold factor in the denominator.\footnote{It is straightforward to observe similar solutions for $\ell>0$ in a very simple amplitude such as a scattering length approximation, $k^{2\ell+1}\cot\delta_\ell=1/a_\ell$.}

In simple coupled $\DDst\SLJc{3}{D}{2}-\DstDst\SLJc{5}{S}{2}$ systems, an additional term $-ig_{D^\ast\bar{D}^\ast}^2 (2k_{D^\ast\bar{D}^\ast})/\sqrt{s}$ arises in the denominator. Adding this results in a very close agreement with the solutions obtained from the amplitudes determined from lattice QCD energies, as shown by the pale blue-green points in Fig.~\ref{fig:fig_kDDst}. Fig.~\ref{fig:fig_kDDst} shows only sheets where $\mathrm{Im}\,k_\DstDst >0$, but poles are also present on sheets with $\mathrm{Im}\,k_\DstDst<0$. In this highly simplified two-channel system, an additional group of five poles arises that are approximately the complex-conjugates in $k_{\DDst}$ of those in Fig.~\ref{fig:fig_kDDst}.

\vspace{3mm}

\begin{figure*}
  \includegraphics[width=0.99\textwidth]{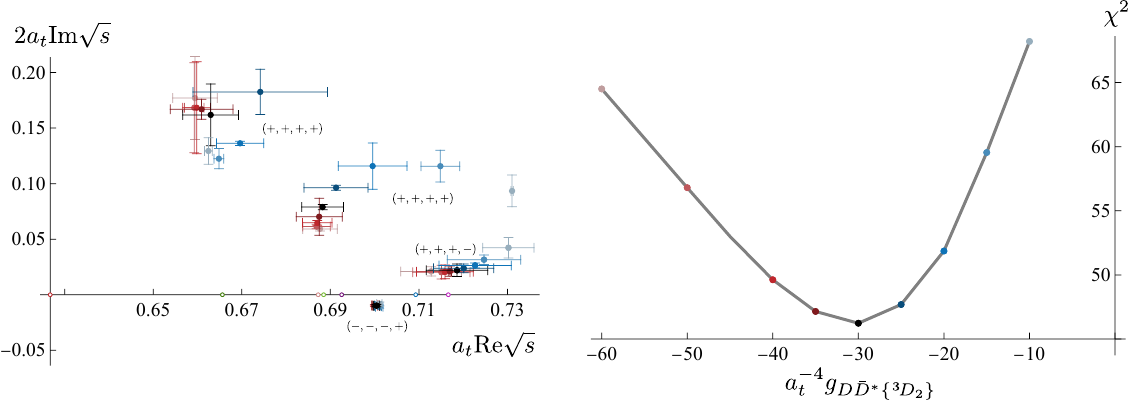}
  \caption{Left: Pole positions in $\sqrt{s}$ from the $2^{++}$ amplitudes as a function of $g_\DDst$ colored according to the values shown in the right figure. Four clusters of poles are shown. The tight cluster below the real axis is the resonance pole on the proximal sheet. The other poles are strongly dependent on the value of $g_\DstDst$. Right: the $\chi^2$ value computed from the $[000] \, E^+$ and $[000] \,T_2^+$ irreps at the same values of $g_\DDst$.}
  \label{fig:pole_motion_gDDst}
\end{figure*}

The sensitivity of the additional poles to the value of $g_{\DDst}$ can be explored. By default we have fixed $g_{\DD}=10\, a_t$, but owing to the coupling-ratio phenomenon, there is very little sensitivity to this choice.   We may now consider in addition fixing $g_{\DDst}$ to a range of values, and redetermine the remaining parameters by $\chi^2$ minimization for each choice. For simplicity we do this using only the energies from the $[000]\,E^{+}$ and $[000]\,T_2^{+}$ irreps. In Fig.~\ref{fig:pole_motion_gDDst}, the result of this procedure is shown. Four clusters of poles are plotted, the resonance pole on the proximal sheet $(-,-,-,+)$ is shown below the real $\sqrt{s}$ axis, and three clusters of spurious poles on the $(+,+,+,\pm)$ sheets are plotted above the axis. The closest of these on the $(+,+,+,+)$ sheet is the equivalent of the physical sheet pole shown in red in Fig.~\ref{fig:fig_kDDst}, and the nearby pole on the $(+,+,+,-)$ sheet is the ``mirror'' obtained by switching to the unphysical $\DstDst$ sheet.
These closest two poles move away from the constrained energy region as $g_\DDst$ is reduced. The resonance pole position on the proximal sheet is very well determined and relatively insensitive to the precise value of $g_\DDst$. The $\chi^2$ is also shown and has a smooth dependence on $g_\DDst$. We see that the energy levels clearly do favor a large value of $g_\DDst$ and the proximity of the physical sheet pole can be associated with this.

Since the narrow resonance we are interested in lies some way above $\DDst$ threshold, we might anticipate that its properties would not be overly sensitive to the barrier behavior at threshold. This can be explored by crudely replacing the $k_{\DDst}^5$ factor with a lower power in the parameterization. This introduces a mismatch with the behavior of the L\"uscher Zeta functions, but in practice this only has a significant effect for energies very close to or more severely \emph{below} threshold. 
By restricting to consideration of the $[000]\, E^+$ and $[000]\, T_2^+$ spectra, which have no energy levels at or below threshold that have significant overlap onto \DDst-like operators, we anticipate that we do not introduce a serious error into the analysis.

\begin{figure*}
  \includegraphics[width=0.99\textwidth]{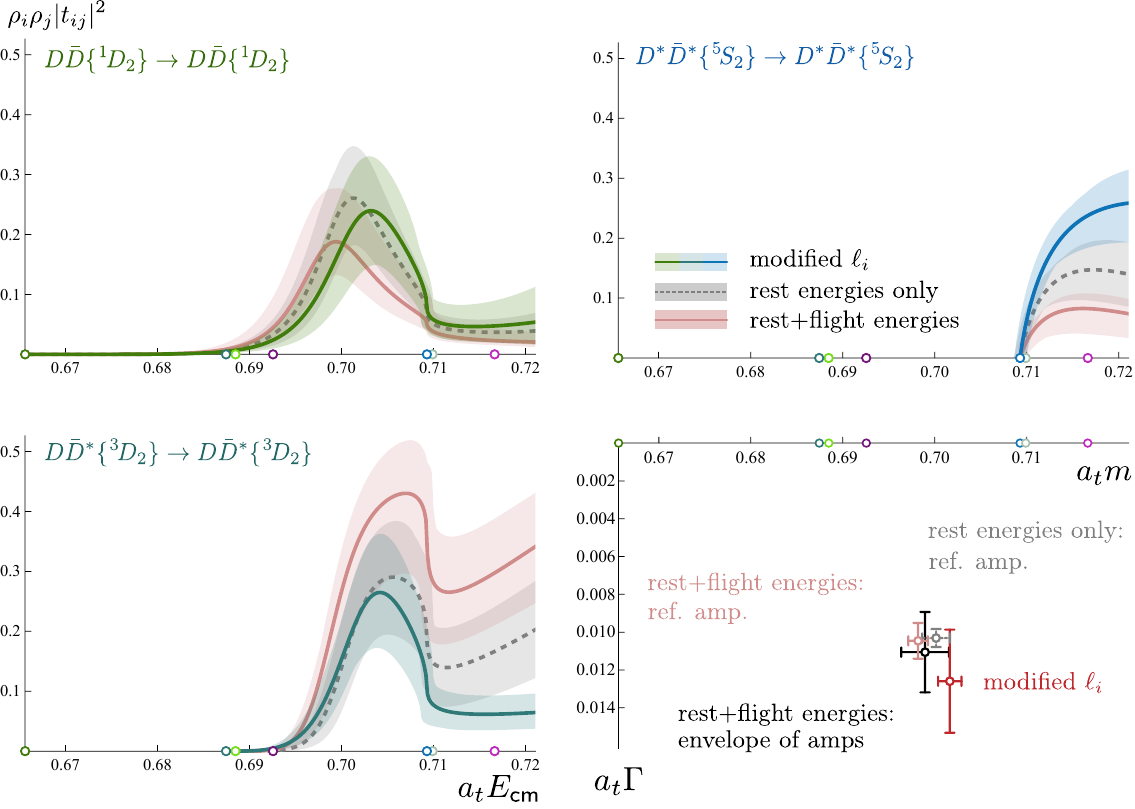}
  \caption{The amplitude and resonance pole position from the amplitude in Eq.~\eqref{eq:fit_2pP_rest_fudged} using the modified threshold factors compared with other amplitudes obtained in this work. In the left and upper panels we show the diagonal $\DD$, $\DDst$ and $\DstDst$ amplitudes compared with the amplitudes given in Eq.~\eqref{eq:fit_2pP_rest} using only rest-frame energies (dotted curves) and Eq.~\eqref{eq:fit_2pP_rest+flight} including also moving frame energies (pink curves). In the lower right panel, a comparison of the resonance pole positions are shown, including the value obtained considering all parameterization variations in table~\ref{tab:J2pP_par_var}.}
  \label{fig:fudged}
\end{figure*}

Using a similar parameterization to the reference parameterization determined from $[000] \, E^{+}$ and $[000] \, T_2^{+}$ energies, as given in Eq.~\ref{eq:fit_2pP_rest}, we artificially modify the $\ell_i$ for $\DDst\SLJc{3}{D}{2}$ terms in Eq.~\ref{eq:kmat_poles_poly} to take value 1 rather than 2. This leads to a term in the denominator with only three powers of momentum rather than five. Describing 47 energy levels results in an amplitude, 
  \begin{center}
  {\small
\begin{tabular}{rll}
  $a_tm = $ & $(0.7037 \pm 0.0011 \pm 0.0001)$ & \multirow{10}{*}{ $\begin{bmatrix*}[r]   1.00 &  -0.17 &  -0.18 &  -0.24 &  -0.01 &  -0.15 &   0.08 &  -0.07 &  -0.16 &   0.01\\
  &  1.00 &  -0.17 &  -0.56 &  -0.00 &   0.11 &   0.27 &  -0.03 &   0.06 &  -0.10\\
  &&  1.00 &   0.29 &  -0.02 &   0.03 &  -0.57 &  -0.06 &  -0.09 &  -0.06\\
  &&&  1.00 &  -0.00 &   0.04 &  -0.32 &   0.07 &   0.10 &  -0.04\\
  &&&&  1.00 &  -0.00 &   0.02 &  -0.02 &   0.01 &   0.00\\
  &&&&&  1.00 &   0.01 &   0.02 &   0.01 &  -0.01\\
  &&&&&&  1.00 &   0.08 &   0.10 &   0.09\\
  &&&&&&&  1.00 &   0.08 &   0.23\\
  &&&&&&&&  1.00 &   0.03\\
  &&&&&&&&&  1.00\end{bmatrix*}$ } \\ 
  ``$g_{\DDst\SLJc{3}{D}{2}}$'' $ = $ & $\mathit{(-4.39 \pm 0.70 \pm 0.17)}$ & \\ 
  $g_{\DsDs\SLJc{1}{D}{2}} = $ & $(-0.32 \pm 3.49 \pm 0.97) \cdot a_t$ & \\
  $g_{\DstDst\SLJc{5}{S}{2}} = $  & $(1.74 \pm 0.22 \pm 0.13) \cdot a_t^{-1}$ & \\
  $g_{\psiom\SLJc{5}{S}{2}} = $  & $(0.00 \pm 0.22 \pm 0.06) \cdot a_t^{-1}$ & \\
  $\gamma_{\eta_c\eta\SLJc{1}{D}{2},\eta_c\eta\SLJc{1}{D}{2}} = $ & $(22.0 \pm 23.9 \pm 7.83) \cdot a_t^4$ & \\
  $\gamma_{\DD\SLJc{1}{D}{2},\DsDs\SLJc{1}{D}{2}} = $ & $(163 \pm 189 \pm 44) \cdot a_t^4$ & \\
  $\gamma_{\psiom\SLJc{5}{S}{2},\psiom\SLJc{5}{S}{2}} = $ & $(-0.88 \pm 0.45 \pm 0.05)$ & \\
  $\gamma_{\psiom\SLJc{3}{D}{2},\psiom\SLJc{3}{D}{2}} = $ & $(561 \pm 513 \pm 132) \cdot a_t^4$ & \\
  $\gamma_{\psi\phi\SLJc{5}{S}{2},\psi\phi\SLJc{5}{S}{2}} = $ & $(1.33 \pm 0.78 \pm 0.04)$ & \\
  $g_{\DD\SLJc{1}{D}{2}}=$ & $10\cdot a_t$ {\bf (fixed)}\\[1.3ex]
  &\multicolumn{2}{l}{ $\chi^2/ N_\mathrm{dof} = \frac{49.1}{47-10} = 1.33$\,.}
\end{tabular}
}
\vspace{-1cm}
\begin{equation}\label{eq:fit_2pP_rest_fudged}\end{equation}
\end{center}
This amplitude and its resonance pole position are shown in Fig.~\ref{fig:fudged}, alongside those of other $2^{++}$ amplitudes given in Eq.~\ref{eq:fit_2pP_rest} and Eq.~\ref{eq:fit_2pP_rest+flight}.

The nearby resonance pole on the proximal sheet appears at $a_t\sqrt{s_0}= (0.7016 \pm 0.0013) - \frac{i}{2}(0.013 \pm 0.003)$, which, as anticipated, is essentially the same location as when the correct $D$-wave barrier behavior was present for \DDst. On the other hand, physical sheet poles are found in completely different locations for this amplitude, with the nearest being at $a_t\sqrt{s_0}= (0.651 \pm 0.010) + \frac{i}{2}(0.169 \pm 0.052)$, which is very far from physical scattering. 

In summary we conclude that the physical sheet poles found in the amplitudes presented in the main text are an artifact of the $D$-wave barrier factors for the \DDst channel, while the narrow resonance pole which is the dominant feature of the $2^{++}$ amplitude is a robust result.

\clearpage
\twocolumngrid
\bibliography{biblio.bib}

\begin{thebibliography}{99}
\expandafter\ifx\csname natexlab\endcsname\relax\def\natexlab#1{#1}\fi
\expandafter\ifx\csname bibnamefont\endcsname\relax
  \def\bibnamefont#1{#1}\fi
\expandafter\ifx\csname bibfnamefont\endcsname\relax
  \def\bibfnamefont#1{#1}\fi
\expandafter\ifx\csname citenamefont\endcsname\relax
  \def\citenamefont#1{#1}\fi
\expandafter\ifx\csname url\endcsname\relax
  \def\url#1{\texttt{#1}}\fi
\expandafter\ifx\csname urlprefix\endcsname\relax\def\urlprefix{URL }\fi
\providecommand{\bibinfo}[2]{#2}
\providecommand{\eprint}[2][]{\url{#2}}

\bibitem[{\citenamefont{Brambilla et~al.}(2020)\citenamefont{Brambilla,
  Eidelman, Hanhart, Nefediev, Shen, Thomas, Vairo, and
  Yuan}}]{Brambilla:2019esw}
\bibinfo{author}{\bibfnamefont{N.}~\bibnamefont{Brambilla}},
  \bibinfo{author}{\bibfnamefont{S.}~\bibnamefont{Eidelman}},
  \bibinfo{author}{\bibfnamefont{C.}~\bibnamefont{Hanhart}},
  \bibinfo{author}{\bibfnamefont{A.}~\bibnamefont{Nefediev}},
  \bibinfo{author}{\bibfnamefont{C.-P.} \bibnamefont{Shen}},
  \bibinfo{author}{\bibfnamefont{C.~E.} \bibnamefont{Thomas}},
  \bibinfo{author}{\bibfnamefont{A.}~\bibnamefont{Vairo}}, \bibnamefont{and}
  \bibinfo{author}{\bibfnamefont{C.-Z.} \bibnamefont{Yuan}},
  \bibinfo{journal}{Phys. Rept.} \textbf{\bibinfo{volume}{873}},
  \bibinfo{pages}{1} (\bibinfo{year}{2020}), \eprint{1907.07583}.

\bibitem[{\citenamefont{Godfrey and Isgur}(1985)}]{Godfrey:1985xj}
\bibinfo{author}{\bibfnamefont{S.}~\bibnamefont{Godfrey}} \bibnamefont{and}
  \bibinfo{author}{\bibfnamefont{N.}~\bibnamefont{Isgur}},
  \bibinfo{journal}{Phys. Rev. D} \textbf{\bibinfo{volume}{32}},
  \bibinfo{pages}{189} (\bibinfo{year}{1985}).

\bibitem[{\citenamefont{Chilikin et~al.}(2017)}]{Belle:2017egg}
\bibinfo{author}{\bibfnamefont{K.}~\bibnamefont{Chilikin}} \bibnamefont{et~al.}
  (\bibinfo{collaboration}{Belle}), \bibinfo{journal}{Phys. Rev. D}
  \textbf{\bibinfo{volume}{95}}, \bibinfo{pages}{112003}
  (\bibinfo{year}{2017}), \eprint{1704.01872}.

\bibitem[{\citenamefont{Uehara et~al.}(2006)}]{Belle:2005rte}
\bibinfo{author}{\bibfnamefont{S.}~\bibnamefont{Uehara}} \bibnamefont{et~al.}
  (\bibinfo{collaboration}{Belle}), \bibinfo{journal}{Phys. Rev. Lett.}
  \textbf{\bibinfo{volume}{96}}, \bibinfo{pages}{082003}
  (\bibinfo{year}{2006}), \eprint{hep-ex/0512035}.

\bibitem[{\citenamefont{Aubert et~al.}(2010)}]{BaBar:2010jfn}
\bibinfo{author}{\bibfnamefont{B.}~\bibnamefont{Aubert}} \bibnamefont{et~al.}
  (\bibinfo{collaboration}{BaBar}), \bibinfo{journal}{Phys. Rev. D}
  \textbf{\bibinfo{volume}{81}}, \bibinfo{pages}{092003}
  (\bibinfo{year}{2010}), \eprint{1002.0281}.

\bibitem[{\citenamefont{Aaij et~al.}(2020)}]{LHCb:2020pxc}
\bibinfo{author}{\bibfnamefont{R.}~\bibnamefont{Aaij}} \bibnamefont{et~al.}
  (\bibinfo{collaboration}{LHCb}), \bibinfo{journal}{Phys. Rev. D}
  \textbf{\bibinfo{volume}{102}}, \bibinfo{pages}{112003}
  (\bibinfo{year}{2020}), \eprint{2009.00026}.

\bibitem[{\citenamefont{Aaij et~al.}(2019)}]{LHCb:2019lnr}
\bibinfo{author}{\bibfnamefont{R.}~\bibnamefont{Aaij}} \bibnamefont{et~al.}
  (\bibinfo{collaboration}{LHCb}), \bibinfo{journal}{JHEP}
  \textbf{\bibinfo{volume}{07}}, \bibinfo{pages}{035} (\bibinfo{year}{2019}),
  \eprint{1903.12240}.

\bibitem[{\citenamefont{Wang et~al.}(2021{\natexlab{a}})\citenamefont{Wang, Li,
  Liang, and Oset}}]{Wang:2020elp}
\bibinfo{author}{\bibfnamefont{E.}~\bibnamefont{Wang}},
  \bibinfo{author}{\bibfnamefont{H.-S.} \bibnamefont{Li}},
  \bibinfo{author}{\bibfnamefont{W.-H.} \bibnamefont{Liang}}, \bibnamefont{and}
  \bibinfo{author}{\bibfnamefont{E.}~\bibnamefont{Oset}},
  \bibinfo{journal}{Phys. Rev. D} \textbf{\bibinfo{volume}{103}},
  \bibinfo{pages}{054008} (\bibinfo{year}{2021}{\natexlab{a}}),
  \eprint{2010.15431}.

\bibitem[{\citenamefont{Deineka et~al.}(2022)\citenamefont{Deineka, Danilkin,
  and Vanderhaeghen}}]{Deineka:2021aeu}
\bibinfo{author}{\bibfnamefont{O.}~\bibnamefont{Deineka}},
  \bibinfo{author}{\bibfnamefont{I.}~\bibnamefont{Danilkin}}, \bibnamefont{and}
  \bibinfo{author}{\bibfnamefont{M.}~\bibnamefont{Vanderhaeghen}},
  \bibinfo{journal}{Phys. Lett. B} \textbf{\bibinfo{volume}{827}},
  \bibinfo{pages}{136982} (\bibinfo{year}{2022}), \eprint{2111.15033}.

\bibitem[{\citenamefont{Aaij et~al.}(2023)}]{LHCb:2022aki}
\bibinfo{author}{\bibfnamefont{R.}~\bibnamefont{Aaij}} \bibnamefont{et~al.}
  (\bibinfo{collaboration}{LHCb}), \bibinfo{journal}{Phys. Rev. Lett.}
  \textbf{\bibinfo{volume}{131}}, \bibinfo{pages}{071901}
  (\bibinfo{year}{2023}), \eprint{2210.15153}.

\bibitem[{\citenamefont{Guo et~al.}(2022)\citenamefont{Guo, Wang, Chen, and
  Liu}}]{Guo:2022zbc}
\bibinfo{author}{\bibfnamefont{D.}~\bibnamefont{Guo}},
  \bibinfo{author}{\bibfnamefont{J.-Z.} \bibnamefont{Wang}},
  \bibinfo{author}{\bibfnamefont{D.-Y.} \bibnamefont{Chen}}, \bibnamefont{and}
  \bibinfo{author}{\bibfnamefont{X.}~\bibnamefont{Liu}},
  \bibinfo{journal}{Phys. Rev. D} \textbf{\bibinfo{volume}{106}},
  \bibinfo{pages}{094037} (\bibinfo{year}{2022}), \eprint{2210.16720}.

\bibitem[{\citenamefont{Ji et~al.}(2022)\citenamefont{Ji, Dong, Albaladejo, Du,
  Guo, Nieves, and Zou}}]{Ji:2022vdj}
\bibinfo{author}{\bibfnamefont{T.}~\bibnamefont{Ji}},
  \bibinfo{author}{\bibfnamefont{X.-K.} \bibnamefont{Dong}},
  \bibinfo{author}{\bibfnamefont{M.}~\bibnamefont{Albaladejo}},
  \bibinfo{author}{\bibfnamefont{M.-L.} \bibnamefont{Du}},
  \bibinfo{author}{\bibfnamefont{F.-K.} \bibnamefont{Guo}},
  \bibinfo{author}{\bibfnamefont{J.}~\bibnamefont{Nieves}}, \bibnamefont{and}
  \bibinfo{author}{\bibfnamefont{B.-S.} \bibnamefont{Zou}}
  (\bibinfo{year}{2022}), \eprint{2212.00631}.

\bibitem[{\citenamefont{Uehara et~al.}(2010)}]{Belle:2009and}
\bibinfo{author}{\bibfnamefont{S.}~\bibnamefont{Uehara}} \bibnamefont{et~al.}
  (\bibinfo{collaboration}{Belle}), \bibinfo{journal}{Phys. Rev. Lett.}
  \textbf{\bibinfo{volume}{104}}, \bibinfo{pages}{092001}
  (\bibinfo{year}{2010}), \eprint{0912.4451}.

\bibitem[{\citenamefont{De~Rujula et~al.}(1977)\citenamefont{De~Rujula, Georgi,
  and Glashow}}]{DeRujula:1976zlg}
\bibinfo{author}{\bibfnamefont{A.}~\bibnamefont{De~Rujula}},
  \bibinfo{author}{\bibfnamefont{H.}~\bibnamefont{Georgi}}, \bibnamefont{and}
  \bibinfo{author}{\bibfnamefont{S.~L.} \bibnamefont{Glashow}},
  \bibinfo{journal}{Phys. Rev. Lett.} \textbf{\bibinfo{volume}{38}},
  \bibinfo{pages}{317} (\bibinfo{year}{1977}).

\bibitem[{\citenamefont{Tornqvist}(1994)}]{Tornqvist:1993ng}
\bibinfo{author}{\bibfnamefont{N.~A.} \bibnamefont{Tornqvist}},
  \bibinfo{journal}{Z. Phys. C} \textbf{\bibinfo{volume}{61}},
  \bibinfo{pages}{525} (\bibinfo{year}{1994}), \eprint{hep-ph/9310247}.

\bibitem[{\citenamefont{Esposito et~al.}(2017)\citenamefont{Esposito, Pilloni,
  and Polosa}}]{Esposito:2016noz}
\bibinfo{author}{\bibfnamefont{A.}~\bibnamefont{Esposito}},
  \bibinfo{author}{\bibfnamefont{A.}~\bibnamefont{Pilloni}}, \bibnamefont{and}
  \bibinfo{author}{\bibfnamefont{A.~D.} \bibnamefont{Polosa}},
  \bibinfo{journal}{Phys. Rept.} \textbf{\bibinfo{volume}{668}},
  \bibinfo{pages}{1} (\bibinfo{year}{2017}), \eprint{1611.07920}.

\bibitem[{\citenamefont{Guo et~al.}(2018)\citenamefont{Guo, Hanhart,
  Mei\ss{}ner, Wang, Zhao, and Zou}}]{Guo:2017jvc}
\bibinfo{author}{\bibfnamefont{F.-K.} \bibnamefont{Guo}},
  \bibinfo{author}{\bibfnamefont{C.}~\bibnamefont{Hanhart}},
  \bibinfo{author}{\bibfnamefont{U.-G.} \bibnamefont{Mei\ss{}ner}},
  \bibinfo{author}{\bibfnamefont{Q.}~\bibnamefont{Wang}},
  \bibinfo{author}{\bibfnamefont{Q.}~\bibnamefont{Zhao}}, \bibnamefont{and}
  \bibinfo{author}{\bibfnamefont{B.-S.} \bibnamefont{Zou}},
  \bibinfo{journal}{Rev. Mod. Phys.} \textbf{\bibinfo{volume}{90}},
  \bibinfo{pages}{015004} (\bibinfo{year}{2018}), \bibinfo{note}{[Erratum:
  Rev.Mod.Phys. 94, 029901 (2022)]}, \eprint{1705.00141}.

\bibitem[{\citenamefont{Albaladejo et~al.}(2022)}]{JPAC:2021rxu}
\bibinfo{author}{\bibfnamefont{M.}~\bibnamefont{Albaladejo}}
  \bibnamefont{et~al.} (\bibinfo{collaboration}{JPAC}), \bibinfo{journal}{Prog.
  Part. Nucl. Phys.} \textbf{\bibinfo{volume}{127}}, \bibinfo{pages}{103981}
  (\bibinfo{year}{2022}), \eprint{2112.13436}.

\bibitem[{\citenamefont{Bicudo}(2022)}]{Bicudo:2022cqi}
\bibinfo{author}{\bibfnamefont{P.}~\bibnamefont{Bicudo}}
  (\bibinfo{year}{2022}), \eprint{2212.07793}.

\bibitem[{\citenamefont{Briceno
  et~al.}(2018{\natexlab{a}})\citenamefont{Briceno, Dudek, and
  Young}}]{Briceno:2017max}
\bibinfo{author}{\bibfnamefont{R.~A.} \bibnamefont{Briceno}},
  \bibinfo{author}{\bibfnamefont{J.~J.} \bibnamefont{Dudek}}, \bibnamefont{and}
  \bibinfo{author}{\bibfnamefont{R.~D.} \bibnamefont{Young}},
  \bibinfo{journal}{Rev. Mod. Phys.} \textbf{\bibinfo{volume}{90}},
  \bibinfo{pages}{025001} (\bibinfo{year}{2018}{\natexlab{a}}),
  \eprint{1706.06223}.

\bibitem[{\citenamefont{Dudek et~al.}(2014)\citenamefont{Dudek, Edwards,
  Thomas, and Wilson}}]{Dudek:2014qha}
\bibinfo{author}{\bibfnamefont{J.~J.} \bibnamefont{Dudek}},
  \bibinfo{author}{\bibfnamefont{R.~G.} \bibnamefont{Edwards}},
  \bibinfo{author}{\bibfnamefont{C.~E.} \bibnamefont{Thomas}},
  \bibnamefont{and} \bibinfo{author}{\bibfnamefont{D.~J.} \bibnamefont{Wilson}}
  (\bibinfo{collaboration}{Hadron Spectrum}), \bibinfo{journal}{Phys. Rev.
  Lett.} \textbf{\bibinfo{volume}{113}}, \bibinfo{pages}{182001}
  (\bibinfo{year}{2014}), \eprint{1406.4158}.

\bibitem[{\citenamefont{Wilson et~al.}(2015{\natexlab{a}})\citenamefont{Wilson,
  Dudek, Edwards, and Thomas}}]{Wilson:2014cna}
\bibinfo{author}{\bibfnamefont{D.~J.} \bibnamefont{Wilson}},
  \bibinfo{author}{\bibfnamefont{J.~J.} \bibnamefont{Dudek}},
  \bibinfo{author}{\bibfnamefont{R.~G.} \bibnamefont{Edwards}},
  \bibnamefont{and} \bibinfo{author}{\bibfnamefont{C.~E.}
  \bibnamefont{Thomas}}, \bibinfo{journal}{Phys. Rev.}
  \textbf{\bibinfo{volume}{D91}}, \bibinfo{pages}{054008}
  (\bibinfo{year}{2015}{\natexlab{a}}), \eprint{1411.2004}.

\bibitem[{\citenamefont{Briceno}(2014)}]{Briceno:2014oea}
\bibinfo{author}{\bibfnamefont{R.~A.} \bibnamefont{Briceno}},
  \bibinfo{journal}{Phys. Rev.} \textbf{\bibinfo{volume}{D89}},
  \bibinfo{pages}{074507} (\bibinfo{year}{2014}), \eprint{1401.3312}.

\bibitem[{\citenamefont{Woss et~al.}(2018)\citenamefont{Woss, Thomas, Dudek,
  Edwards, and Wilson}}]{Woss:2018irj}
\bibinfo{author}{\bibfnamefont{A.}~\bibnamefont{Woss}},
  \bibinfo{author}{\bibfnamefont{C.~E.} \bibnamefont{Thomas}},
  \bibinfo{author}{\bibfnamefont{J.~J.} \bibnamefont{Dudek}},
  \bibinfo{author}{\bibfnamefont{R.~G.} \bibnamefont{Edwards}},
  \bibnamefont{and} \bibinfo{author}{\bibfnamefont{D.~J.}
  \bibnamefont{Wilson}}, \bibinfo{journal}{JHEP} \textbf{\bibinfo{volume}{07}},
  \bibinfo{pages}{043} (\bibinfo{year}{2018}), \eprint{1802.05580}.

\bibitem[{\citenamefont{Woss et~al.}(2019)\citenamefont{Woss, Thomas, Dudek,
  Edwards, and Wilson}}]{Woss:2019hse}
\bibinfo{author}{\bibfnamefont{A.~J.} \bibnamefont{Woss}},
  \bibinfo{author}{\bibfnamefont{C.~E.} \bibnamefont{Thomas}},
  \bibinfo{author}{\bibfnamefont{J.~J.} \bibnamefont{Dudek}},
  \bibinfo{author}{\bibfnamefont{R.~G.} \bibnamefont{Edwards}},
  \bibnamefont{and} \bibinfo{author}{\bibfnamefont{D.~J.}
  \bibnamefont{Wilson}}, \bibinfo{journal}{Phys. Rev. D}
  \textbf{\bibinfo{volume}{100}}, \bibinfo{pages}{054506}
  (\bibinfo{year}{2019}), \eprint{1904.04136}.

\bibitem[{\citenamefont{Woss et~al.}(2021)\citenamefont{Woss, Dudek, Edwards,
  Thomas, and Wilson}}]{Woss:2020ayi}
\bibinfo{author}{\bibfnamefont{A.~J.} \bibnamefont{Woss}},
  \bibinfo{author}{\bibfnamefont{J.~J.} \bibnamefont{Dudek}},
  \bibinfo{author}{\bibfnamefont{R.~G.} \bibnamefont{Edwards}},
  \bibinfo{author}{\bibfnamefont{C.~E.} \bibnamefont{Thomas}},
  \bibnamefont{and} \bibinfo{author}{\bibfnamefont{D.~J.} \bibnamefont{Wilson}}
  (\bibinfo{collaboration}{Hadron Spectrum}), \bibinfo{journal}{Phys. Rev. D}
  \textbf{\bibinfo{volume}{103}}, \bibinfo{pages}{054502}
  (\bibinfo{year}{2021}), \eprint{2009.10034}.

\bibitem[{\citenamefont{Lang et~al.}(2015)\citenamefont{Lang, Leskovec, Mohler,
  and Prelovsek}}]{Lang:2015sba}
\bibinfo{author}{\bibfnamefont{C.~B.} \bibnamefont{Lang}},
  \bibinfo{author}{\bibfnamefont{L.}~\bibnamefont{Leskovec}},
  \bibinfo{author}{\bibfnamefont{D.}~\bibnamefont{Mohler}}, \bibnamefont{and}
  \bibinfo{author}{\bibfnamefont{S.}~\bibnamefont{Prelovsek}},
  \bibinfo{journal}{JHEP} \textbf{\bibinfo{volume}{09}}, \bibinfo{pages}{089}
  (\bibinfo{year}{2015}), \eprint{1503.05363}.

\bibitem[{\citenamefont{Prelovsek et~al.}(2021)\citenamefont{Prelovsek,
  Collins, Mohler, Padmanath, and Piemonte}}]{Prelovsek:2020eiw}
\bibinfo{author}{\bibfnamefont{S.}~\bibnamefont{Prelovsek}},
  \bibinfo{author}{\bibfnamefont{S.}~\bibnamefont{Collins}},
  \bibinfo{author}{\bibfnamefont{D.}~\bibnamefont{Mohler}},
  \bibinfo{author}{\bibfnamefont{M.}~\bibnamefont{Padmanath}},
  \bibnamefont{and} \bibinfo{author}{\bibfnamefont{S.}~\bibnamefont{Piemonte}},
  \bibinfo{journal}{JHEP} \textbf{\bibinfo{volume}{06}}, \bibinfo{pages}{035}
  (\bibinfo{year}{2021}), \eprint{2011.02542}.

\bibitem[{\citenamefont{Wilson~et al}(2023)}]{Wilson:2023_short}
\bibinfo{author}{\bibfnamefont{D.~J.} \bibnamefont{Wilson~et al}}
  (\bibinfo{year}{2023}).

\bibitem[{\citenamefont{Liu et~al.}(2012)\citenamefont{Liu, Moir, Peardon,
  Ryan, Thomas, Vilaseca, Dudek, Edwards, Joo, and Richards}}]{Liu:2012ze}
\bibinfo{author}{\bibfnamefont{L.}~\bibnamefont{Liu}},
  \bibinfo{author}{\bibfnamefont{G.}~\bibnamefont{Moir}},
  \bibinfo{author}{\bibfnamefont{M.}~\bibnamefont{Peardon}},
  \bibinfo{author}{\bibfnamefont{S.~M.} \bibnamefont{Ryan}},
  \bibinfo{author}{\bibfnamefont{C.~E.} \bibnamefont{Thomas}},
  \bibinfo{author}{\bibfnamefont{P.}~\bibnamefont{Vilaseca}},
  \bibinfo{author}{\bibfnamefont{J.~J.} \bibnamefont{Dudek}},
  \bibinfo{author}{\bibfnamefont{R.~G.} \bibnamefont{Edwards}},
  \bibinfo{author}{\bibfnamefont{B.}~\bibnamefont{Joo}}, \bibnamefont{and}
  \bibinfo{author}{\bibfnamefont{D.~G.} \bibnamefont{Richards}}
  (\bibinfo{collaboration}{Hadron Spectrum}), \bibinfo{journal}{JHEP}
  \textbf{\bibinfo{volume}{07}}, \bibinfo{pages}{126} (\bibinfo{year}{2012}),
  \eprint{1204.5425}.

\bibitem[{\citenamefont{Edwards et~al.}(2008)\citenamefont{Edwards, Joo, and
  Lin}}]{Edwards:2008ja}
\bibinfo{author}{\bibfnamefont{R.~G.} \bibnamefont{Edwards}},
  \bibinfo{author}{\bibfnamefont{B.}~\bibnamefont{Joo}}, \bibnamefont{and}
  \bibinfo{author}{\bibfnamefont{H.-W.} \bibnamefont{Lin}},
  \bibinfo{journal}{Phys. Rev.} \textbf{\bibinfo{volume}{D78}},
  \bibinfo{pages}{054501} (\bibinfo{year}{2008}), \eprint{0803.3960}.

\bibitem[{\citenamefont{Lin et~al.}(2009)}]{Lin:2008pr}
\bibinfo{author}{\bibfnamefont{H.-W.} \bibnamefont{Lin}} \bibnamefont{et~al.}
  (\bibinfo{collaboration}{Hadron Spectrum}), \bibinfo{journal}{Phys. Rev.}
  \textbf{\bibinfo{volume}{D79}}, \bibinfo{pages}{034502}
  (\bibinfo{year}{2009}), \eprint{0810.3588}.

\bibitem[{\citenamefont{Peardon et~al.}(2009)\citenamefont{Peardon, Bulava,
  Foley, Morningstar, Dudek, Edwards, Joo, Lin, Richards, and
  Juge}}]{Peardon:2009gh}
\bibinfo{author}{\bibfnamefont{M.}~\bibnamefont{Peardon}},
  \bibinfo{author}{\bibfnamefont{J.}~\bibnamefont{Bulava}},
  \bibinfo{author}{\bibfnamefont{J.}~\bibnamefont{Foley}},
  \bibinfo{author}{\bibfnamefont{C.}~\bibnamefont{Morningstar}},
  \bibinfo{author}{\bibfnamefont{J.}~\bibnamefont{Dudek}},
  \bibinfo{author}{\bibfnamefont{R.~G.} \bibnamefont{Edwards}},
  \bibinfo{author}{\bibfnamefont{B.}~\bibnamefont{Joo}},
  \bibinfo{author}{\bibfnamefont{H.-W.} \bibnamefont{Lin}},
  \bibinfo{author}{\bibfnamefont{D.~G.} \bibnamefont{Richards}},
  \bibnamefont{and} \bibinfo{author}{\bibfnamefont{K.~J.} \bibnamefont{Juge}}
  (\bibinfo{collaboration}{Hadron Spectrum}), \bibinfo{journal}{Phys. Rev.}
  \textbf{\bibinfo{volume}{D80}}, \bibinfo{pages}{054506}
  (\bibinfo{year}{2009}), \eprint{0905.2160}.

\bibitem[{\citenamefont{Edwards et~al.}(2013)\citenamefont{Edwards, Mathur,
  Richards, and Wallace}}]{Edwards:2012fx}
\bibinfo{author}{\bibfnamefont{R.~G.} \bibnamefont{Edwards}},
  \bibinfo{author}{\bibfnamefont{N.}~\bibnamefont{Mathur}},
  \bibinfo{author}{\bibfnamefont{D.~G.} \bibnamefont{Richards}},
  \bibnamefont{and} \bibinfo{author}{\bibfnamefont{S.~J.}
  \bibnamefont{Wallace}} (\bibinfo{collaboration}{Hadron Spectrum}),
  \bibinfo{journal}{Phys. Rev. D} \textbf{\bibinfo{volume}{87}},
  \bibinfo{pages}{054506} (\bibinfo{year}{2013}), \eprint{1212.5236}.

\bibitem[{\citenamefont{Dudek et~al.}(2012)\citenamefont{Dudek, Edwards, and
  Thomas}}]{Dudek:2012gj}
\bibinfo{author}{\bibfnamefont{J.~J.} \bibnamefont{Dudek}},
  \bibinfo{author}{\bibfnamefont{R.~G.} \bibnamefont{Edwards}},
  \bibnamefont{and} \bibinfo{author}{\bibfnamefont{C.~E.}
  \bibnamefont{Thomas}}, \bibinfo{journal}{Phys. Rev.}
  \textbf{\bibinfo{volume}{D86}}, \bibinfo{pages}{034031}
  (\bibinfo{year}{2012}), \eprint{1203.6041}.

\bibitem[{\citenamefont{Michael}(1985)}]{Michael:1985ne}
\bibinfo{author}{\bibfnamefont{C.}~\bibnamefont{Michael}},
  \bibinfo{journal}{Nucl. Phys.} \textbf{\bibinfo{volume}{B259}},
  \bibinfo{pages}{58} (\bibinfo{year}{1985}).

\bibitem[{\citenamefont{Luscher and Wolff}(1990)}]{Luscher:1990ck}
\bibinfo{author}{\bibfnamefont{M.}~\bibnamefont{Luscher}} \bibnamefont{and}
  \bibinfo{author}{\bibfnamefont{U.}~\bibnamefont{Wolff}},
  \bibinfo{journal}{Nucl. Phys. B} \textbf{\bibinfo{volume}{339}},
  \bibinfo{pages}{222} (\bibinfo{year}{1990}).

\bibitem[{\citenamefont{Blossier et~al.}(2009)\citenamefont{Blossier,
  Della~Morte, von Hippel, Mendes, and Sommer}}]{Blossier:2009kd}
\bibinfo{author}{\bibfnamefont{B.}~\bibnamefont{Blossier}},
  \bibinfo{author}{\bibfnamefont{M.}~\bibnamefont{Della~Morte}},
  \bibinfo{author}{\bibfnamefont{G.}~\bibnamefont{von Hippel}},
  \bibinfo{author}{\bibfnamefont{T.}~\bibnamefont{Mendes}}, \bibnamefont{and}
  \bibinfo{author}{\bibfnamefont{R.}~\bibnamefont{Sommer}},
  \bibinfo{journal}{JHEP} \textbf{\bibinfo{volume}{04}}, \bibinfo{pages}{094}
  (\bibinfo{year}{2009}), \eprint{0902.1265}.

\bibitem[{\citenamefont{Dudek et~al.}(2010)\citenamefont{Dudek, Edwards,
  Peardon, Richards, and Thomas}}]{Dudek:2010wm}
\bibinfo{author}{\bibfnamefont{J.~J.} \bibnamefont{Dudek}},
  \bibinfo{author}{\bibfnamefont{R.~G.} \bibnamefont{Edwards}},
  \bibinfo{author}{\bibfnamefont{M.~J.} \bibnamefont{Peardon}},
  \bibinfo{author}{\bibfnamefont{D.~G.} \bibnamefont{Richards}},
  \bibnamefont{and} \bibinfo{author}{\bibfnamefont{C.~E.}
  \bibnamefont{Thomas}}, \bibinfo{journal}{Phys. Rev.}
  \textbf{\bibinfo{volume}{D82}}, \bibinfo{pages}{034508}
  (\bibinfo{year}{2010}), \eprint{1004.4930}.

\bibitem[{\citenamefont{Dudek et~al.}(2009)\citenamefont{Dudek, Edwards,
  Peardon, Richards, and Thomas}}]{Dudek:2009qf}
\bibinfo{author}{\bibfnamefont{J.~J.} \bibnamefont{Dudek}},
  \bibinfo{author}{\bibfnamefont{R.~G.} \bibnamefont{Edwards}},
  \bibinfo{author}{\bibfnamefont{M.~J.} \bibnamefont{Peardon}},
  \bibinfo{author}{\bibfnamefont{D.~G.} \bibnamefont{Richards}},
  \bibnamefont{and} \bibinfo{author}{\bibfnamefont{C.~E.}
  \bibnamefont{Thomas}}, \bibinfo{journal}{Phys. Rev. Lett.}
  \textbf{\bibinfo{volume}{103}}, \bibinfo{pages}{262001}
  (\bibinfo{year}{2009}), \eprint{0909.0200}.

\bibitem[{\citenamefont{Thomas et~al.}(2012)\citenamefont{Thomas, Edwards, and
  Dudek}}]{Thomas:2011rh}
\bibinfo{author}{\bibfnamefont{C.~E.} \bibnamefont{Thomas}},
  \bibinfo{author}{\bibfnamefont{R.~G.} \bibnamefont{Edwards}},
  \bibnamefont{and} \bibinfo{author}{\bibfnamefont{J.~J.} \bibnamefont{Dudek}},
  \bibinfo{journal}{Phys. Rev.} \textbf{\bibinfo{volume}{D85}},
  \bibinfo{pages}{014507} (\bibinfo{year}{2012}), \eprint{1107.1930}.

\bibitem[{\citenamefont{Briceno
  et~al.}(2018{\natexlab{b}})\citenamefont{Briceno, Dudek, Edwards, and
  Wilson}}]{Briceno:2017qmb}
\bibinfo{author}{\bibfnamefont{R.~A.} \bibnamefont{Briceno}},
  \bibinfo{author}{\bibfnamefont{J.~J.} \bibnamefont{Dudek}},
  \bibinfo{author}{\bibfnamefont{R.~G.} \bibnamefont{Edwards}},
  \bibnamefont{and} \bibinfo{author}{\bibfnamefont{D.~J.}
  \bibnamefont{Wilson}}, \bibinfo{journal}{Phys. Rev.}
  \textbf{\bibinfo{volume}{D97}}, \bibinfo{pages}{054513}
  (\bibinfo{year}{2018}{\natexlab{b}}), \eprint{1708.06667}.

\bibitem[{\citenamefont{Hansen et~al.}(2021)\citenamefont{Hansen, Brice\~no,
  Edwards, Thomas, and Wilson}}]{Hansen:2020otl}
\bibinfo{author}{\bibfnamefont{M.~T.} \bibnamefont{Hansen}},
  \bibinfo{author}{\bibfnamefont{R.~A.} \bibnamefont{Brice\~no}},
  \bibinfo{author}{\bibfnamefont{R.~G.} \bibnamefont{Edwards}},
  \bibinfo{author}{\bibfnamefont{C.~E.} \bibnamefont{Thomas}},
  \bibnamefont{and} \bibinfo{author}{\bibfnamefont{D.~J.} \bibnamefont{Wilson}}
  (\bibinfo{collaboration}{Hadron Spectrum}), \bibinfo{journal}{Phys. Rev.
  Lett.} \textbf{\bibinfo{volume}{126}}, \bibinfo{pages}{012001}
  (\bibinfo{year}{2021}), \eprint{2009.04931}.

\bibitem[{\citenamefont{Moir et~al.}(2016)\citenamefont{Moir, Peardon, Ryan,
  Thomas, and Wilson}}]{Moir:2016srx}
\bibinfo{author}{\bibfnamefont{G.}~\bibnamefont{Moir}},
  \bibinfo{author}{\bibfnamefont{M.}~\bibnamefont{Peardon}},
  \bibinfo{author}{\bibfnamefont{S.~M.} \bibnamefont{Ryan}},
  \bibinfo{author}{\bibfnamefont{C.~E.} \bibnamefont{Thomas}},
  \bibnamefont{and} \bibinfo{author}{\bibfnamefont{D.~J.}
  \bibnamefont{Wilson}}, \bibinfo{journal}{JHEP} \textbf{\bibinfo{volume}{10}},
  \bibinfo{pages}{011} (\bibinfo{year}{2016}), \eprint{1607.07093}.

\bibitem[{\citenamefont{Cheung et~al.}(2021)\citenamefont{Cheung, Thomas,
  Wilson, Moir, Peardon, and Ryan}}]{Cheung:2020mql}
\bibinfo{author}{\bibfnamefont{G.~K.~C.} \bibnamefont{Cheung}},
  \bibinfo{author}{\bibfnamefont{C.~E.} \bibnamefont{Thomas}},
  \bibinfo{author}{\bibfnamefont{D.~J.} \bibnamefont{Wilson}},
  \bibinfo{author}{\bibfnamefont{G.}~\bibnamefont{Moir}},
  \bibinfo{author}{\bibfnamefont{M.}~\bibnamefont{Peardon}}, \bibnamefont{and}
  \bibinfo{author}{\bibfnamefont{S.~M.} \bibnamefont{Ryan}}
  (\bibinfo{collaboration}{Hadron Spectrum}), \bibinfo{journal}{JHEP}
  \textbf{\bibinfo{volume}{02}}, \bibinfo{pages}{100} (\bibinfo{year}{2021}),
  \eprint{2008.06432}.

\bibitem[{\citenamefont{Lang and Wilson}(2022)}]{Lang:2022elg}
\bibinfo{author}{\bibfnamefont{N.}~\bibnamefont{Lang}} \bibnamefont{and}
  \bibinfo{author}{\bibfnamefont{D.~J.} \bibnamefont{Wilson}}
  (\bibinfo{collaboration}{Hadron Spectrum}) (\bibinfo{year}{2022}),
  \eprint{2205.05026}.

\bibitem[{\citenamefont{Briceno et~al.}(2017)\citenamefont{Briceno, Dudek,
  Edwards, and Wilson}}]{Briceno:2016mjc}
\bibinfo{author}{\bibfnamefont{R.~A.} \bibnamefont{Briceno}},
  \bibinfo{author}{\bibfnamefont{J.~J.} \bibnamefont{Dudek}},
  \bibinfo{author}{\bibfnamefont{R.~G.} \bibnamefont{Edwards}},
  \bibnamefont{and} \bibinfo{author}{\bibfnamefont{D.~J.}
  \bibnamefont{Wilson}}, \bibinfo{journal}{Phys. Rev. Lett.}
  \textbf{\bibinfo{volume}{118}}, \bibinfo{pages}{022002}
  (\bibinfo{year}{2017}), \eprint{1607.05900}.

\bibitem[{\citenamefont{Piemonte et~al.}(2019)\citenamefont{Piemonte, Collins,
  Mohler, Padmanath, and Prelovsek}}]{Piemonte:2019cbi}
\bibinfo{author}{\bibfnamefont{S.}~\bibnamefont{Piemonte}},
  \bibinfo{author}{\bibfnamefont{S.}~\bibnamefont{Collins}},
  \bibinfo{author}{\bibfnamefont{D.}~\bibnamefont{Mohler}},
  \bibinfo{author}{\bibfnamefont{M.}~\bibnamefont{Padmanath}},
  \bibnamefont{and}
  \bibinfo{author}{\bibfnamefont{S.}~\bibnamefont{Prelovsek}},
  \bibinfo{journal}{Phys. Rev. D} \textbf{\bibinfo{volume}{100}},
  \bibinfo{pages}{074505} (\bibinfo{year}{2019}), \eprint{1905.03506}.

\bibitem[{\citenamefont{Xing et~al.}(2022)\citenamefont{Xing, Liang, Liu, Sun,
  and Yang}}]{Xing:2022ijm}
\bibinfo{author}{\bibfnamefont{H.}~\bibnamefont{Xing}},
  \bibinfo{author}{\bibfnamefont{J.}~\bibnamefont{Liang}},
  \bibinfo{author}{\bibfnamefont{L.}~\bibnamefont{Liu}},
  \bibinfo{author}{\bibfnamefont{P.}~\bibnamefont{Sun}}, \bibnamefont{and}
  \bibinfo{author}{\bibfnamefont{Y.-B.} \bibnamefont{Yang}}
  (\bibinfo{year}{2022}), \eprint{2210.08555}.

\bibitem[{\citenamefont{Cheung et~al.}(2017)\citenamefont{Cheung, Thomas,
  Dudek, and Edwards}}]{Cheung:2017tnt}
\bibinfo{author}{\bibfnamefont{G.~K.~C.} \bibnamefont{Cheung}},
  \bibinfo{author}{\bibfnamefont{C.~E.} \bibnamefont{Thomas}},
  \bibinfo{author}{\bibfnamefont{J.~J.} \bibnamefont{Dudek}}, \bibnamefont{and}
  \bibinfo{author}{\bibfnamefont{R.~G.} \bibnamefont{Edwards}}
  (\bibinfo{collaboration}{Hadron Spectrum}), \bibinfo{journal}{JHEP}
  \textbf{\bibinfo{volume}{11}}, \bibinfo{pages}{033} (\bibinfo{year}{2017}),
  \eprint{1709.01417}.

\bibitem[{\citenamefont{Woss et~al.}(2020)\citenamefont{Woss, Wilson, and
  Dudek}}]{Woss:2020cmp}
\bibinfo{author}{\bibfnamefont{A.~J.} \bibnamefont{Woss}},
  \bibinfo{author}{\bibfnamefont{D.~J.} \bibnamefont{Wilson}},
  \bibnamefont{and} \bibinfo{author}{\bibfnamefont{J.~J.} \bibnamefont{Dudek}}
  (\bibinfo{collaboration}{Hadron Spectrum}), \bibinfo{journal}{Phys. Rev. D}
  \textbf{\bibinfo{volume}{101}}, \bibinfo{pages}{114505}
  (\bibinfo{year}{2020}), \eprint{2001.08474}.

\bibitem[{\citenamefont{Gayer et~al.}(2021)\citenamefont{Gayer, Lang, Ryan,
  Tims, Thomas, and Wilson}}]{Gayer:2021xzv}
\bibinfo{author}{\bibfnamefont{L.}~\bibnamefont{Gayer}},
  \bibinfo{author}{\bibfnamefont{N.}~\bibnamefont{Lang}},
  \bibinfo{author}{\bibfnamefont{S.~M.} \bibnamefont{Ryan}},
  \bibinfo{author}{\bibfnamefont{D.}~\bibnamefont{Tims}},
  \bibinfo{author}{\bibfnamefont{C.~E.} \bibnamefont{Thomas}},
  \bibnamefont{and} \bibinfo{author}{\bibfnamefont{D.~J.} \bibnamefont{Wilson}}
  (\bibinfo{collaboration}{Hadron Spectrum}), \bibinfo{journal}{JHEP}
  \textbf{\bibinfo{volume}{07}}, \bibinfo{pages}{123} (\bibinfo{year}{2021}),
  \eprint{2102.04973}.

\bibitem[{\citenamefont{Michael}(1994)}]{Michael:1993yj}
\bibinfo{author}{\bibfnamefont{C.}~\bibnamefont{Michael}},
  \bibinfo{journal}{Phys. Rev. D} \textbf{\bibinfo{volume}{49}},
  \bibinfo{pages}{2616} (\bibinfo{year}{1994}), \eprint{hep-lat/9310026}.

\bibitem[{\citenamefont{Michael and McKerrell}(1995)}]{Michael:1994sz}
\bibinfo{author}{\bibfnamefont{C.}~\bibnamefont{Michael}} \bibnamefont{and}
  \bibinfo{author}{\bibfnamefont{A.}~\bibnamefont{McKerrell}},
  \bibinfo{journal}{Phys. Rev. D} \textbf{\bibinfo{volume}{51}},
  \bibinfo{pages}{3745} (\bibinfo{year}{1995}), \eprint{hep-lat/9412087}.

\bibitem[{\citenamefont{Dowdall et~al.}(2019)\citenamefont{Dowdall, Davies,
  Horgan, Lepage, Monahan, Shigemitsu, and Wingate}}]{Dowdall:2019bea}
\bibinfo{author}{\bibfnamefont{R.~J.} \bibnamefont{Dowdall}},
  \bibinfo{author}{\bibfnamefont{C.~T.~H.} \bibnamefont{Davies}},
  \bibinfo{author}{\bibfnamefont{R.~R.} \bibnamefont{Horgan}},
  \bibinfo{author}{\bibfnamefont{G.~P.} \bibnamefont{Lepage}},
  \bibinfo{author}{\bibfnamefont{C.~J.} \bibnamefont{Monahan}},
  \bibinfo{author}{\bibfnamefont{J.}~\bibnamefont{Shigemitsu}},
  \bibnamefont{and} \bibinfo{author}{\bibfnamefont{M.}~\bibnamefont{Wingate}},
  \bibinfo{journal}{Phys. Rev. D} \textbf{\bibinfo{volume}{100}},
  \bibinfo{pages}{094508} (\bibinfo{year}{2019}), \eprint{1907.01025}.

\bibitem[{\citenamefont{Bruno and Sommer}(2023)}]{Bruno:2022mfy}
\bibinfo{author}{\bibfnamefont{M.}~\bibnamefont{Bruno}} \bibnamefont{and}
  \bibinfo{author}{\bibfnamefont{R.}~\bibnamefont{Sommer}},
  \bibinfo{journal}{Comput. Phys. Commun.} \textbf{\bibinfo{volume}{285}},
  \bibinfo{pages}{108643} (\bibinfo{year}{2023}), \eprint{2209.14188}.

\bibitem[{\citenamefont{Dudek et~al.}(2013)\citenamefont{Dudek, Edwards, and
  Thomas}}]{Dudek:2012xn}
\bibinfo{author}{\bibfnamefont{J.~J.} \bibnamefont{Dudek}},
  \bibinfo{author}{\bibfnamefont{R.~G.} \bibnamefont{Edwards}},
  \bibnamefont{and} \bibinfo{author}{\bibfnamefont{C.~E.} \bibnamefont{Thomas}}
  (\bibinfo{collaboration}{Hadron Spectrum}), \bibinfo{journal}{Phys. Rev.}
  \textbf{\bibinfo{volume}{D87}}, \bibinfo{pages}{034505}
  (\bibinfo{year}{2013}), \bibinfo{note}{[Erratum: Phys.
  Rev.D90,no.9,099902(2014)]}, \eprint{1212.0830}.

\bibitem[{\citenamefont{Wilson et~al.}(2015{\natexlab{b}})\citenamefont{Wilson,
  Briceno, Dudek, Edwards, and Thomas}}]{Wilson:2015dqa}
\bibinfo{author}{\bibfnamefont{D.~J.} \bibnamefont{Wilson}},
  \bibinfo{author}{\bibfnamefont{R.~A.} \bibnamefont{Briceno}},
  \bibinfo{author}{\bibfnamefont{J.~J.} \bibnamefont{Dudek}},
  \bibinfo{author}{\bibfnamefont{R.~G.} \bibnamefont{Edwards}},
  \bibnamefont{and} \bibinfo{author}{\bibfnamefont{C.~E.}
  \bibnamefont{Thomas}}, \bibinfo{journal}{Phys. Rev.}
  \textbf{\bibinfo{volume}{D92}}, \bibinfo{pages}{094502}
  (\bibinfo{year}{2015}{\natexlab{b}}), \eprint{1507.02599}.

\bibitem[{\citenamefont{Pelaez}(2016)}]{Pelaez:2015qba}
\bibinfo{author}{\bibfnamefont{J.~R.} \bibnamefont{Pelaez}},
  \bibinfo{journal}{Phys. Rept.} \textbf{\bibinfo{volume}{658}},
  \bibinfo{pages}{1} (\bibinfo{year}{2016}), \eprint{1510.00653}.

\bibitem[{\citenamefont{Rodas et~al.}(2023{\natexlab{a}})\citenamefont{Rodas,
  Dudek, and Edwards}}]{Rodas:2023twk}
\bibinfo{author}{\bibfnamefont{A.}~\bibnamefont{Rodas}},
  \bibinfo{author}{\bibfnamefont{J.~J.} \bibnamefont{Dudek}}, \bibnamefont{and}
  \bibinfo{author}{\bibfnamefont{R.~G.} \bibnamefont{Edwards}}
  (\bibinfo{year}{2023}{\natexlab{a}}), \eprint{2304.03762}.

\bibitem[{\citenamefont{Workman et~al.}(2022)}]{ParticleDataGroup:2022pth}
\bibinfo{author}{\bibfnamefont{R.~L.} \bibnamefont{Workman}}
  \bibnamefont{et~al.} (\bibinfo{collaboration}{Particle Data Group}),
  \bibinfo{journal}{PTEP} \textbf{\bibinfo{volume}{2022}},
  \bibinfo{pages}{083C01} (\bibinfo{year}{2022}).

\bibitem[{\citenamefont{Dudek et~al.}(2016)\citenamefont{Dudek, Edwards, and
  Wilson}}]{Dudek:2016cru}
\bibinfo{author}{\bibfnamefont{J.~J.} \bibnamefont{Dudek}},
  \bibinfo{author}{\bibfnamefont{R.~G.} \bibnamefont{Edwards}},
  \bibnamefont{and} \bibinfo{author}{\bibfnamefont{D.~J.} \bibnamefont{Wilson}}
  (\bibinfo{collaboration}{Hadron Spectrum}), \bibinfo{journal}{Phys. Rev.}
  \textbf{\bibinfo{volume}{D93}}, \bibinfo{pages}{094506}
  (\bibinfo{year}{2016}), \eprint{1602.05122}.

\bibitem[{\citenamefont{Morgan}(1992)}]{Morgan:1992ge}
\bibinfo{author}{\bibfnamefont{D.}~\bibnamefont{Morgan}},
  \bibinfo{journal}{Nucl. Phys. A} \textbf{\bibinfo{volume}{543}},
  \bibinfo{pages}{632} (\bibinfo{year}{1992}).

\bibitem[{\citenamefont{Morgan and Pennington}(1993)}]{Morgan:1993td}
\bibinfo{author}{\bibfnamefont{D.}~\bibnamefont{Morgan}} \bibnamefont{and}
  \bibinfo{author}{\bibfnamefont{M.~R.} \bibnamefont{Pennington}},
  \bibinfo{journal}{Phys. Rev. D} \textbf{\bibinfo{volume}{48}},
  \bibinfo{pages}{1185} (\bibinfo{year}{1993}).

\bibitem[{\citenamefont{Hamilton and Spearman}(1961)}]{HAMILTON1961172}
\bibinfo{author}{\bibfnamefont{J.}~\bibnamefont{Hamilton}} \bibnamefont{and}
  \bibinfo{author}{\bibfnamefont{T.}~\bibnamefont{Spearman}},
  \bibinfo{journal}{Annals of Physics} \textbf{\bibinfo{volume}{12}},
  \bibinfo{pages}{172} (\bibinfo{year}{1961}), ISSN \bibinfo{issn}{0003-4916},
  \urlprefix\url{https://www.sciencedirect.com/science/article/pii/0003491661900021}.

\bibitem[{\citenamefont{Blankenbecler et~al.}(1961)\citenamefont{Blankenbecler,
  Goldberger, MacDowell, and Treiman}}]{PhysRev.123.692}
\bibinfo{author}{\bibfnamefont{R.}~\bibnamefont{Blankenbecler}},
  \bibinfo{author}{\bibfnamefont{M.~L.} \bibnamefont{Goldberger}},
  \bibinfo{author}{\bibfnamefont{S.~W.} \bibnamefont{MacDowell}},
  \bibnamefont{and} \bibinfo{author}{\bibfnamefont{S.~B.}
  \bibnamefont{Treiman}}, \bibinfo{journal}{Phys. Rev.}
  \textbf{\bibinfo{volume}{123}}, \bibinfo{pages}{692} (\bibinfo{year}{1961}),
  \urlprefix\url{https://link.aps.org/doi/10.1103/PhysRev.123.692}.

\bibitem[{\citenamefont{Raposo and Hansen}(2023)}]{Raposo:2023nex}
\bibinfo{author}{\bibfnamefont{A.~B.~a.} \bibnamefont{Raposo}}
  \bibnamefont{and} \bibinfo{author}{\bibfnamefont{M.~T.}
  \bibnamefont{Hansen}}, \bibinfo{journal}{PoS}
  \textbf{\bibinfo{volume}{LATTICE2022}}, \bibinfo{pages}{051}
  (\bibinfo{year}{2023}), \eprint{2301.03981}.

\bibitem[{\citenamefont{Du et~al.}(2023)\citenamefont{Du, Filin, Baru, Dong,
  Epelbaum, Guo, Hanhart, Nefediev, Nieves, and Wang}}]{Du:2023hlu}
\bibinfo{author}{\bibfnamefont{M.-L.} \bibnamefont{Du}},
  \bibinfo{author}{\bibfnamefont{A.}~\bibnamefont{Filin}},
  \bibinfo{author}{\bibfnamefont{V.}~\bibnamefont{Baru}},
  \bibinfo{author}{\bibfnamefont{X.-K.} \bibnamefont{Dong}},
  \bibinfo{author}{\bibfnamefont{E.}~\bibnamefont{Epelbaum}},
  \bibinfo{author}{\bibfnamefont{F.-K.} \bibnamefont{Guo}},
  \bibinfo{author}{\bibfnamefont{C.}~\bibnamefont{Hanhart}},
  \bibinfo{author}{\bibfnamefont{A.}~\bibnamefont{Nefediev}},
  \bibinfo{author}{\bibfnamefont{J.}~\bibnamefont{Nieves}}, \bibnamefont{and}
  \bibinfo{author}{\bibfnamefont{Q.}~\bibnamefont{Wang}}
  (\bibinfo{year}{2023}), \eprint{2303.09441}.

\bibitem[{\citenamefont{Padmanath and Prelovsek}(2022)}]{Padmanath:2022cvl}
\bibinfo{author}{\bibfnamefont{M.}~\bibnamefont{Padmanath}} \bibnamefont{and}
  \bibinfo{author}{\bibfnamefont{S.}~\bibnamefont{Prelovsek}},
  \bibinfo{journal}{Phys. Rev. Lett.} \textbf{\bibinfo{volume}{129}},
  \bibinfo{pages}{032002} (\bibinfo{year}{2022}), \eprint{2202.10110}.

\bibitem[{\citenamefont{Hudspith and Mohler}(2022)}]{Hudspith:2021iqu}
\bibinfo{author}{\bibfnamefont{R.~J.} \bibnamefont{Hudspith}} \bibnamefont{and}
  \bibinfo{author}{\bibfnamefont{D.}~\bibnamefont{Mohler}},
  \bibinfo{journal}{Phys. Rev. D} \textbf{\bibinfo{volume}{106}},
  \bibinfo{pages}{034508} (\bibinfo{year}{2022}), \eprint{2112.01997}.

\bibitem[{\citenamefont{Barnes et~al.}(1997)\citenamefont{Barnes, Close, Page,
  and Swanson}}]{Barnes:1996ff}
\bibinfo{author}{\bibfnamefont{T.}~\bibnamefont{Barnes}},
  \bibinfo{author}{\bibfnamefont{F.~E.} \bibnamefont{Close}},
  \bibinfo{author}{\bibfnamefont{P.~R.} \bibnamefont{Page}}, \bibnamefont{and}
  \bibinfo{author}{\bibfnamefont{E.~S.} \bibnamefont{Swanson}},
  \bibinfo{journal}{Phys. Rev. D} \textbf{\bibinfo{volume}{55}},
  \bibinfo{pages}{4157} (\bibinfo{year}{1997}), \eprint{hep-ph/9609339}.

\bibitem[{\citenamefont{Barnes et~al.}(2005)\citenamefont{Barnes, Godfrey, and
  Swanson}}]{Barnes:2005pb}
\bibinfo{author}{\bibfnamefont{T.}~\bibnamefont{Barnes}},
  \bibinfo{author}{\bibfnamefont{S.}~\bibnamefont{Godfrey}}, \bibnamefont{and}
  \bibinfo{author}{\bibfnamefont{E.~S.} \bibnamefont{Swanson}},
  \bibinfo{journal}{Phys. Rev. D} \textbf{\bibinfo{volume}{72}},
  \bibinfo{pages}{054026} (\bibinfo{year}{2005}), \eprint{hep-ph/0505002}.

\bibitem[{\citenamefont{Albaladejo et~al.}(2015)\citenamefont{Albaladejo, Guo,
  Hidalgo-Duque, Nieves, and Valderrama}}]{Albaladejo:2015dsa}
\bibinfo{author}{\bibfnamefont{M.}~\bibnamefont{Albaladejo}},
  \bibinfo{author}{\bibfnamefont{F.~K.} \bibnamefont{Guo}},
  \bibinfo{author}{\bibfnamefont{C.}~\bibnamefont{Hidalgo-Duque}},
  \bibinfo{author}{\bibfnamefont{J.}~\bibnamefont{Nieves}}, \bibnamefont{and}
  \bibinfo{author}{\bibfnamefont{M.~P.} \bibnamefont{Valderrama}},
  \bibinfo{journal}{Eur. Phys. J. C} \textbf{\bibinfo{volume}{75}},
  \bibinfo{pages}{547} (\bibinfo{year}{2015}), \eprint{1504.00861}.

\bibitem[{\citenamefont{Baru et~al.}(2016)\citenamefont{Baru, Epelbaum, Filin,
  Hanhart, Mei\ss{}ner, and Nefediev}}]{Baru:2016iwj}
\bibinfo{author}{\bibfnamefont{V.}~\bibnamefont{Baru}},
  \bibinfo{author}{\bibfnamefont{E.}~\bibnamefont{Epelbaum}},
  \bibinfo{author}{\bibfnamefont{A.~A.} \bibnamefont{Filin}},
  \bibinfo{author}{\bibfnamefont{C.}~\bibnamefont{Hanhart}},
  \bibinfo{author}{\bibfnamefont{U.-G.} \bibnamefont{Mei\ss{}ner}},
  \bibnamefont{and} \bibinfo{author}{\bibfnamefont{A.~V.}
  \bibnamefont{Nefediev}}, \bibinfo{journal}{Phys. Lett. B}
  \textbf{\bibinfo{volume}{763}}, \bibinfo{pages}{20} (\bibinfo{year}{2016}),
  \eprint{1605.09649}.

\bibitem[{\citenamefont{Maiani et~al.}(2005)\citenamefont{Maiani, Piccinini,
  Polosa, and Riquer}}]{Maiani:2004vq}
\bibinfo{author}{\bibfnamefont{L.}~\bibnamefont{Maiani}},
  \bibinfo{author}{\bibfnamefont{F.}~\bibnamefont{Piccinini}},
  \bibinfo{author}{\bibfnamefont{A.~D.} \bibnamefont{Polosa}},
  \bibnamefont{and} \bibinfo{author}{\bibfnamefont{V.}~\bibnamefont{Riquer}},
  \bibinfo{journal}{Phys. Rev. D} \textbf{\bibinfo{volume}{71}},
  \bibinfo{pages}{014028} (\bibinfo{year}{2005}), \eprint{hep-ph/0412098}.

\bibitem[{\citenamefont{Maiani et~al.}(2014)\citenamefont{Maiani, Piccinini,
  Polosa, and Riquer}}]{Maiani:2014aja}
\bibinfo{author}{\bibfnamefont{L.}~\bibnamefont{Maiani}},
  \bibinfo{author}{\bibfnamefont{F.}~\bibnamefont{Piccinini}},
  \bibinfo{author}{\bibfnamefont{A.~D.} \bibnamefont{Polosa}},
  \bibnamefont{and} \bibinfo{author}{\bibfnamefont{V.}~\bibnamefont{Riquer}},
  \bibinfo{journal}{Phys. Rev. D} \textbf{\bibinfo{volume}{89}},
  \bibinfo{pages}{114010} (\bibinfo{year}{2014}), \eprint{1405.1551}.

\bibitem[{\citenamefont{Esposito et~al.}(2022)\citenamefont{Esposito, Maiani,
  Pilloni, Polosa, and Riquer}}]{Esposito:2021vhu}
\bibinfo{author}{\bibfnamefont{A.}~\bibnamefont{Esposito}},
  \bibinfo{author}{\bibfnamefont{L.}~\bibnamefont{Maiani}},
  \bibinfo{author}{\bibfnamefont{A.}~\bibnamefont{Pilloni}},
  \bibinfo{author}{\bibfnamefont{A.~D.} \bibnamefont{Polosa}},
  \bibnamefont{and} \bibinfo{author}{\bibfnamefont{V.}~\bibnamefont{Riquer}},
  \bibinfo{journal}{Phys. Rev. D} \textbf{\bibinfo{volume}{105}},
  \bibinfo{pages}{L031503} (\bibinfo{year}{2022}), \eprint{2108.11413}.

\bibitem[{\citenamefont{Nebreda and Pelaez.}(2010)}]{Nebreda:2010wv}
\bibinfo{author}{\bibfnamefont{J.}~\bibnamefont{Nebreda}} \bibnamefont{and}
  \bibinfo{author}{\bibfnamefont{J.~R.} \bibnamefont{Pelaez.}},
  \bibinfo{journal}{Phys. Rev.} \textbf{\bibinfo{volume}{D81}},
  \bibinfo{pages}{054035} (\bibinfo{year}{2010}), \eprint{1001.5237}.

\bibitem[{\citenamefont{Bolton et~al.}(2016)\citenamefont{Bolton, Briceno, and
  Wilson}}]{Bolton:2015psa}
\bibinfo{author}{\bibfnamefont{D.~R.} \bibnamefont{Bolton}},
  \bibinfo{author}{\bibfnamefont{R.~A.} \bibnamefont{Briceno}},
  \bibnamefont{and} \bibinfo{author}{\bibfnamefont{D.~J.}
  \bibnamefont{Wilson}}, \bibinfo{journal}{Phys. Lett. B}
  \textbf{\bibinfo{volume}{757}}, \bibinfo{pages}{50} (\bibinfo{year}{2016}),
  \eprint{1507.07928}.

\bibitem[{\citenamefont{Wilson et~al.}(2019)\citenamefont{Wilson, Briceno,
  Dudek, Edwards, and Thomas}}]{Wilson:2019wfr}
\bibinfo{author}{\bibfnamefont{D.~J.} \bibnamefont{Wilson}},
  \bibinfo{author}{\bibfnamefont{R.~A.} \bibnamefont{Briceno}},
  \bibinfo{author}{\bibfnamefont{J.~J.} \bibnamefont{Dudek}},
  \bibinfo{author}{\bibfnamefont{R.~G.} \bibnamefont{Edwards}},
  \bibnamefont{and} \bibinfo{author}{\bibfnamefont{C.~E.}
  \bibnamefont{Thomas}}, \bibinfo{journal}{Phys. Rev. Lett.}
  \textbf{\bibinfo{volume}{123}}, \bibinfo{pages}{042002}
  (\bibinfo{year}{2019}), \eprint{1904.03188}.

\bibitem[{\citenamefont{Molina and Ruiz~de Elvira}(2020)}]{Molina:2020qpw}
\bibinfo{author}{\bibfnamefont{R.}~\bibnamefont{Molina}} \bibnamefont{and}
  \bibinfo{author}{\bibfnamefont{J.}~\bibnamefont{Ruiz~de Elvira}},
  \bibinfo{journal}{JHEP} \textbf{\bibinfo{volume}{11}}, \bibinfo{pages}{017}
  (\bibinfo{year}{2020}), \eprint{2005.13584}.

\bibitem[{\citenamefont{Rodas et~al.}(2023{\natexlab{b}})\citenamefont{Rodas,
  Dudek, and Edwards}}]{Rodas:2023gma}
\bibinfo{author}{\bibfnamefont{A.}~\bibnamefont{Rodas}},
  \bibinfo{author}{\bibfnamefont{J.~J.} \bibnamefont{Dudek}}, \bibnamefont{and}
  \bibinfo{author}{\bibfnamefont{R.~G.} \bibnamefont{Edwards}}
  (\bibinfo{collaboration}{Hadron Spectrum}), \bibinfo{journal}{Phys. Rev. D}
  \textbf{\bibinfo{volume}{108}}, \bibinfo{pages}{034513}
  (\bibinfo{year}{2023}{\natexlab{b}}), \eprint{2303.10701}.

\bibitem[{\citenamefont{Johnson and Dudek}(2021)}]{Johnson:2020ilc}
\bibinfo{author}{\bibfnamefont{C.~T.} \bibnamefont{Johnson}} \bibnamefont{and}
  \bibinfo{author}{\bibfnamefont{J.~J.} \bibnamefont{Dudek}}
  (\bibinfo{collaboration}{Hadron Spectrum}), \bibinfo{journal}{Phys. Rev. D}
  \textbf{\bibinfo{volume}{103}}, \bibinfo{pages}{074502}
  (\bibinfo{year}{2021}), \eprint{2012.00518}.

\bibitem[{\citenamefont{Wang et~al.}(2022)}]{Belle:2021nuv}
\bibinfo{author}{\bibfnamefont{X.~L.} \bibnamefont{Wang}} \bibnamefont{et~al.}
  (\bibinfo{collaboration}{Belle}), \bibinfo{journal}{Phys. Rev. D}
  \textbf{\bibinfo{volume}{105}}, \bibinfo{pages}{112011}
  (\bibinfo{year}{2022}), \eprint{2105.06605}.

\bibitem[{\citenamefont{Pennington and Wilson}(2007)}]{Pennington:2007xr}
\bibinfo{author}{\bibfnamefont{M.~R.} \bibnamefont{Pennington}}
  \bibnamefont{and} \bibinfo{author}{\bibfnamefont{D.~J.}
  \bibnamefont{Wilson}}, \bibinfo{journal}{Phys. Rev. D}
  \textbf{\bibinfo{volume}{76}}, \bibinfo{pages}{077502}
  (\bibinfo{year}{2007}), \eprint{0704.3384}.

\bibitem[{\citenamefont{Guo and Meissner}(2012)}]{Guo:2012tv}
\bibinfo{author}{\bibfnamefont{F.-K.} \bibnamefont{Guo}} \bibnamefont{and}
  \bibinfo{author}{\bibfnamefont{U.-G.} \bibnamefont{Meissner}},
  \bibinfo{journal}{Phys. Rev. D} \textbf{\bibinfo{volume}{86}},
  \bibinfo{pages}{091501} (\bibinfo{year}{2012}), \eprint{1208.1134}.

\bibitem[{\citenamefont{Wang et~al.}(2021{\natexlab{b}})\citenamefont{Wang,
  Liang, and Oset}}]{Wang:2019evy}
\bibinfo{author}{\bibfnamefont{E.}~\bibnamefont{Wang}},
  \bibinfo{author}{\bibfnamefont{W.-H.} \bibnamefont{Liang}}, \bibnamefont{and}
  \bibinfo{author}{\bibfnamefont{E.}~\bibnamefont{Oset}},
  \bibinfo{journal}{Eur. Phys. J. A} \textbf{\bibinfo{volume}{57}},
  \bibinfo{pages}{38} (\bibinfo{year}{2021}{\natexlab{b}}),
  \eprint{1902.06461}.

\bibitem[{\citenamefont{Dong et~al.}(2021)\citenamefont{Dong, Guo, and
  Zou}}]{Dong:2021bvy}
\bibinfo{author}{\bibfnamefont{X.-K.} \bibnamefont{Dong}},
  \bibinfo{author}{\bibfnamefont{F.-K.} \bibnamefont{Guo}}, \bibnamefont{and}
  \bibinfo{author}{\bibfnamefont{B.-S.} \bibnamefont{Zou}},
  \bibinfo{journal}{Commun. Theor. Phys.} \textbf{\bibinfo{volume}{73}},
  \bibinfo{pages}{125201} (\bibinfo{year}{2021}), \eprint{2108.02673}.

\bibitem[{\citenamefont{Edwards and Joo}(2005)}]{Edwards:2004sx}
\bibinfo{author}{\bibfnamefont{R.~G.} \bibnamefont{Edwards}} \bibnamefont{and}
  \bibinfo{author}{\bibfnamefont{B.}~\bibnamefont{Joo}},
  \bibinfo{journal}{Nucl. Phys. Proc. Suppl.} \textbf{\bibinfo{volume}{140}},
  \bibinfo{pages}{832} (\bibinfo{year}{2005}), \bibinfo{note}{[,832(2004)]},
  \eprint{hep-lat/0409003}.

\bibitem[{\citenamefont{Clark et~al.}(2010)\citenamefont{Clark, Babich, Barros,
  Brower, and Rebbi}}]{Clark:2009wm}
\bibinfo{author}{\bibfnamefont{M.~A.} \bibnamefont{Clark}},
  \bibinfo{author}{\bibfnamefont{R.}~\bibnamefont{Babich}},
  \bibinfo{author}{\bibfnamefont{K.}~\bibnamefont{Barros}},
  \bibinfo{author}{\bibfnamefont{R.~C.} \bibnamefont{Brower}},
  \bibnamefont{and} \bibinfo{author}{\bibfnamefont{C.}~\bibnamefont{Rebbi}},
  \bibinfo{journal}{Comput. Phys. Commun.} \textbf{\bibinfo{volume}{181}},
  \bibinfo{pages}{1517} (\bibinfo{year}{2010}), \eprint{0911.3191}.

\bibitem[{\citenamefont{Babich et~al.}(2010{\natexlab{a}})\citenamefont{Babich,
  Clark, and Joo}}]{Babich:2010mu}
\bibinfo{author}{\bibfnamefont{R.}~\bibnamefont{Babich}},
  \bibinfo{author}{\bibfnamefont{M.~A.} \bibnamefont{Clark}}, \bibnamefont{and}
  \bibinfo{author}{\bibfnamefont{B.}~\bibnamefont{Joo}}, in
  \emph{\bibinfo{booktitle}{{SC 10 (Supercomputing 2010) New Orleans,
  Louisiana, November 13-19, 2010}}} (\bibinfo{year}{2010}{\natexlab{a}}),
  \eprint{1011.0024},
  \urlprefix\url{http://www1.jlab.org/Ul/publications/view_pub.cfm?pub_id=10186}.

\bibitem[{\citenamefont{Clark et~al.}(2016)\citenamefont{Clark, Joo,
  Strelchenko, Cheng, Gambhir, and Brower}}]{Clark:SC2016}
\bibinfo{author}{\bibfnamefont{K.}~\bibnamefont{Clark}},
  \bibinfo{author}{\bibfnamefont{B.}~\bibnamefont{Joo}},
  \bibinfo{author}{\bibfnamefont{A.}~\bibnamefont{Strelchenko}},
  \bibinfo{author}{\bibfnamefont{M.}~\bibnamefont{Cheng}},
  \bibinfo{author}{\bibfnamefont{A.}~\bibnamefont{Gambhir}}, \bibnamefont{and}
  \bibinfo{author}{\bibfnamefont{R.}~\bibnamefont{Brower}}, in
  \emph{\bibinfo{booktitle}{{Proceedings of SC 16 (Supercomputing 2016) Salt
  Lake City, Utah, November 2016}}} (\bibinfo{year}{2016}).

\bibitem[{\citenamefont{Jo\'o et~al.}(2013)\citenamefont{Jo\'o, Kalamkar,
  Vaidyanathan, Smelyanskiy, Pamnany, Lee, Dubey, and Watson}}]{ISC13Phi}
\bibinfo{author}{\bibfnamefont{B.}~\bibnamefont{Jo\'o}},
  \bibinfo{author}{\bibfnamefont{D.}~\bibnamefont{Kalamkar}},
  \bibinfo{author}{\bibfnamefont{K.}~\bibnamefont{Vaidyanathan}},
  \bibinfo{author}{\bibfnamefont{M.}~\bibnamefont{Smelyanskiy}},
  \bibinfo{author}{\bibfnamefont{K.}~\bibnamefont{Pamnany}},
  \bibinfo{author}{\bibfnamefont{V.}~\bibnamefont{Lee}},
  \bibinfo{author}{\bibfnamefont{P.}~\bibnamefont{Dubey}}, \bibnamefont{and}
  \bibinfo{author}{\bibfnamefont{W.}~\bibnamefont{Watson}}, in
  \emph{\bibinfo{booktitle}{Supercomputing}}, edited by
  \bibinfo{editor}{\bibfnamefont{J.}~\bibnamefont{Kunkel}},
  \bibinfo{editor}{\bibfnamefont{T.}~\bibnamefont{Ludwig}}, \bibnamefont{and}
  \bibinfo{editor}{\bibfnamefont{H.}~\bibnamefont{Meuer}}
  (\bibinfo{publisher}{Springer Berlin Heidelberg}, \bibinfo{year}{2013}), vol.
  \bibinfo{volume}{7905} of \emph{\bibinfo{series}{Lecture Notes in Computer
  Science}}, pp. \bibinfo{pages}{40--54}, ISBN
  \bibinfo{isbn}{978-3-642-38749-4},
  \urlprefix\url{http://dx.doi.org/10.1007/978-3-642-38750-0_4}.

\bibitem[{\citenamefont{Jo\'o}()}]{MGProtoDownload}
\bibinfo{author}{\bibfnamefont{B.}~\bibnamefont{Jo\'o}},
  \emph{\bibinfo{title}{{MG\_PROTO: A Multigrid Library for QCD}}},
  \bibinfo{howpublished}{\url{https://github.com/JeffersonLab/mg_proto/}}.

\bibitem[{\citenamefont{Osborn et~al.}(2010)\citenamefont{Osborn, Babich,
  Brannick, Brower, Clark, Cohen, and Rebbi}}]{Osborn:2010mb}
\bibinfo{author}{\bibfnamefont{J.~C.} \bibnamefont{Osborn}},
  \bibinfo{author}{\bibfnamefont{R.}~\bibnamefont{Babich}},
  \bibinfo{author}{\bibfnamefont{J.}~\bibnamefont{Brannick}},
  \bibinfo{author}{\bibfnamefont{R.~C.} \bibnamefont{Brower}},
  \bibinfo{author}{\bibfnamefont{M.~A.} \bibnamefont{Clark}},
  \bibinfo{author}{\bibfnamefont{S.~D.} \bibnamefont{Cohen}}, \bibnamefont{and}
  \bibinfo{author}{\bibfnamefont{C.}~\bibnamefont{Rebbi}},
  \bibinfo{journal}{PoS} \textbf{\bibinfo{volume}{LATTICE2010}},
  \bibinfo{pages}{037} (\bibinfo{year}{2010}), \eprint{1011.2775}.

\bibitem[{\citenamefont{Babich et~al.}(2010{\natexlab{b}})\citenamefont{Babich,
  Brannick, Brower, Clark, Manteuffel, McCormick, Osborn, and
  Rebbi}}]{Babich:2010qb}
\bibinfo{author}{\bibfnamefont{R.}~\bibnamefont{Babich}},
  \bibinfo{author}{\bibfnamefont{J.}~\bibnamefont{Brannick}},
  \bibinfo{author}{\bibfnamefont{R.~C.} \bibnamefont{Brower}},
  \bibinfo{author}{\bibfnamefont{M.~A.} \bibnamefont{Clark}},
  \bibinfo{author}{\bibfnamefont{T.~A.} \bibnamefont{Manteuffel}},
  \bibinfo{author}{\bibfnamefont{S.~F.} \bibnamefont{McCormick}},
  \bibinfo{author}{\bibfnamefont{J.~C.} \bibnamefont{Osborn}},
  \bibnamefont{and} \bibinfo{author}{\bibfnamefont{C.}~\bibnamefont{Rebbi}},
  \bibinfo{journal}{Phys. Rev. Lett.} \textbf{\bibinfo{volume}{105}},
  \bibinfo{pages}{201602} (\bibinfo{year}{2010}{\natexlab{b}}),
  \eprint{1005.3043}.

\bibitem[{\citenamefont{Chen et~al.}(2023)\citenamefont{Chen, Edwards, and
  Mao}}]{Chen:2023zyy}
\bibinfo{author}{\bibfnamefont{J.}~\bibnamefont{Chen}},
  \bibinfo{author}{\bibfnamefont{R.~G.} \bibnamefont{Edwards}},
  \bibnamefont{and} \bibinfo{author}{\bibfnamefont{W.}~\bibnamefont{Mao}}, in
  \emph{\bibinfo{booktitle}{{Platform for Advanced Scientific Computing}}}
  (\bibinfo{year}{2023}).

\bibitem[{\citenamefont{Ledoit and Wolf}(2004)}]{LEDOIT2004365}
\bibinfo{author}{\bibfnamefont{O.}~\bibnamefont{Ledoit}} \bibnamefont{and}
  \bibinfo{author}{\bibfnamefont{M.}~\bibnamefont{Wolf}},
  \bibinfo{journal}{Journal of Multivariate Analysis}
  \textbf{\bibinfo{volume}{88}}, \bibinfo{pages}{365} (\bibinfo{year}{2004}),
  ISSN \bibinfo{issn}{0047-259X},
  \urlprefix\url{https://www.sciencedirect.com/science/article/pii/S0047259X03000964}.

\bibitem[{\citenamefont{Rinaldi et~al.}(2019)\citenamefont{Rinaldi, Syritsyn,
  Wagman, Buchoff, Schroeder, and Wasem}}]{Rinaldi:2018osy}
\bibinfo{author}{\bibfnamefont{E.}~\bibnamefont{Rinaldi}},
  \bibinfo{author}{\bibfnamefont{S.}~\bibnamefont{Syritsyn}},
  \bibinfo{author}{\bibfnamefont{M.~L.} \bibnamefont{Wagman}},
  \bibinfo{author}{\bibfnamefont{M.~I.} \bibnamefont{Buchoff}},
  \bibinfo{author}{\bibfnamefont{C.}~\bibnamefont{Schroeder}},
  \bibnamefont{and} \bibinfo{author}{\bibfnamefont{J.}~\bibnamefont{Wasem}},
  \bibinfo{journal}{Phys. Rev. Lett.} \textbf{\bibinfo{volume}{122}},
  \bibinfo{pages}{162001} (\bibinfo{year}{2019}), \eprint{1809.00246}.

\end{thebibliography}
\bibliographystyle{apsrev}

\end{document}